\RequirePackage{fix-cm}
\documentclass[twocolumn]{svjour3}          
\smartqed  
\usepackage{rotating,graphicx,color,mathptmx,longtable,amssymb}
\usepackage{latexsym}
\usepackage{hyperref}
\usepackage[normalem]{ulem}
\usepackage{amsmath}    
\usepackage{amssymb}    

\usepackage{chngcntr}
\counterwithout{equation}{section}
\counterwithout{figure}{section}

\newcommand{\mmax}{M_G^{\mathit{max}}}
  
\journalname{EPJA}
\begin{document}

\title{Equations of state for hot neutron stars - II. The role of exotic particle degrees of freedom}

\titlerunning{Finite temperature equations of states with exotica}        

\author{Adriana R. Raduta}

\authorrunning{Raduta} 

\institute{Adriana R. Raduta \at
  National Institute for Physics and Nuclear Engineering (IFIN-HH),
  RO-077125 Bucharest, Romania \\
  \email{araduta@nipne.ro}  
}

\date{Received: date / Accepted: date}

\maketitle

\begin{abstract}
Explosive astrophysical systems - such as supernovae or compact star binary
mergers - provide conditions where exotic degrees of freedom can be populated.
Within the covariant density functional theory of nuclear matter we build
several general purpose equations of state which, in addition to the baryonic
octet, account for $\Delta(1232)$ resonance states. The thermodynamic
stability of $\Delta$-admixed nuclear matter is investigated in the limiting
case of vanishing temperature for charge fractions $Y_Q=0.01$ and $Y_Q=0.5$ and
wide ranges of the coupling constants to the scalar and vector mesonic fields.
General purpose equation of state models with exotica
presently available on the \textsc{CompOSE} database are further reviewed;
for a selection of them we then investigate
thermal properties for thermodynamic conditions relevant for core-collapse
supernovae and binary neutron star mergers. Modifications induced by
hyperons, $\Delta(1232)$, $K^-$, pions and quarks are discussed.
\end{abstract}

\section{Introduction}
\label{sec:Intro}

In astrophysical events like core-collapse supernovae (CCSN)
\cite{Janka_PhysRep_2007,Mezzacappa2015,SRO_PRC_2017,Connor2018ApJ,Burrows2020MNRAS} and
binary neutron star (BNS) mergers \cite{Shibata_11,Rosswog_15,Baiotti_2017,Endrizzi_PRD_2018,Ruiz2020,Prakash_PRD_2021,Most_2022}
as well as in the evolution of proto-neutron
stars (PNS) \cite{Pons_ApJ_1999,Pascal_MNRAS_2022} and formation of stellar black holes (BH)
\cite{Sumiyoshi_2007,Fischer_2009,OConnor_2011,hempel12}
high density and high temperature states
of matter are populated. Under these circumstances the appearance
of non-nucleonic degrees of freedom (d.o.f.) - such as hyperons, meson
condensates and quarks - is highly probable. From the energetic view point
these exotic phases are the natural consequence of the Pauli principle
and are expected to nucleate whenever the chemical potential of neutrons
exceeds a certain threshold.
Limited current information on the equation of state (EoS) does nevertheless
not allow to ascertain the occurrence of any of these species and even less
to identify the thermodynamic conditions and astrophysical environments
where one species or another will dominate over nucleonic matter.

Signatures of exotic matter are typically sought after by comparing
results of numerical simulations obtained by accounting for
various extra particle d.o.f. with those obtained when only nucleons are considered.
As such hyperons and hadrons to quark phase transitions have been found
to affect the dynamics, neutrino signals and gravitational wave (GW) emissions.

Sekiguchi et al.~\cite{Sekiguchi_PRL_2011} has shown that the onset of
$\Lambda$-hyperons
modifies the GWs emitted by the hypermassive neutron star (HMNS) formed
after BNS as well as the surrounding torus. GW were found to be modified in two
respects. Firstly, the amplitude of quasiperiodic GW at the black hole
formation is damped. Second, the characteristic GW frequency increases with time.
According to Radice et al.~\cite{Radice_ApJL_2017} the production of hyperons makes waveforms
louder and alter the amplitude modulation and phase evolution;
other modifications concern the compactness and binding energy of the merger remnant,
higher when hyperons are present.
Bauswein et al.~\cite{Bauswein_PRL_2019,Blacker_PRD_2020} showed that hadron-quark phase transitions occurring
during the post-merger phase modify the relation between the frequency
of dominant gravitational-waves and the tidal deformability during the inspiral.
Patterns of inspiral and postmerger GW signals have been confronted by Most et al. \cite{Most_PRL_2019}, who shown that the latter has the potential to bare clear signatures of phase transitions. 
Weith et al.~\cite{Weih_PRL_2020} demonstrated that the postmerger GW signal exhibits two
distinct fundamental frequencies before and after the phase transition.
More precisely, if the collapse into a BH is triggered by the phase transition,
the frequencies of GW emitted by the HMHS increase as the collapse proceeds;
if, at variance, the phase transition develops after the merger and a metastable object
with a quark-matter core is formed, the  transition to a signal at higher
frequencies is more smooth and the largest values are smaller than in the first scenario.
Modification of post-merger GW frequencies by the quark deconfinement phase transition is confirmed
also by Prakash et al. \cite{Prakash_PRD_2021} who have additionally studied modifications of
dynamical ejecta, remnant accretion disk masses, $r$-process nucleosynthesis yields and
electromagnetic counterparts.

Effects of pions, hyperons and hadron-quark phase transition in stellar core collapse have been
identified as well.
According to Sagert et al. \cite{Sagert_PRL_2009}, phase transitions occurring during the
early postbounce evolution might produce a second shock wave. If this is strong enough a delayed
supernova explosion is triggered and an additional neutrino-peak generated.
Zha et al. \cite{Zha_PRL_2020} have proved that if, due to the QCD phase transition,
the PNS collapses following the core-collapse, a loud burst of GW is formed. Its properties -
amplitude, frequency range, duration, energy - are different than those of normally found
postbounce GW.
According to Lin et al. \cite{Lin_2022} the neutrino signal emitted by these
failed CCSN is oscillatory.
Effects of pions and $\Lambda$-hyperons, including a possible strangeness-driven phase transition
on the evolution of core collapse and formation of stellar BH have been considered in
Peres et al.~\cite{Peres_PRD_2013}.
It was found that the EoS softening results in reduced (increased) values of central density (temperature)
at bounce; increased radii of the homologous core and PNS; steeper increase of central density during
postbounce; faster collapse into BH. The magnitude of modifications increases with abundances of
exotic species. Phase transitions additionally induce core oscillations.


The consequences of phase transitions, non-monotonic density-dependence
of the sound speed and numerical accidents in the surface of thermodynamic potentials have
been discussed by Aloy et al.~\cite{Aloy_MNRAS_2019}. Among their results we remind the
supersonic behavior of the infalling matter in connection with a phase transition.

Numerical simulations of these phenomena require detailed information on composition, thermodynamic
and, preferably also, microscopic properties of matter over wide domains of baryonic density
$10^{-14}~{\rm fm}^{-3} \leq n_B \leq 1.5~{\rm fm}^{-3}$,
electron fraction $0 \leq Y_e=n_e/n_B \leq 0.6$ and temperature $0 \leq T \leq 100~{\rm MeV}$.
The first so-called general purpose EoS tables for astrophysical use have been made available
by Lattimer and Swesty~\cite{LS_NPA_1991} and Shen et al. \cite{STOS_PTP_1998,STOS_NPA_1998} and
accounted only for nucleonic d.o.f. For almost two decades they were the only ones to exist.
As such more "finite-temperature" EoS have been built heuristically by the so-called $\Gamma$-law, which
consists in supplementing cold EoS with ideal-gas thermal contributions - a method still in use, see
Sec. \ref{sec:FiniteT}.
Starting with 2010 the situation improved significantly and today almost one hundred such EoS exist.
Out of them 36 account also for one or several exotic particles.
These EoS tables have been obtained within different theoretical frameworks;
rely on various baryon-baryon interactions;
implement different constraints derived from nuclear physics,
astrophysics and {\em ab initio} calculations;
for a review, see \cite{Oertel_RMP_2017,Burgio_PPNP_2021}.
Though the collection of existing models is still far from exhausting the huge parameter space
associated with the uncertainties in effective baryonic interactions at high densities
and isospin asymmetries, it allowed to identify how particle abundances depend on
thermodynamic conditions; the extend in which exotic d.o.f.
impact on thermodynamic state variables or derived quantities like
adiabatic index or speed of sound;
whether correlations exist with EoS of nuclear matter;
investigate properties of isentropic PNS and the relation with their
nucleonic counterparts~\cite{Oertel_PRC_2012,Peres_PRD_2013,Oertel_EPJA_2016,Malfatti_PRC_2019,Raduta_MNRAS_2020,Khadkikar_PRC_2021,Wei_PRC_2021,Sedrakian_EPJA_2022}.

The first aim of this work is to present three general purpose EoS models that account for the baryonic octet and
$\Delta(1232)$-resonances. They are derived in the framework of the covariant density functional theory (CDFT),
are available on the \textsc{CompOSE} site~\footnote{\url{https://compose.obspm.fr/}}~\cite{Typel_2013,Typel_2022}
and - to our knowledge - are the first publicly
available EoS models which account for $\Delta$s.
Accounting for $\Delta$s, in principle, one may face first order phase transition(s) in thermodynamic
conditions relevant for astrophysics \cite{Raduta_PLB_2021}. Our EoS models are built by giving to the coupling
constants to meson fields values in accord with present constraints. When the contribution of the neutralizing
electron gas is accounted for none of these models manifests instabilities. Though, when only baryonic matter
is considered, one of the models shows $\Delta$-driven instabilities.
 
The second motivation is to provide a better understanding of the finite-$T$ behavior of EoS models with exotic d.o.f.
To this end we review general purpose EoS tables with exotica presently available on
the \textsc{CompOSE} site.
Then, for a collection of EoS models which allow for different exotic d.o.f. and have the property that
general purpose EoS tables for purely nucleonic matter exist as well a comparative study is performed
for a series of thermal properties. These two aspects make present work be the follow-up of our previous
paper devoted to EoS with nucleonic d.o.f. \cite{Raduta_EPJA_2021}, called in the following Paper I.

The paper in organized as follows. General purpose EoS with exotica presently available on the \textsc{CompOSE} site
are reviewed in Sec. \ref{sec:Models}.
Sec. \ref{sec:CDFT} offers an overview of CDFT models with density-dependent couplings.
The role of hyperons and $\Delta$-resonances at finite-T is analyzed in Sec. \ref{sec:ND} by considering
simplified mixtures of nucleons and
the $\Lambda$-hyperons and, respectively, $\Delta$s. 
Thermodynamic stability of cold $\Delta$-admixed nucleonic matter at $T=0$ is investigated in Sec. \ref{sec:ND}. 
Thermal properties of a large collection of models, including the newly proposed ones, are addressed in
Sec. \ref{sec:FiniteT}. In order to better highlight the role of various particle d.o.f.
or the way in which the
hadron to quark phase transition is implemented 
none of the currently used constraints on parameters of nuclear matter at saturation or properties
of NS is used as a filter.

Throughout this paper we use the natural units with $c=\hslash=k_B=G=1$.

\section{General purpose EoS models with exotic particle degrees of freedom on \textsc{CompOSE} }

\label{sec:Models}

In this section we provide an overview 
of models with exotic particle degrees of freedom for which
general purpose EoS tables are available on the
\textsc{CompOSE} site.
General purpose tables cover large ranges of baryon number densities $n_B$,
temperatures $T$, and electron fractions $Y_e$
and are ready to use in numerical simulations of CCSN, PNS and BNS mergers.
Contributions of electron-positron and photon gases, treated as ideal
Fermi and Bose gases, are included in all tables assuming that
the net electron fraction equals the baryon charge fraction, $Y_e=Y_Q$, and
that the different sectors are in thermal equilibrium. For some models
tables corresponding to pure baryonic matter are also provided.

The standard format for the name of a particular EoS table takes the
form XXX(YYY)zzz. XXX thereby indicates the initials of the 
authors in the original publication(s) proposing the corresponding
model. 
YYY represents the name of the interaction and provides information
on the effective nucleon potential and non-nucleonic particle degrees of freedom.
zzz offers extra information on coupling constants in the exotic sector, if tables
are provided for more than one XXX(YYY) model.

\begin{table*}
  \caption{List of general purpose EoS tables with exotic degrees of freedom, available on \textsc{Compose}.
    For each EoS model we provide information on: considered degrees of freedom;
    maximum mass of cold $\beta$-equilibrated NS ($\mmax$);
    radius of canonical $1.4M_{\odot}$ NS ($R_{14}$);
    radius of a $2.072M_{\odot}$ NS ($R_{2.072}$);
    limits of combined tidal deformability $\tilde \Lambda=16\left[
      \left( M_1+12 M_2\right) M_1^4 \Lambda_1 + \left( M_2+12
      M_1\right) M_2^4 \Lambda_2 \right]/13 \left(M_1 +M_2\right)^5$
    corresponding to the GW170817 event with an estimated total mass
    $M_T=2.73^{+0.04}_{-0.01} M_{\odot}$ and a mass ratio range $0.73\leq q=M_2/M_1 \leq 1$.
    Present astrophysical constraints on EoS regard: i) the lower limit of maximum gravitational mass
    \cite{Demorest_Nature_2010,Antoniadis2013,Arzoumanian_ApJSS_2018,Cromartie2020,Fonseca_2021};
    ii) radii of canonical mass NS \cite{Miller_2019,Riley_2019};
    iii) radii of a massive NS \cite{Miller_may2021,Riley_may2021};
    iv) range for the tidal deformability obtained from the GW170817 event \cite{Abbott_PRL119_161101,Abbott_ApJ2017ApJ_L12,Abbott_2019}.
    As in Paper I models in tension with constraint i) $M \geq 2.01 - 0.04 M_{\odot}$ \cite{Antoniadis2013}
    are marked in bold.
    Values outside the ranges
    ii) $11.80 \leq R(1.4M_{\odot}) \leq 13.10$ km \cite{Miller_may2021},
    iii) $11.41~{\rm km} \leq R(2.072M_{\odot})\leq 13.69~{\rm km}$ \cite{Riley_may2021},
    iv) $110 \leq \tilde \Lambda \leq 800$ \cite{Abbott_2019}
    are also marked in bold.
    For $\mmax$ and $R_{14}$ provided are the values on \textsc{CompOSE}
    or the original publications.
    As in Paper I $R_{2.072}$, $\tilde \Lambda(q=0.73)$ and $\tilde \Lambda(q=1)$ are calculated by using
    for the crust the EoS models by \cite{NV_1973} and \cite{HDZ_1989}.
    n.a. (not available) means that quantities could not be calculated from the table, the reason being
    that, for baryonic densities exceeding a certain value, $Y_e$ of $\beta$-equilibrated matter
    is lower than the lowest value in the table.
    Other notations are: $q$ stands for $u$, $d$, $s$ quarks;
    $\Lambda$ denotes the $\Lambda$-hyperon;
    $\Delta$ is the $\Delta(1232)$ resonance;
    $Y$ generically denotes the $\Lambda$, $\Sigma$ and $\Xi$ hyperons;
    $\pi$ and $K$ respectively stand for pions and kaons.
  }
  \label{tab:models} 
\begin{tabular}{lccccccl}
\hline\noalign{\smallskip}
model &  non-nucleonic & $\mmax$ & $R_{14}$ & $R_{2.072}$ & $\tilde \Lambda$  & $\tilde \Lambda$ & Ref.  \\
& d.o.f & ($M_{\odot}$) & (km)  & (km) & q=0.73 & q=1  \\
\noalign{\smallskip}\hline\noalign{\smallskip}

{\bf GROM(LS220L) no low densities} & $\Lambda$ & {\bf 1.91} & 12.4 & - & 576 & 498 & \cite{LS_NPA_1991,Oertel_PRC_2012}\\
{\bf GROM(LS220L) with low densities} & $\Lambda$ & {\bf 1.91} & 12.4 & - & 576 & 498 & \cite{LS_NPA_1991,Oertel_PRC_2012}\\
{\bf STOS(TM1L)} & $\Lambda$ & {\bf 1.90} & {\bf 14.4} & - & {\bf 1366} & {\bf 1283} & \cite{STOS_PTP_1998,STOS_NPA_1998,STOS_ApJSS_2011}  \\
{\bf IOTSY(TM1Y-30)} & Y & {\bf 1.63} & {\bf 14.3} & - & - & {\bf 1258}  & \cite{Ishizuka_JPG_2008} \\
{\bf IOTSY(TM1Y0)} & Y & {\bf 1.64} & {\bf 14.3} & - & {\bf 1361} & {\bf 1286} & \cite{Ishizuka_JPG_2008} \\
{\bf IOTSY(TM1Y30)} & Y & {\bf 1.64} & {\bf 14.3} & - & {\bf 1362} & {\bf 1286} & \cite{Ishizuka_JPG_2008} \\
{\bf IOTSY(TM1Y90)} & Y & {\bf 1.64} & {\bf 14.3} & - & {\bf 1362} & {\bf 1286} & \cite{Ishizuka_JPG_2008} \\
{\bf IOTSY(TM1Ypi-30)} & $Y, \pi$ & {\bf 1.66} & {\bf 13.6} & - & {\bf 844} & 781 & \cite{Ishizuka_JPG_2008} \\
{\bf IOTSY(TM1Ypi0)} & $Y, \pi$ & {\bf 1.66} & {\bf 13.6} & - & {\bf 858} & 778 & \cite{Ishizuka_JPG_2008} \\
{\bf IOTSY(TM1Ypi30)} & $Y, \pi$ & {\bf 1.66} & {\bf 13.6} & - & {\bf 858} & 778 & \cite{Ishizuka_JPG_2008} \\
{\bf IOTSY(TM1Ypi90)} & $Y, \pi$ & {\bf 1.66} & {\bf 13.6} & - & {\bf 858} & 778 & \cite{Ishizuka_JPG_2008} \\
SFHPST(TM1B139) & q & 2.08 & 12.6 & - & n.a. & n.a. & \cite{STOS_PTP_1998,STOS_NPA_1998,Sagert_PRL_2009,Sagert_JPG_2010,Fischer_ApJ_2011}\\
SFHPST(TM1B145) & q & 2.01 & 13.0 & - & n.a. & n.a. & \cite{STOS_PTP_1998,STOS_NPA_1998,Sagert_PRL_2009,Sagert_JPG_2010,Fischer_ApJ_2011}\\
{\bf SFHPST(TM1B155)} & q & {\bf 1.70} & {\bf 10.7} & - & n.a. & n.a. & \cite{STOS_PTP_1998,STOS_NPA_1998,Sagert_PRL_2009,Sagert_JPG_2010,Fischer_ApJ_2011}\\
{\bf SFHPST(TM1B165)} & q & {\bf 1.51} & {\bf 9.1} & - & n.a. & n.a. &  \cite{STOS_PTP_1998,STOS_NPA_1998,Sagert_PRL_2009,Sagert_JPG_2010,Fischer_ApJ_2011}\\
FOP(SFHoY) & Y & 1.99 & 11.9 & - & 401 & 366 &  \cite{Hempel_NPA_2010,SHF_ApJ_2013,FOP_2017} \\
{\bf BHB(DD2L)} & $\Lambda$ & {\bf 1.95} & {\bf 13.2} & - & 787 & 757 &  \cite{Hempel_NPA_2010,BHB_2014} \\
BHB(DD2Lphi) & $\Lambda$ & 2.10 & {\bf 13.2} & 12.2 & 790 & 757 & \cite{Hempel_NPA_2010,BHB_2014} \\
OMHN(DD2Y) & Y & 2.03 & {\bf 13.2} & - & 787 & 756 & \cite{Hempel_NPA_2010,Marques_PRC_2017} \\
MBB(DD2K) & $K^-$ & 2.19 & {\bf 13.2} & 13.0 & 798 & 758 & \cite{Hempel_NPA_2010,Malik_EPJA_2021}\\
MBB(BHBLphiK) & $\Lambda$, $K^-$ & 2.05 & {\bf 13.2} & - & 788 & 755 & \cite{Hempel_NPA_2010,Malik_ApJ_2021}\\
SDGTT(QMC-A) & Y & 2.08 & 13.0 & n.a. & n.a. & n.a. & \cite{Guichon_PLB_1988,Guichon_PPNP_2018,Stone_MNRAS_2021} \\
R(DD2YDelta)(1.1;1.1;1.0) & Y$\Delta$ & 2.04 & 12.97 & - & 697 & 643 & \cite{Hempel_NPA_2010}, this work \\
R(DD2YDelta)(1.2;1.1;1.0) & Y$\Delta$ & 2.05 & 12.30 & - & 470 & 434 & \cite{Hempel_NPA_2010}, this work \\
R(DD2YDelta)(1.2;1.3;1.0) & Y$\Delta$ & 2.03 & {\bf 13.25} & - & 786 & 757 & \cite{Hempel_NPA_2010}, this work \\
DNS(CMF) & Y, q & 2.1 & {\bf 14.0} & 12.6 & {\bf 1114} & {\bf 1043} & \cite{DNS_PRC_2015,Dexheimer_PASA_2017} \\
BBKF(DD2F-SF)1.1 & q & 2.13 & 12.2 & {\bf 10.7} & 507 & 467 & \cite{Bastian_PRC_2017,Bauswein_PRL_2019,Bastian_PRD_2021} \\
BBKF(DD2F-SF)1.2 & q & 2.15 & 12.2 & 11.4 & 501 & 473 & \cite{Bastian_PRC_2017,Bauswein_PRL_2019,Bastian_PRD_2021} \\
BBKF(DD2F-SF)1.3 & q & 2.02 & 12.2 & - & 512 & 467 & \cite{Bastian_PRC_2017,Bauswein_PRL_2019,Bastian_PRD_2021} \\
BBKF(DD2F-SF)1.4 & q & 2.02 & 12.2 & - & 516 & 467 & \cite{Bastian_PRC_2017,Bauswein_PRL_2019,Bastian_PRD_2021} \\
BBKF(DD2F-SF)1.5 & q & 2.03 & 12.2 & - &  488 & 467 &  \cite{Bastian_PRC_2017,Bauswein_PRL_2019,Bastian_PRD_2021} \\
BBKF(DD2F-SF)1.6 & q & 2.00 & 12.2 & - & 513 & 467 & \cite{Bastian_PRC_2017,Bauswein_PRL_2019,Bastian_PRD_2021} \\
BBKF(DD2F-SF)1.7 & q & 2.11 & 12.2 & {\bf 11.2} & 514 & 467 & \cite{Bastian_PRC_2017,Bastian_PRD_2021} \\ 
BBKF(DD2-SF)1.8 & q & 2.06 & {\bf 11.0} & - & 218 & 180 & \cite{Bastian_PRC_2017,Bastian_PRD_2021} \\
BBKF(DD2-SF)1.9 & q & 2.17 & {\bf 11.3} & {\bf 11.2} & 228 & 196 & \cite{Bastian_PRC_2017,Bastian_PRD_2021} \\
\noalign{\smallskip}\hline
\end{tabular}
\end{table*}

\begin{table*}
  \caption{List of nucleonic effective interactions on which general
    purpose EoS tables in Table \ref{tab:models} are built.
    For each interaction we specify the properties of symmetric nuclear matter at
    saturation density ($n_{\mathit{sat}}$): energy per nucleon ($E_{\mathit{sat}}$);
    compression modulus ($K_{\mathit{sat}}$); skewness ($Q_{\mathit{sat}}$); symmetry
    energy ($J_{sym}$); slope ($L_{sym}$); curvature ($K_{sym}$) and skewness ($Q_{sym}$) of
    the symmetry energy.} 
\label{tab:effint}       
\begin{tabular}{lccccccccl}
\hline\noalign{\smallskip}
 int. & $n_{\mathit{sat}}$ & $E_{\mathit{sat}}$ & $K_{\mathit{sat}}$ & $Q_{\mathit{sat}}$ & $J_{sym}$ & $L_{sym}$ & $K_{sym}$ & $Q_{sym}$ &  Ref.  \\
& (fm$^{-3}$) & (MeV) & (MeV) & (MeV) & (MeV) & (MeV) & (MeV) & (MeV)   \\
 \noalign{\smallskip}\hline\noalign{\smallskip}
 LS220  & 0.155 & -16.64 & 219.85 & -410.80 & 28.61 & 73.81 & -24.04 & 96.17 &  \cite{LS_NPA_1991} \\
TM1     & 0.145      & -16.26 & 281.16 &-285.22   & 36.89 & 110.79 &  33.63 & -66.54 & \cite{TM1} \\
SFHo    & 0.158      & -16.19 & 245.4  &-467.8    & 31.57 &  47.10 & -205.5 & n.a.   & \cite{SHF_ApJ_2013} \\
DD2     & 0.149      & -16.02 & 242.72 & 168.65   & 31.67 &  55.04 & -93.23 & 598.14 & \cite{DD2} \\
DD2F    & 0.149      & -16.02 & 242.72 & 168.65   & 31.67 &  55.04 & -93.23 & 598.14 & \cite{DD2F} \\
CMF     & 0.15       & -16.0  & 300    & 281      & 30    &  88   &   27   & n.a.    &  \cite{DS_ApJ_2008,DNS_PRC_2015} \\
QMC-A   & 0.156      & -16.2  & 292    & n.a.     & 28.5  &  54.0 &  n.a.  & n.a.    & \cite{Stone_MNRAS_2021}\\ 
\noalign{\smallskip}\hline
\end{tabular}
\end{table*}

Table \ref{tab:models} lists the presently available tables, along with information on the
exotic particles and properties of corresponding cold $\beta$-equilibrated NS.
With the exception of GROM(LS220L), which relies on a non-relativistic Skyrme-like effective interaction,
all the models belong to CDFT.
The nuclear matter (NM) parameters of underlying nucleonic effective interactions are provided
in Table \ref{tab:effint}.
As with the exception of DNS(CMF)~\cite{DNS_PRC_2015,Dexheimer_PASA_2017} and
BBKF(DD2F-SF)~\cite{Bastian_PRC_2017,Bauswein_PRL_2019,Bastian_PRD_2021}
\textsc{CompOSE} provides also general purpose EoS for purely nucleonic matter and these have been discussed
at length in Paper I here we shall focus only on exotic d.o.f. and the supra-saturation regime.

\begin{table*}
  \caption{Details on CDFT models on which EoS tables have been built: list of meson couplings and/or
    their type (col. 6); list of hidden-strangeness mesons (if any) (col. 7);
    quark flavor symmetry group on which vertices of vector mesons to exotica have been fixed (col. 8).
    DD stays for density-dependent.
    For FOP(SFHoY) $i=1,...,6$; $j=1,...,3$.    
    $U_X^{(N)}\left(n_{sat}\right)$, eq. (\ref{eq:Ujk}), stays for the well depth of the particle $X$ at rest
    in symmetric saturated nuclear matter.
  }
 \label{tab:UYN} 
\begin{tabular}{lccccccccl}
\hline\noalign{\smallskip}
model & $U_{\Lambda}^{(N)}\left(n_{sat}\right)$ & $U_{\Sigma}^{(N)}\left(n_{sat}\right)$ & $U_{\Xi}^{(N)}\left(n_{sat}\right)$ & $U_{\Delta}^{(N)}\left(n_{sat}\right)$ & meson     & hidden str.  &  flavor sym. & Ref. \\
      & [MeV]            & [MeV]           & [MeV]         & [MeV]           & couplings &  mesons      &       \\         
\noalign{\smallskip}\hline\noalign{\smallskip}
STOS(TM1L)     & -30     &  -              &  -            &   -             & $\sigma^3$, $\sigma^4$, $\omega^4$ & -  &  SU(6)        &\cite{STOS_ApJSS_2011}\\
IOTSY(TM1Y-30) & -30     &  -30            &   -15         &   -             & $\sigma^3$, $\sigma^4$, $\omega^4$ & $\sigma^*$, $\phi$ & SU(3) & \cite{Ishizuka_JPG_2008} \\
IOTSY(TM1Y0)   & -30     &    0            &   -15         &   -             & $\sigma^3$, $\sigma^4$, $\omega^4$ & $\sigma^*$, $\phi$ & SU(3) & \cite{Ishizuka_JPG_2008}  \\
IOTSY(TM1Y30)  & -30     &   30            &   -15         &   -             & $\sigma^3$, $\sigma^4$, $\omega^4$ & $\sigma^*$, $\phi$ & SU(3) & \cite{Ishizuka_JPG_2008} \\
IOTSY(TM1Y90)  & -30     &   90            &   -15         &   -             & $\sigma^3$, $\sigma^4$, $\omega^4$ & $\sigma^*$, $\phi$ & SU(3) & \cite{Ishizuka_JPG_2008} \\
IOTSY(TM1Ypi-30)& -30   &  -30            &   -15         &   -             & $\sigma^3$, $\sigma^4$, $\omega^4$ & $\sigma^*$, $\phi$ & SU(3) & \cite{Ishizuka_JPG_2008} \\
IOTSY(TM1Ypi0)  & -30   &    0            &   -15         &   -             & $\sigma^3$, $\sigma^4$, $\omega^4$ & $\sigma^*$, $\phi$ & SU(3) & \cite{Ishizuka_JPG_2008} \\
IOTSY(TM1Ypi30) & -30   &   30            &   -15         &   -             & $\sigma^3$, $\sigma^4$, $\omega^4$ & $\sigma^*$, $\phi$ & SU(3) & \cite{Ishizuka_JPG_2008} \\
IOTSY(TM1Ypi90) & -30   &   90            &   -15         &   -             & $\sigma^3$, $\sigma^4$, $\omega^4$ & $\sigma^*$, $\phi$ & SU(3) & \cite{Ishizuka_JPG_2008} \\
FOP(SFHoY)      & -30   &  30             &   -14         &   -             & $\sigma^3$, $\sigma^4$, $\omega^4$, $\sigma^i \rho^2$, $\omega^{2j} \rho^2$         &   $\phi$    & SU(6)      &  \cite{FOP_2017} \\
BHB(DD2L)       & -30   & -               &   -           &   -             & DD        & - & SU(6)      &  \cite{BHB_2014}\\
BHB(DD2Lphi)    & -30   & -               &   -           &   -             & DD       & $\phi$ & SU(6)      &  \cite{BHB_2014}\\
OMHN(DD2Y)      & -30   &  30             &   -18         &   -             & DD       &  $\sigma^*$, $\phi$ & SU(6)      &  \cite{Marques_PRC_2017}\\
SDGTT(QMC-A)          & -28 & -0.96       &   -12.7       & -               & QMC    &  -                   & - & \cite{Stone_MNRAS_2021} \\
R(DD2YDelta)(1.1;1.1;1.0) & -28 & 30          &   -20         & -83             & DD      & $\phi$ & SU(6)      &  this work\\
R(DD2YDelta)(1.2;1.1;1.0) & -28 & 30          &   -20         & -124            & DD     & $\phi$ & SU(6)      &  this work  \\
R(DD2YDelta)(1.2;1.3;1.0) & -28 & 30          &   -20         & -57             & DD      & $\phi$ & SU(6)      &  this work  \\
DNS(CMF)        & -28   &   5             &   -18         &   -             & $\sigma^4$, $\delta^4$, $\omega^4$, $\sigma^2 \delta^2$ & $\sigma^*$, $\phi$ & SU(3)      &  \cite{Dexheimer_PASA_2017} \\
\noalign{\smallskip}\hline
\end{tabular}
\end{table*}

Experimental challenges due to the short lifetime of hyperons as well as low intensity
  beams make that only a few hundreds scattering events exist for $N\Lambda$ and $N\Sigma$;
  only a few scattering events exist for $N\Xi$; no scattering data is available for $YY$.
  Unsufficient constraints on the hyperon-nucleon potentials are reflected in different
  predictions of various models in spite of accurate description of scattering phase shifts
  \cite{Rijken_PTPS_2010,Nagels_PRC_2019,Haidenbauer_EPJA_2020}.
  Realistic two-body baryon–baryon interactions that describe the scattering data in free
  space along with phenomenological three body forces
  have been exploited in microscopic approaches in order to extract the EoS of hyperon
  admixed NS matter.
  Brueckner-Hartree-Fock calculations of hypernuclear matter in \cite{Schulze_PRC_2006,Burgio_PRC_2011,Rijken_EPJA_2016,Kochankovski_EPJA_2022} proved to be unable to
  produce maximum NS mass of $\approx 2M_{\odot}$, which was in part explained by a three body interaction not repulsive enough; for more, see \cite{Burgio_PPNP_2021}.

An alternative solution to extract information on the $YN$ and $YY$ interactions relies on
  the study of hypernuclei, bound systems composed of nucleons and one or more hyperons.
So far more than 40 single-$\Lambda$ nuclei and a few double-$\Lambda$ and
single-$\Xi$ nuclei have been produced and studied.
The first estimation of the well depth of the $\Lambda$-hyperon in nuclear matter,
$U_{\Lambda}^{(N)}(n_{sat}) \approx -30~{\rm MeV}$,
has been obtained from the extrapolation at $A \to \infty$, where $A$ is the hypernucleus mass number,
of the experimental binding energy of single-$\Lambda$ hypernuclei \cite{Millener1988}.
Calibration of $\sigma \Lambda$ and $\omega \Lambda$ vertices on
experimental data of binding energies of nuclei with a variable number of nucleons
and one $\Lambda$, realized by solving the Dirac equations of the
nucleons and the hyperon, confirm this value \cite{Fortin_PRC_2017}.
\cite{Fortin_PRC_2017} also shows that $U_{\Lambda}^{(N)}$ is little affected by
the assumed nucleonic effective interaction and flavor symmetry arguments used for the vector mesons.
The method in \cite{Fortin_PRC_2017} was further employed in \cite{Fortin_PRD_2020} to calibrate
the coupling constants of the $\Xi$-hyperon to the experimental binding energy of the single-$\Xi$
nuclei $_{\Xi^-}^{15}$C and $_{\Xi^-}^{12}$Be.
The thus obtained values $-18.8~{\rm MeV} \lesssim U_{\Xi}^{(N)}(n_{sat}) \lesssim -14.6~{\rm MeV}$,
which depend on the nucleon effective interaction, are in fair agreement with the value deduced from inclusive
$\left(K^-, K^+ \right)$ spectra,
$U_{\Xi}^{(N)}(n_{sat}) \approx -15~{\rm MeV}$ \cite{Gal_RMP_2016} and underestimate
the more attractive $U_{\Xi}^{(N)}(n_{sat}) \lesssim -20~{\rm MeV}$ obtained from
$\Xi^-p \to \Lambda \Lambda$ two body capture events in $^{12}$C and $^{14}$N \cite{Friedman_PLB_2021}.
Experimental data on strong-interaction
level shifts, widths and yields collected from $\Sigma^-$ atoms
and inclusive $\left(\pi^-, K^+ \right)$ spectra on medium to heavy
targets indicate a repulsive but loosely constrained $\Sigma N$ potential
$U_{\Sigma}^{(N)}(n_{sat}) \approx 30\pm 20~{\rm MeV}$~\cite{Millener1988,Gal_RMP_2016}.

$-U_{\Lambda}^{(\Lambda)}(n_{sat})$ is extracted by identification with the $\Lambda \Lambda$ bond energy,
$\Delta B_{\Lambda \Lambda}=B_{\Lambda \Lambda}\left(_{\Lambda \Lambda}^AZ\right)-2B_{\Lambda}\left(_{\Lambda}^{A-1}Z \right)$, where $B\left( ^A Z\right)$ stays for the binding energy of the nucleus $^AZ$, 
in double $\Lambda$ hypernuclei. The initial value, $\Delta B_{\Lambda \Lambda}=1.01 \pm 0.2^{+0.18}_{-0.11}~{\rm MeV}$
obtained in the KEK event \cite{Takahashi_PRL_2001} was revised to $0.67 \pm 0.17~{\rm MeV}$ \cite{Ahn_PRC_2013}
due to a change in the value of
the $\Xi^-$ mass. The small value of this attractive potential makes it that it is overlooked
in most EoS model calculations.

For EoS models which account for hyperons, Table \ref{tab:UYN} lists the values of well depths in
symmetric saturated matter on which the coupling constant to the $\sigma$ meson has been fixed, the meson couplings
and the flavor symmetry group assumed for deciding the strengths of vector mesons. One can see that
all models are in fair agreement or marginally consistent
with present constraints on $U_{\Lambda}^{(N)}$ and $U_{\Xi}^{(N)}$.
The wide range explored by $U_{\Sigma}^{(N)}$ allows one to inspect the consequences entailed by uncertainties
in the $N\Sigma$ interaction. So far this aspect was systematically addressed only for cold $\beta$-equilibrated
NS~\cite{Fortin_PRD_2020,Fortin_PRD_2021}.
Out of the considered models only IOTSY models \cite{Ishizuka_JPG_2008} account for $YY$ potentials. 
The assumed values are:
$U_{\Sigma}^{(\Sigma)} \approx U_{\Lambda}^{(\Sigma)} \approx U_{\Sigma}^{(\Lambda)}\approx 2 U_{\Lambda}^{(\Lambda)} \approx -40~{\rm MeV}$.

Some of IOTSY models \cite{Ishizuka_JPG_2008} also account for free thermal pions. Modification of pion masses due
to the interaction is disregarded. $\pi^{-,0,+}$ number densities account for $s$-wave pion condensation.
This holds when the absolute value of pion chemical potentials equals the pion mass.
Pion chemical potentials are calculated as $\mu_{\pi^+}=\mu_Q$, $\mu_{\pi^0}=0$, $\mu_{\pi^-}=-\mu_Q$,
where $\mu_Q=\mu_p-\mu_n$ is the charge chemical potential.

MBB(DD2K) \cite{Malik_EPJA_2021} accounts for thermal (anti)kaons and a Bose–Einstein condensate of $K^-$ mesons.
The phase transition from the nuclear to antikaon condensed phase is second-order.
Nucleons in the antikaon condensed phase behave differently than those in the hadronic phase.
Kaon-nucleon interaction is considered in the same footing as the
nucleon–nucleon interaction. Kaon-vector meson coupling constants are fixed by flavor symmetry arguments.
The scalar coupling constant is determined from the real part of $K^-$ optical potential at the saturation density,
$U_{K^-}^{(N)}=-120~{\rm MeV}$.
MBB(BHBLphiK) \cite{Malik_ApJ_2021} additionally accounts for $\Lambda$-hyperons.

BBKF(DD2-SF) and BBKF(DD2F-SF) models \cite{Bastian_PRD_2021} rely on a two-phase approach for
quark-hadron phase transitions, which leads by construction to a first-order phase transition.
In the hadron sector only nucleonic d.o.f. are considered; in the quark sector 
only up and down quarks are allowed.
The quark confinement is modeled within the string-flip model~\cite{Bastian_PRC_2017}.
The fact that hadrons are composite particles made of quarks is not considered in this description.
As frequently done in astrophysics, each phase is derived independently and the two phases are then
merged through a mixed phase construction.
The solution adopted in this work assumes thermal, mechanical and baryonic chemical equilibrium between
pure hadron and quark phases. The equality of charge chemical potentials
between coexisting phases is replaced, for convenience, by the equality of charge fractions.
It is argued that this simplification has only little effect on thermodynamics due to
the close values of charge fraction in the coexisting phases, which is a peculiarity of this model.
The neutralizing electron gas is added after the phase construction is realized, which means that
it does not affect phase coexistence.
Throughout this work we shall refer to this phase-construction as approximate Gibbs construction;
the attribute "approximate" is a reminder that the equality of charge chemical potentials is not exactly
fulfilled, as it should be in the case of a Gibbs construction.
We remind that the Gibbs construction is the generalization of Maxwell construction for systems with more
than one conserved charge.
The quark model is much more affected by temperature than the hadron model, which can be explained by the lower
masses of quarks. With rising temperature the model features a
monotonic decrease of transition pressure and (baryon) chemical potential.
For more, see Sec. \ref{sec:FiniteT}.

In the hadronic-quark SFHPST-models \cite{Sagert_PRL_2009,Sagert_JPG_2010} the hadronic sector is treated
within CDFT while for the quark sector the MIT bag model is employed.
In the hadron sector only nucleonic d.o.f. are considered;
up, down and strange quarks are considered in the quark sector.
The onset of the hadron-quark phase transition and the width of the phase coexistence domain are
governed by the model parameters, i.e. strange quark mass, bag constant and strong coupling constant.
By increasing the temperature the phase coexistence domain shifts to lower densities without significant
modification of width.
  The phase transition is implemented based on a Glendenning construction~\cite{Glendenning_PRD_1992}.
  At variance with the Gibbs construction which implies thermal, mechanical and chemical equilibrium of all
  conserved charges
  calculated at the boundaries of the coexistence phase
  here the equilibrium is imposed between different phases in the phase coexistence domain.
  As a consequence the pressure does no longer stay constant in the mixed phase but varies
  continuously with the proportion of phases in equilibrium; the conserved charges can be shared in different
  concentrations in each phase; the energy density is not a linear function of the proportion of phases.
  Moreover charge neutrality is not assumed to hold for each of the coexisting phases but for their mixture.
  The advantage of the Glendenning construction over the Gibbs one consists in that external fields
  sensitive to density, as the gravitational field, will not separate the phases~\cite{Glendenning_PRD_1992}.
Characteristic features of SFHPST-models, different from those of
BBKF-models, will be discussed in Sec. \ref{sec:FiniteT}. 

The SU(3) Chiral Mean Field (CMF) model \cite{Dexheimer_PASA_2017} is another example of EoS model which
allows for hadrons and quarks. For the hadronic phase, which accounts for the whole baryonic octet,
a non-linear realization of the SU(3) sigma model with an explicit chiral symmetry breaking term is used.
The particle d.o.f. populated under various thermodynamic conditions change from hadrons to quarks
and vice-versa through the introduction of an extra field $\Phi$
in the effective masses of baryons and quarks. Hadrons are included as quasi-particle d.o.f. in a
chemically equilibrated mixture with quarks, which results in full miscibility of hadrons and quarks.

The microscopic SDGTT(QMC-A) model \cite{Stone_MNRAS_2021} is based on the effective relativistic mean field
quark-meson-coupling (QMC) model \cite{Guichon_PPNP_2018}. At variance with CDFT models, in QMC
the interaction between baryons takes place self-consistently between valence quarks, confined in non-overlapping
baryons. The effect of the dense medium surrounding the baryons is modeled
by the dynamics of quarks inside the individual particles. The parameters of the quark bag model,
representing baryons, are adjusted to reproduce the masses of the baryon octet in free space.
The meson fields are coupled to the quarks.

The three R(DD2YDelta)-models in Table \ref{tab:models} are proposed in this paper and are the first
publicly available general purpose EoS models accounting for the baryonic octet and $\Delta$-resonances.
These EoS tables have been obtained in the framework of CDFT assuming that $\Delta$s preserve
their vacuum masses and have vanishing widths.
The numbers in parenthesis correspond to the coupling constants of the $\Delta$ to $\sigma$, $\omega$
and $\rho$-mesonic fields, expressed as ratios with respect to couplings of nucleons.
These values span a certain range in the parameter space associated to the nucleon-$\Delta$ interaction;
for a discussion on present constraints, see Sec. \ref{sec:ND}.
For the inhomogeneous matter at sub-saturation densities the HS(DD2) data on \textsc{CompOSE}
are used. They have been obtained in the framework of the extended Nuclear Statistical Equilibrium (NSE) model
by Hempel and Schaffner-Bielich \cite{Hempel_NPA_2010}.
The transition from inhomogeneous to homogeneous matter is done for fixed values of $T$ and $Y_Q$ by
minimizing the free energy.
Data in Table \ref{tab:models} show that models R(DD2YDelta)(1.1;1.1;1.0) and R(DD2YDelta)(1.2;1.1;1.0) are consistent
with available constraints from observation of compact stars, specifically, massive NS;
radii and masses inferences by NICER; range of tidal deformability derived from GW170817.
R(DD2YDelta)(1.2;1.3;1.0) has a more modest performance when confronted with astrophysics data as
it provides for $R_{14}$ a value exceeding the maximum value extracted from observations on
PSR J0030+0451 \cite{Miller_2019,Riley_2019}.
Compliance with constraints from nuclear physics experiments and
ab initio calculations of nuclear matter is warranted
by the use of DD2 effective interaction \cite{Fortin_PRC_2016}; compliance with hyper-nuclear physics
data is met by the values of $U_Y^{(N)}$ on which meson coupling constants have been tuned. 
Also available are tables for purely baryonic matter.
The domains of temperature, baryonic number density and charge fraction covered
by these EoS tables and the corresponding numbers of mash points are provided in Table \ref{tab:grids}
in the Appendix A.
The table corresponding to $\left(1.2, 1.1, 1.0 \right)$ is unique in the sense that
pure baryonic matter manifests a small $\Delta$-driven instability domain, see Sec. \ref{sec:ND}.
This instabilities are suppressed by the electron gas,
which means that numerical simulations of astrophysics phenomena should not present signals
typically associated with phase transitions. Provided that the values of coupling constants comply with
experimental constraints, this collection of models can be extended such as to include also
models that present instabilities.

\section{Covariant density functional models for baryonic matter}
\label{sec:CDFT}

Covariant density functional or, equivalently, relativistic mean field (RMF)
models have been successfully used to describe infinite nuclear matter,
finite nuclei and stellar matter. They rely on the Walecka model \cite{Walecka_1974},
which treats nucleons as fundamental particle d. o. f. that interact
with each other through the exchange of scalar and vector mesons.
The scalar isoscalar $\sigma$ and vector isoscalar $\omega$ fields correspond
to these mesons and account for the attractive and repulsive parts of the nuclear
interaction.

With the aim of better describing an ever increasing amount of data the original
model was further developed by accounting for other mesons,
e.g. the vector-isovector $\rho$, scalar-isovector $\delta$ and, if strange baryons are considered,
also hidden-strangeness scalar-isoscalar $\sigma^*$ and vector-isoscalar $\phi$ mesons.
Models with mixed interaction terms and density dependent coupling constants have been
proposed as well.
For a review, see \cite{Dutra_RMF_PRC_2014}.

As the data generated for this paper and discussed in Sec. \ref{sec:HeavyBaryons},
\ref{sec:ND} and partially \ref{sec:FiniteT}, 
have been obtained in the framework of density dependent models in the
following we shall review equations pertaining to this class of models.
For the sake of brevity we shall limit ourselves to $\sigma$, $\omega$, $\rho$, $\phi$
mesonic fields. These are the only mesons considered in
Sec. \ref{sec:HeavyBaryons}, \ref{sec:ND} as well as the new general purpose EoS with hyperons and
$\Delta$s that we present here and made available on the \textsc{CompOSE} database.
We nevertheless mention that density dependent RMF models which account for $\delta$
  and/or $\sigma^*$ mesons exist as well. Examples in this sense are offered by
  \cite{Gaitanos_NPA_2004,Roca-Maza_PRC_2011} and \cite{Fortin_PRC_2017,Sedrakian_EPJA_2022}
  for the scalar-isovector $\delta$ and hidden-strangeness scalar-isoscalar
  $\sigma^*$ mesons, respectively.
  The scalar-isovector $\delta$ meson gives rise to attraction in the isovector channel, while
  the vector-isovector $\rho$ meson is responsible for repulsion;
  improves the neutron-proton effective mass splitting with respect 
  to predictions of relativistic Brueckner Hartree-Fock calculations.
  The hidden-strangeness scalar-isoscalar $\sigma^*$ meson regulates, together with the
  vector-isoscalar $\phi$ meson, the $YY$ interaction.
  The scalar-isovector $\delta$ meson mostly affects the high-density
  behavior of the EoS while the scalar $\sigma^*$ field is dominant at low densities.
  The role of the $\delta$-meson in stellar matter was addressed in
  \cite{Menezes_PRC_2004}, where different hadronic contents, temperature profiles and lepton
  fractions have been considered. According to these authors, who employed a non-linear RMF model,
  inclusion of $\delta$ makes the EoS of cold and $\beta$-equilibrated nucleonic matter
  stiffer while the opposite effect is obtained for hyperonic matter;
  smaller modifications are obtained for protoneutron stars with trapped neutrinos.
  We expect RMF models with density dependent couplings to agree with above cited results
  only for cold and $\beta$-equilibrated matter.

At finite temperature, $T$, the scalar and number densities of a baryon $i$ are given by
\begin{align}
  n_i^s&= \frac{2 J_i+1}{2\pi^2} \int_o^{\infty} \frac{k^2 m^*_i}{E_i(k)}
  \left[f_{\mathrm{FD}}(E_i(k)-\mu^*_i)+ f_{\mathrm{FD}}(E_i(k)+\mu^*_i)  \right] dk~,
  \label{eq:ns}\\
  n_i&= \frac{2 J_i+1}{2\pi^2} \int_o^{\infty} k^2\left[f_{\mathrm{FD}}(E_i(k)-\mu^*_i)- f_{\mathrm{FD}}(E_i(k)+\mu^*_i)  \right] dk~,
  \label{eq:n}
\end{align}
where $k$, $E_i(k)=\sqrt{k^2+m^{*,2}_i}$,
$m^*_i$ and  $\mu^*_i$ 
respectively stand for the wave number, kinetic part of the single particle energy,
Dirac effective mass and effective chemical potential and
$f_{FD}\left(\epsilon\right)=1/\left[ 1+\exp\left(\epsilon/T\right)\right]$
represents the Fermi-Dirac distribution function;
$\left(2 J_i+1 \right)$ represents the
spin degeneracy factor of the $i$ baryon.
Dirac effective mass and effective chemical potential
are linked to bare mass and chemical potential by
\begin{align}
  m^*_i=m_i+\Sigma_{S;i}~,
  \label{eq:meffD} \\
  \mu^*_i=\mu_i-\Sigma_{V;i}-\Sigma_R~.
  \label{eq:mueff}
\end{align}
with the scalar $\Sigma_{S;i}$ and vector $\Sigma_{V;i}$ self-energies and
rearrangement term $\Sigma_R$ expressed by
\begin{eqnarray}
  \Sigma_{S;i} &=&-g_{\sigma i} \bar \sigma~,
  \label{eq:Ss}\\
  \Sigma_{V;i}&=&g_{\omega i} \bar \omega + g_{\rho i} t_{3i} \bar \rho +g_{\phi i} \bar \phi~,
  \label{eq:Sv}\\
  \Sigma_R&=&\sum_j \left(
  \frac{\partial g_{\omega j}}{\partial n_j} \bar \omega n_j +
  t_{3 j} \frac{\partial g_{\rho j}}{\partial n_j} \bar \rho n_j +
  \frac{\partial g_{\phi j}}{\partial n_j} \bar \phi n_j -
  \frac{\partial g_{\sigma j}}{\partial n_j} \bar \sigma n_j^s
  \right)~.\nonumber
  \label{eq:SR}\\
\end{eqnarray}
$t_{3i}$ represents the third component of isospin of baryon $i$ with the convention that
$t_{3p}=1/2$.
As non-linear interaction terms are disregarded, the mean-field expectation values of
meson fields are given by
\begin{subequations}
\begin{eqnarray}
  m_{\sigma}^2 \bar \sigma=\sum_{i} g_{\sigma i} n_i^s, 
  \label{eq:sigma}\\
  m_{\omega}^2 \bar \omega=\sum_{i} g_{\omega i} n_i,
  \label{eq:omega}\\
  m_{\phi}^2 \bar \phi=\sum_{i} g_{\phi i} n_i, 
  \label{eq:phi}\\
  m_{\rho}^2 \bar \rho=\sum_{i} g_{\rho i} t_{3i} n_i,
  \label{eq:rho}
\end{eqnarray}
\end{subequations}
where $m_m$ with $m=\sigma, \omega, \rho, \phi$, stays for the meson masses.

Baryonic energy density and pressure can be cast as sums between a kinetic and
an interaction term,
\begin{eqnarray}
  e_{baryon}=e_{kin}+e_{meson}~,\label{eq:ebar}\\
  p_{baryon}=p_{kin}+p_{meson}+p_{rearrang} \label{eq:pbar}~.
\end{eqnarray}
In eq. (\ref{eq:pbar}) an additional "rearrangement" term exists,
\begin{equation}
  p_{rearrang}=n_B \Sigma_R~,
\end{equation}
due to the couplings' density-dependence.

The kinetic terms in eqs. (\ref{eq:ebar}) and (\ref{eq:pbar}) account for kinetic contributions
of every species,
\begin{eqnarray}
  e_{kin}&=&\sum_i \frac{2 J_i+1}{2 \pi^2} \int_0^{\infty}
  dk k^2 E_i(k) \nonumber\\
  &\cdot &\left[f_{FD}\left(E_i(k)-\mu^*_i \right)
    + f_{FD}\left(E_i(k)+\mu^*_i \right) \right]~,
  \label{eq:ekin}
  \nonumber\\
  \\
  p_{kin}&=&\frac 13 \sum_i \frac{2 J_i+1}{2 \pi^2} \int_0^{\infty}
  \frac{dk k^4}{E_i(k)} \nonumber\\
  &\cdot &
  \left[f_{FD}\left(E_i(k)-\mu^*_i \right)
    + f_{FD}\left(E_i(k)+\mu^*_i \right) \right]~.
  \label{eq:pkin}
  \nonumber
  \\
\end{eqnarray}

Interaction terms exclusively depend on average values of mesonic fields
\begin{eqnarray}
  e_{meson}&=&\frac{m_{\sigma}^2}{2}\bar\sigma^2+
  \frac{m_{\omega}^2}{2}\bar\omega^2+
  \frac{m_{\phi}^2}{2}\bar\phi^2+
  \frac{m_{\rho}^2}{2}\bar\rho^2~,
  \label{eq:emeson}
  \\
  p_{meson}&=&-\frac{m_{\sigma}^2}{2}\bar\sigma^2+
  \frac{m_{\omega}^2}{2}\bar\omega^2+
  \frac{m_{\phi}^2}{2}\bar\phi^2+
  \frac{m_{\rho}^2}{2}\bar\rho^2~.
  \label{eq:pmeson}
\end{eqnarray}

The entropy density of baryonic matter writes
\begin{eqnarray}
  s_{baryon}&=&-\sum_i \frac{2 J_i+1}{2 \pi^2} \int_0^{\infty}
  dk k^2  \nonumber\\
  &\cdot &
  \Bigg\{\Bigl[
    f_{FD}\left(E_i(k)-\mu^*_i \right) \ln f_{FD}\left(E_i(k)-\mu^*_i \right)
    \nonumber\\
    &+&\bar f_{FD}\left(E_i(k)-\mu^*_i \right)
    \ln \bar f_{FD} \left(E_i(k)-\mu^*_i  \right) \Bigr] +
    \left(\mu^*_i \rightarrow -\mu^*_i \right)
   \Biggr\}~.
\end{eqnarray}
Together with other thermodynamic quantities it verifies
the thermodynamic identity
\begin{equation}
e_{baryon}=T s_{baryon}-p+\sum_i \mu_i n_i~.
\end{equation}

Matter composition is determined, at arbitrary thermodynamic conditions,
by conservation of mass and charge,
\begin{eqnarray}
  n_B=\sum_i n_i,\\
  n_Q=\sum_i q_i n_i,
\end{eqnarray}
along with chemical equilibrium. For the case of hyperon and $\Delta$-admixed nucler matter
chemical equilibrium conditions write,
\begin{eqnarray}
  \mu_n=\mu_{\Lambda}=\mu_{\Xi^0}=\mu_{\Sigma^0}=\mu_{\Delta^0}=\mu_B,\\
  \mu_{\Sigma^-}=\mu_{\Xi^-}=\mu_{\Delta^-}=\mu_B-\mu_Q,\\
  \mu_p=\mu_{\Sigma^+}=\mu_{\Delta^+}=\mu_B+\mu_Q,\\
  \mu_{\Delta^{++}}=\mu_B+2\mu_Q,
 \end{eqnarray} 
where $\mu_B$ and $\mu_Q$ stand for baryon and charge chemical potentials.
Note that $\beta$-equilibrium is not assumed.
Non-conservation of strangeness assumes $\mu_S=0$.

Finally the interaction potential of particle $j$ in $k$-particle matter
is given by,
\begin{eqnarray}
  U_j^{(k)}&=&m^*_j-m_j+\mu_j-\mu^*_j,\label{eq:Ujk}\\
  &=&\Sigma_{S;j} + \Sigma_{V;j} + \Sigma_R.
  \label{eq:U}
\end{eqnarray}

As customarily in the literature throughout this paper couplings of heavy baryons to mesonic fields
will be expressed as ratios to the corresponding couplings of nucleons.
We shall also assume that the density dependencies of couplings of heavy baryons to mesonic fields are
identical to the ones of nucleons.

\section{Heavy Baryons at finite-$T$: the cases of
  $\Lambda$- and $\Delta$-admixed nuclear matter}
\label{sec:HeavyBaryons}

In the following we shall analyze the $T$- and $n_B$-dependence
of various thermodynamic and microscopic quantities along with matter composition.
The study will be performed for temperatures in the range $0.5 \leq T \leq 40~{\rm MeV}$,
neutron rich matter with $Y_Q=0.3$ and
two values of baryonic particle number density,
$n_B=0.4~{\rm fm}^{-3}$ and $0.6~{\rm fm}^{-3}$.
The role of strange baryons and nucleonic resonances will be highlighted by confronting
results of $\Lambda$- and $\Delta$-admixed nuclear matter to results of purely nucleonic matter
$\left(N \right)$. For brevity in some circumstances only the characteristics of
dominant species will be illustrated. The employed nucleonic interaction is
DD2~\cite{DD2}.

\begin{figure*}
\includegraphics[width=0.5\columnwidth]{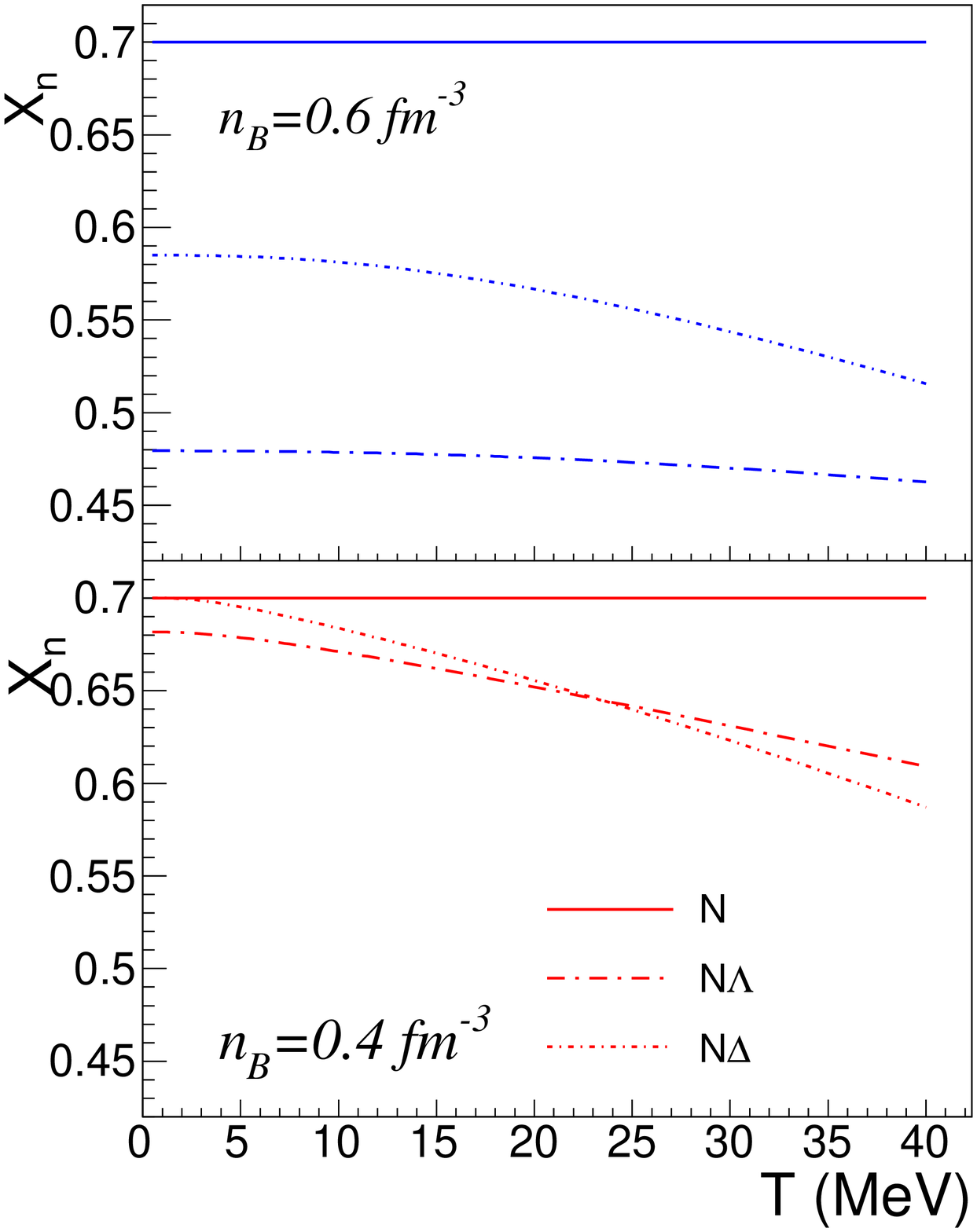}
\includegraphics[width=0.5\columnwidth]{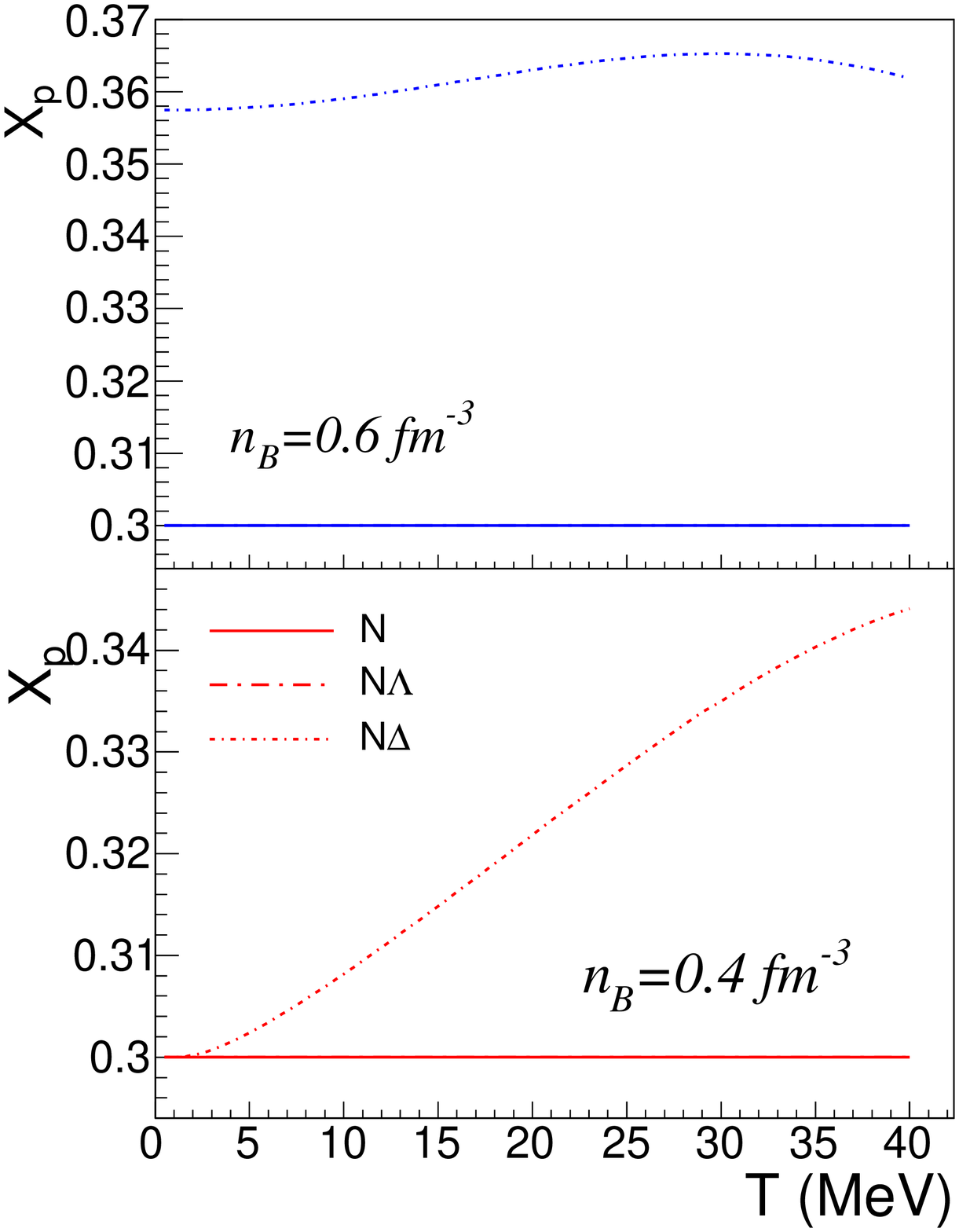}
\includegraphics[width=0.5\columnwidth]{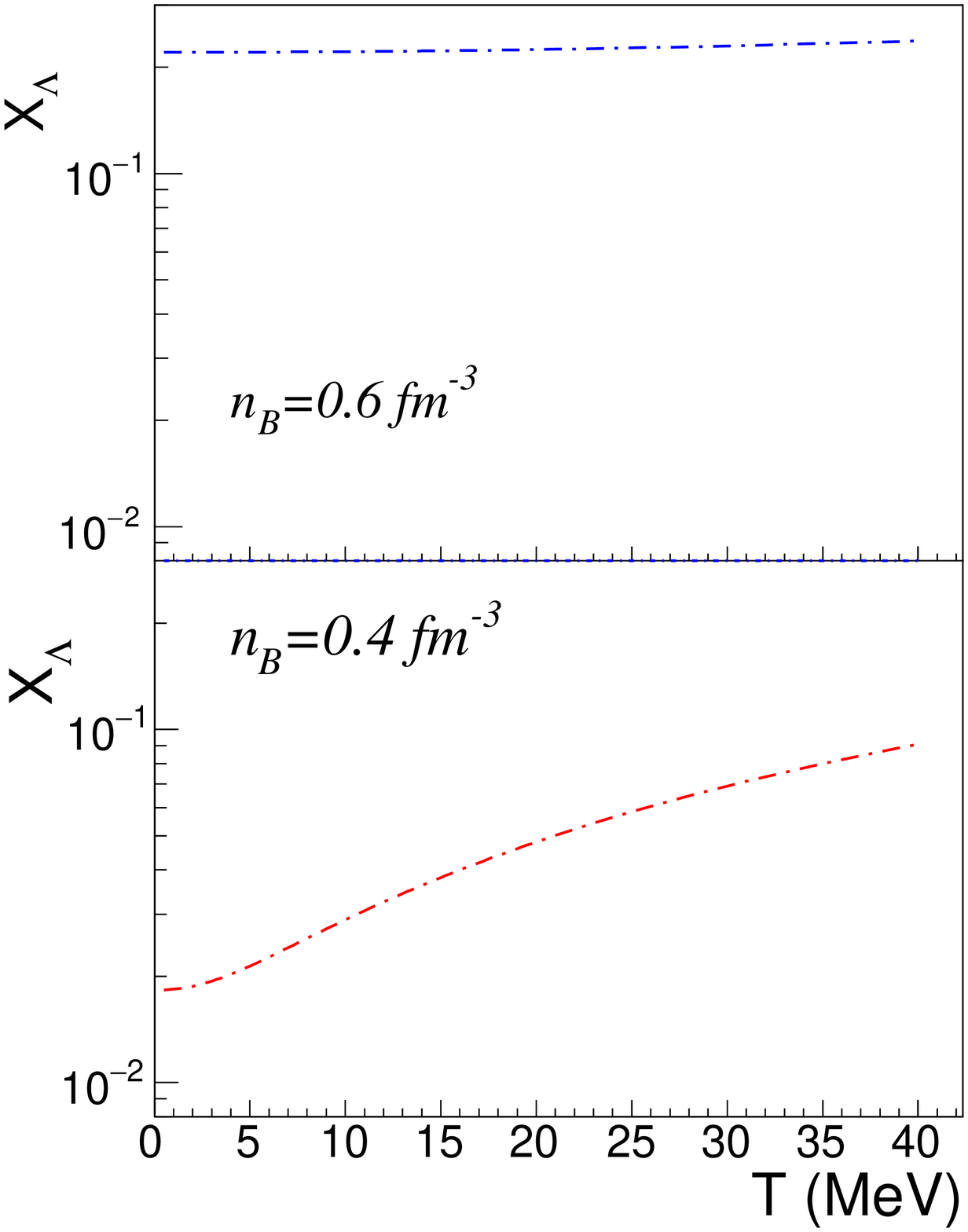}
\includegraphics[width=0.5\columnwidth]{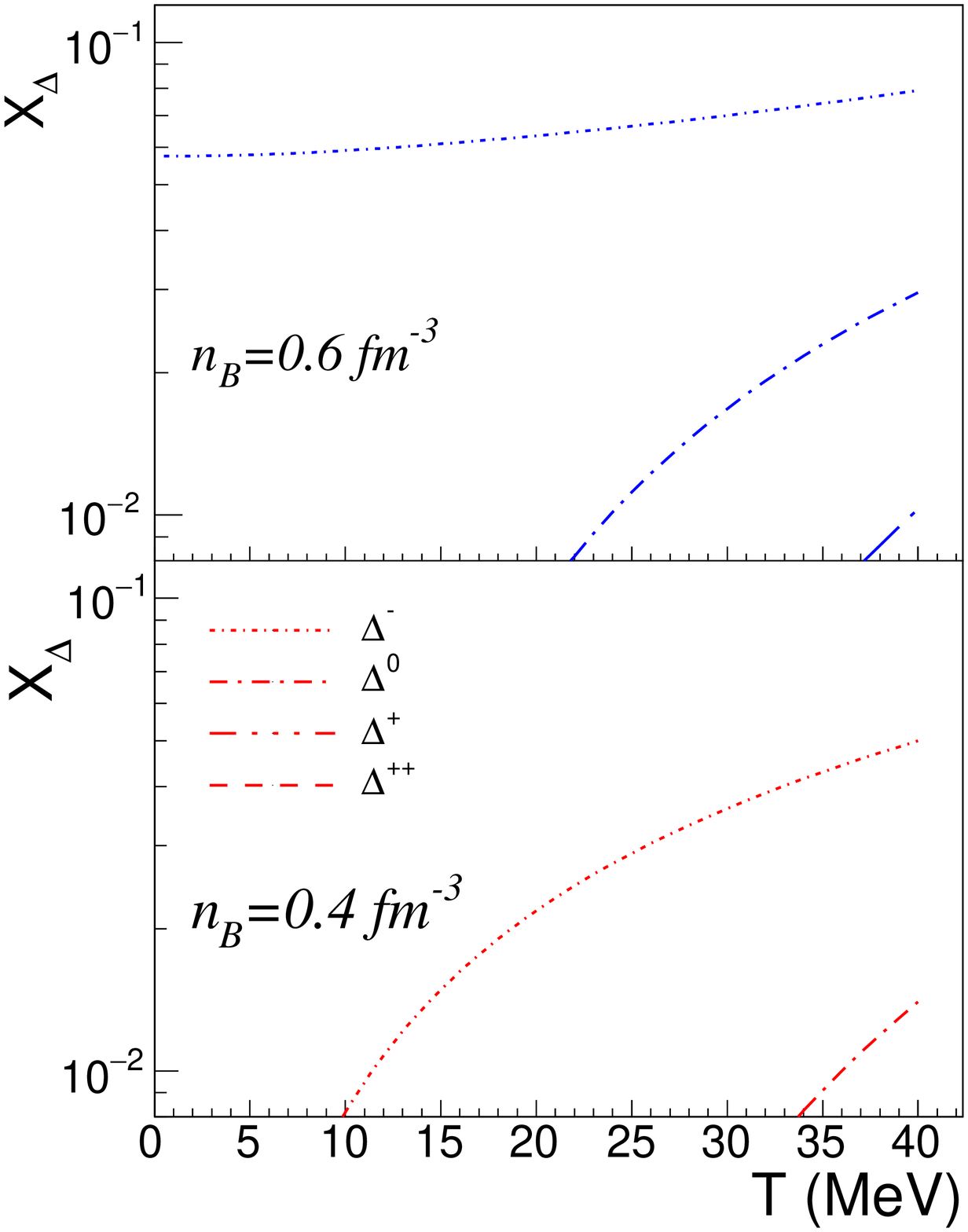}
\caption{$T$-dependence of relative particle abundances.
  Results corresponding to $N \Lambda$ and $N \Delta$-matter
  are confronted with those of nuclear matter ($N$) with $Y_Q=0.3$
  at $n_B=0.4~{\rm fm}^{-3}$ and $0.6~{\rm fm}^{-3}$.
}
\label{fig:Xi} 
\end{figure*}

The $n_B$ and $T$-dependence of neutron, proton, $\Lambda$ and $\Delta$s
relative abundances is depicted in Fig. \ref{fig:Xi}.
Abundances of exotic species strongly depend on thermodynamic
conditions and as a rule of thumb increase with $T$.
At low (high) densities the abundances of $\Lambda$ and $\Delta^-$, which is
the dominant member of the $J=3/2$ quadruplet,
show significant (negligible) dependence on $T$.
Mass conservation and chemical equilibrium condition $\mu_n=\mu_{\Lambda}$
make that nucleation of $\Lambda$ entails a certain diminish
of neutron particle number densities, while no modification is seen in what regards $X_p$.
The situation of $\Delta$-admixed nuclear matter is more complex. The onset
of $\Delta^-$ reduces $X_n$ and enhances $X_p$. The first effect
comes from mass conservation, while the latter is due to charge conservation
and opposite charges of the dominant charged species, $p$ and $\Delta^-$.
For the simple situations considered here one notices that abundances of exotic particles
also increase with $n_B$.

\begin{figure}
\includegraphics[width=0.49\columnwidth]{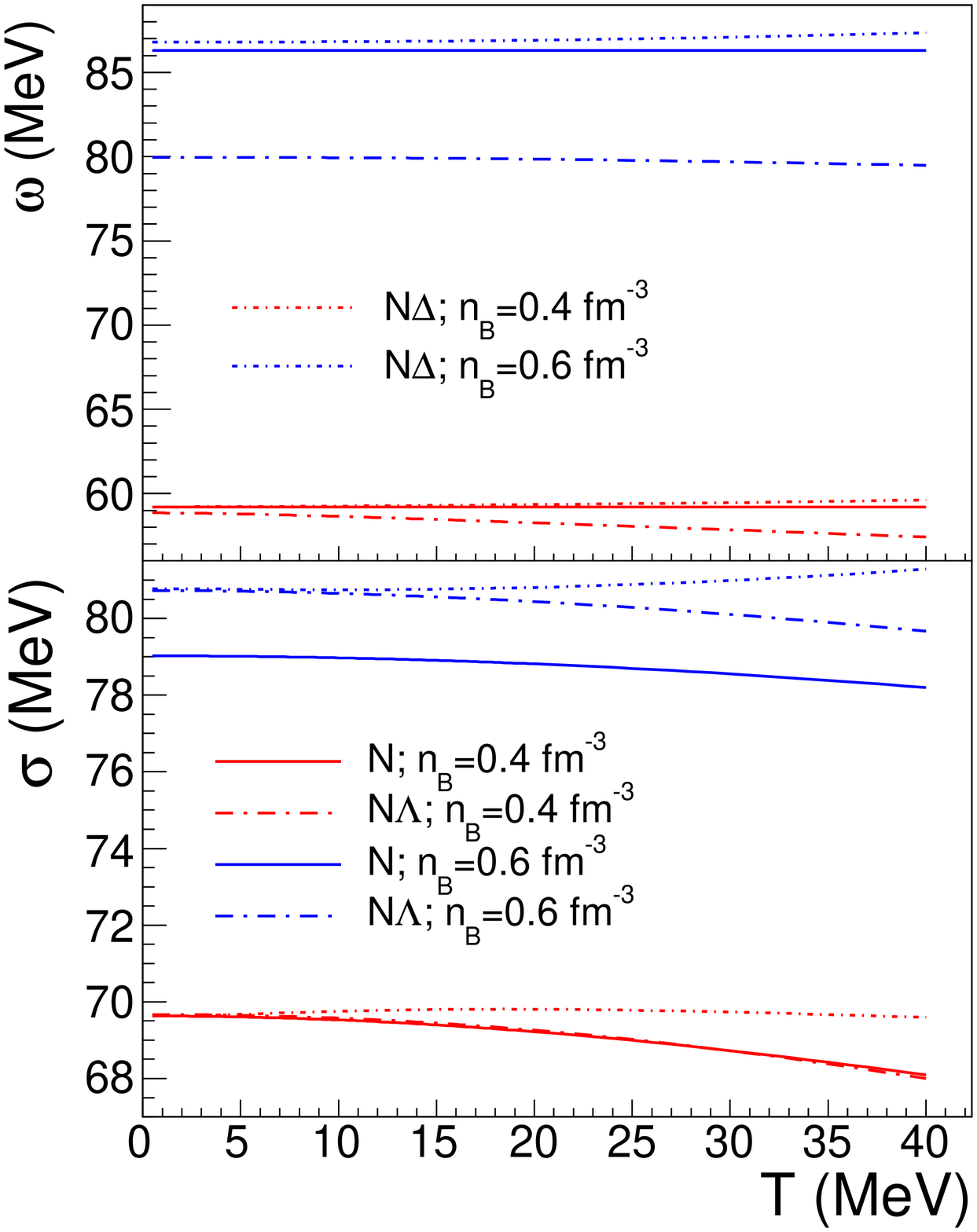}
\includegraphics[width=0.49\columnwidth]{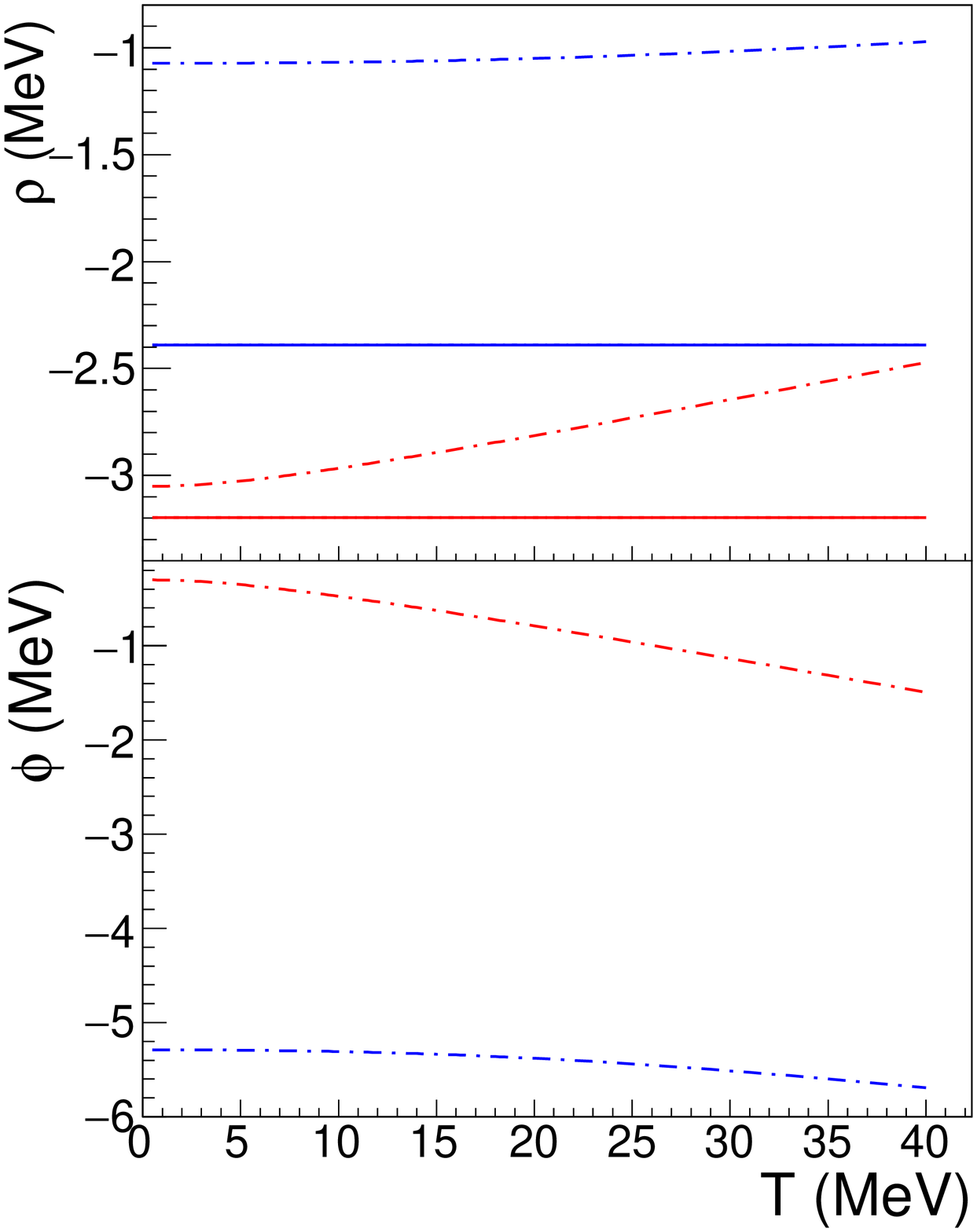}
\caption{$T$-dependence of mesonic fields.
  Results corresponding to $N$, $N \Lambda$ and $N \Delta$-matter
  with $Y_Q=0.3$ at $n_B=0.4~{\rm fm}^{-3}$ and $0.6~{\rm fm}^{-3}$.
}
\label{fig:fields} 
\end{figure}

\begin{figure}
\includegraphics[width=0.49\columnwidth]{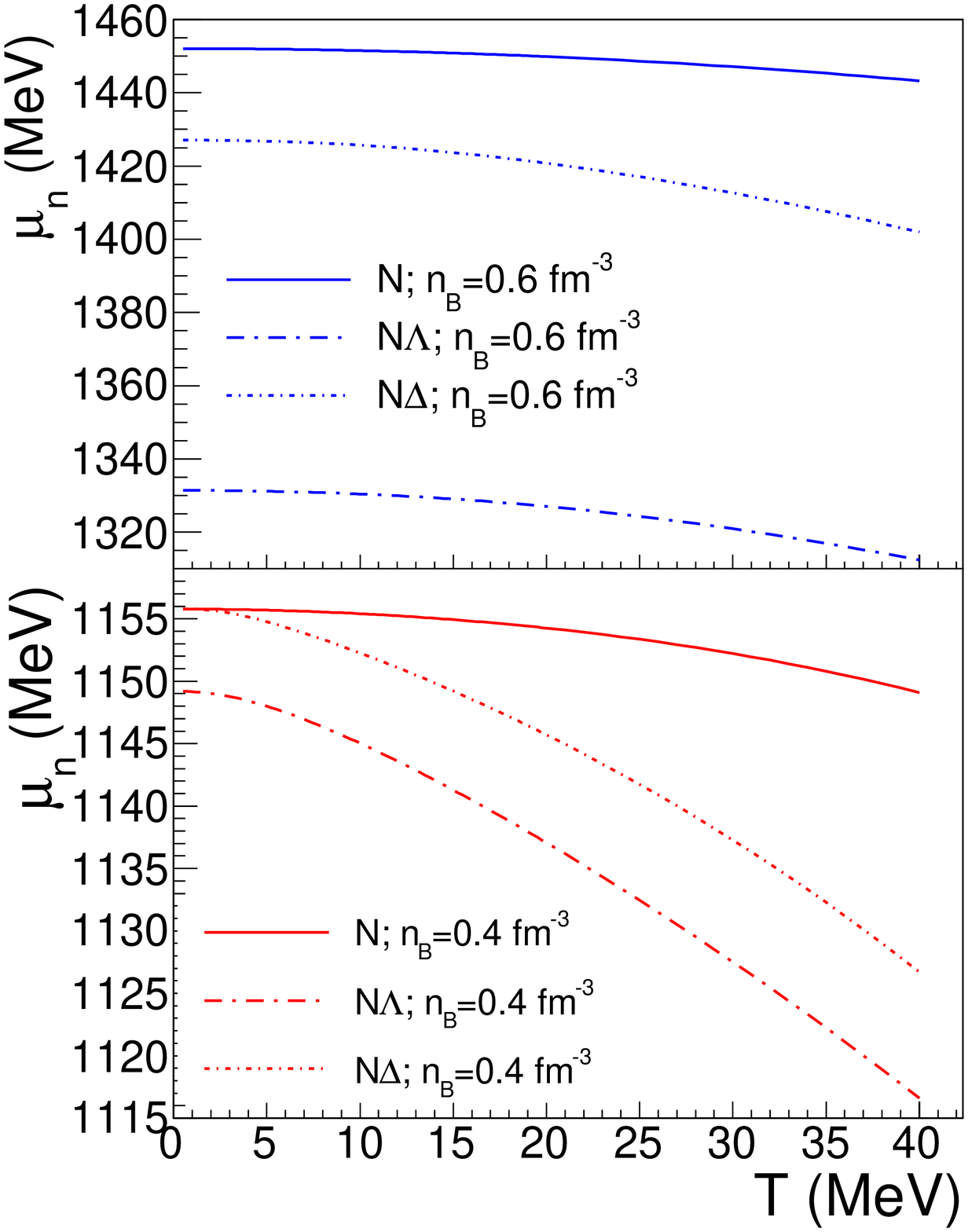}
\includegraphics[width=0.49\columnwidth]{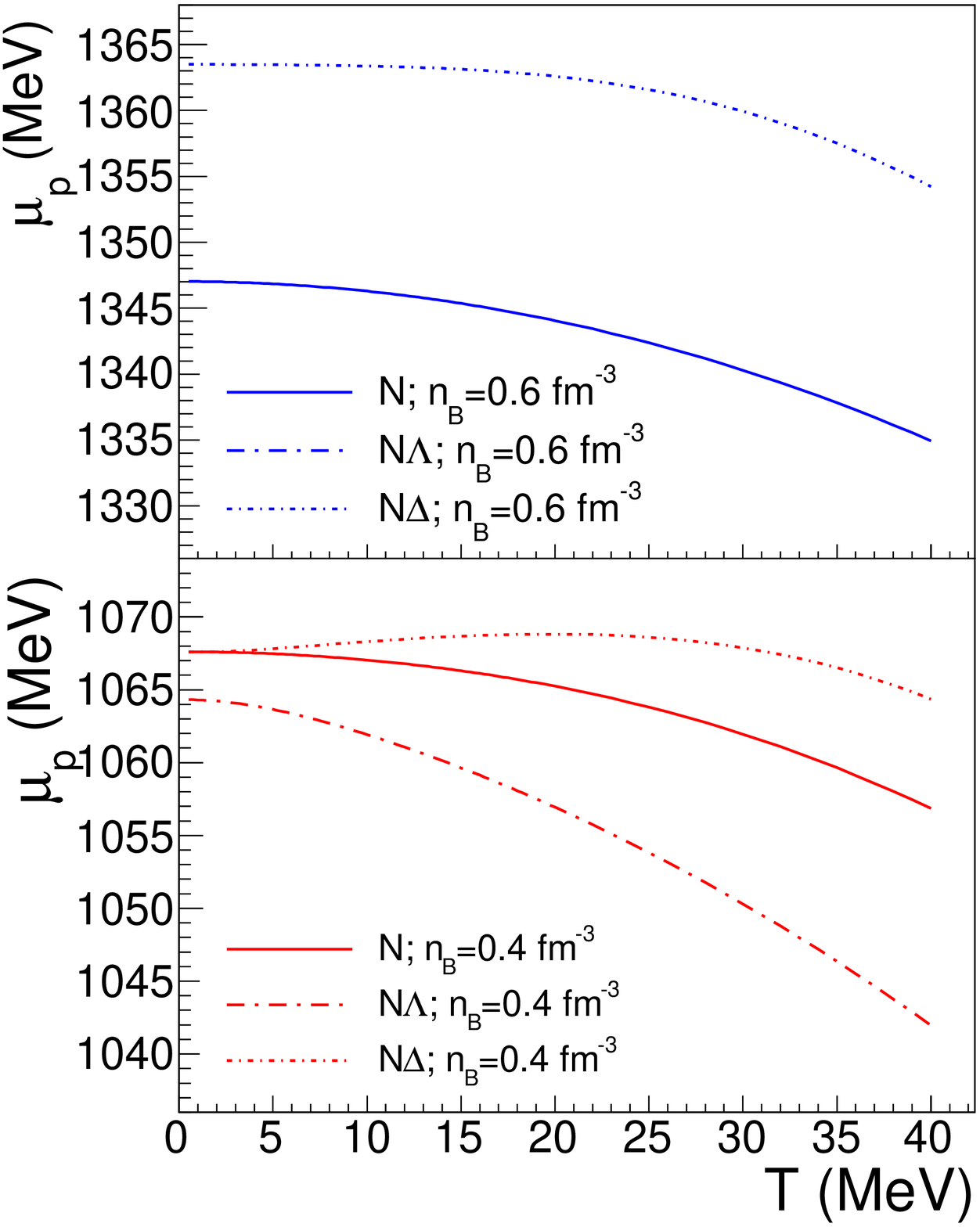}
\caption{Neutron and proton chemical potentials as function of $T$
  in nuclear matter $\left(N \right)$ as well as
  $N \Lambda$ and $N \Delta$-matter with $Y_Q=0.3$
  at $n_B=0.4~{\rm fm}^{-3}$ and $0.6~{\rm fm}^{-3}$.
}
\label{fig:mus} 
\end{figure}

The $n_B$ and $T$-dependence of mesonic fields is addressed in Fig. \ref{fig:fields}.
The following remarks are in order:

{\em (a) $N$-matter:} Being defined in terms of number densities of different
particle degrees of freedom
the average values of vector isoscalar $\omega$ and vector isovector $\rho$ mesonic fields
do not depend on temperature.
The decrease of $\bar \sigma$ with $T$ is due to the fact that, for any species $i$,
the maximum contribution to the integrals in eq. (\ref{eq:sigma}) is given by the $k$-values
close to the solution of $\sqrt{k_i^2+m_{i}^{*2}}=\mu^{*}_{i}$ and that 
neutron and proton effective chemical potentials decrease with $T$, see Fig. \ref{fig:mus}.

{\em (b) $N \Lambda$-matter:} As already discussed, high $T$-values favor the production
of $\Lambda$s at the cost of neutrons, which explains that $|\bar \phi|$ augments with $T$ and
$|\bar \rho|$ diminishes with $T$.
The latter feature suggests that, for fixed $Y_Q$, hot matter is more isospin symmetric
than cold matter, in agreement with Fig. \ref{fig:Xi}.
Similarly to the case of $\bar \sigma(T)$ in $N$-matter and for the same reason
$\bar \sigma$ and $\bar \omega$ decrease with $T$.

{\em (c) $N \Delta$-matter:} Similarly to $\Lambda$s, also the production of $\Delta$s is
favored by high temperatures and densities. Under the considered conditions,
$\Delta^-$ is by far the most abundant species as its chemical potential, $\mu_{\Delta^-}=\mu_B-\mu_Q$,
with $\mu_Q<0$ is the largest.
As $X_{\Delta} \ll X_{\Lambda}$, the average values of all mesonic fields in $N \Delta$ matter differ
less with respect to their values in pure nucleonic matter than it was the case with $N \Lambda$.  
Qualitative differences of $\bar \sigma(T)$ and $\bar \omega (T)$ in $N \Delta$ with respect to
$N \Lambda$ 
stem from opposite modifications of $X_n$ and $X_p$, with the latter prevailing over the first.

$\bar \sigma$ and $\bar \omega$ augment with density as both scalar and vector number densities
do so; the $n_B$-decrease of $g_{\rho B}$ explains the $n_B$-decrease of $|\bar \rho|$, the negative
values being due to the fact that the matter is neutron rich.
As the $\phi$ meson couples only to hyperons, $\bar \phi \neq 0$ only in $\left(N \Lambda \right)$ matter.
Its absolute value increases with $n_B$ as high densities favor production of strange particles.

\begin{figure}
\includegraphics[width=0.49\columnwidth]{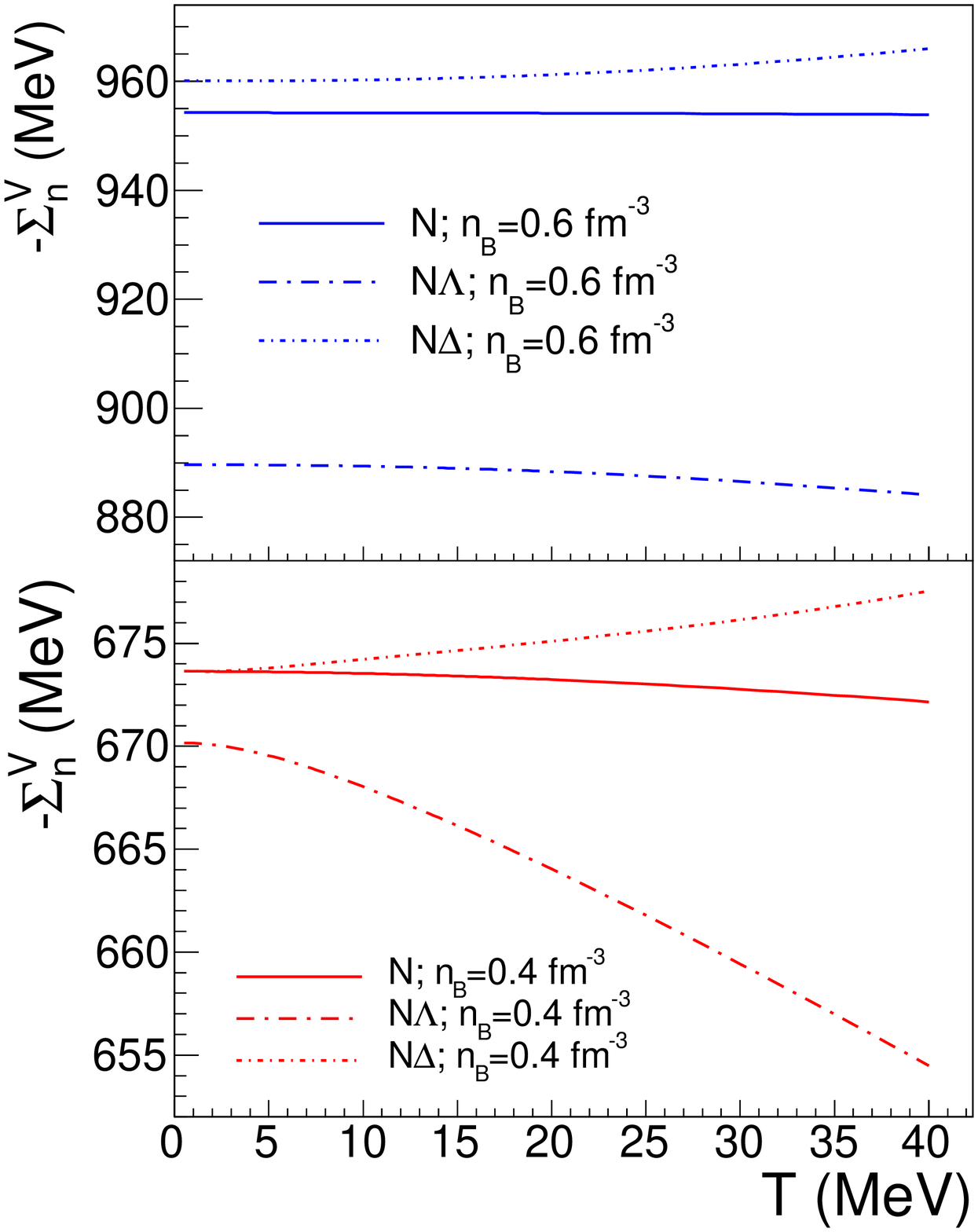}
\includegraphics[width=0.49\columnwidth]{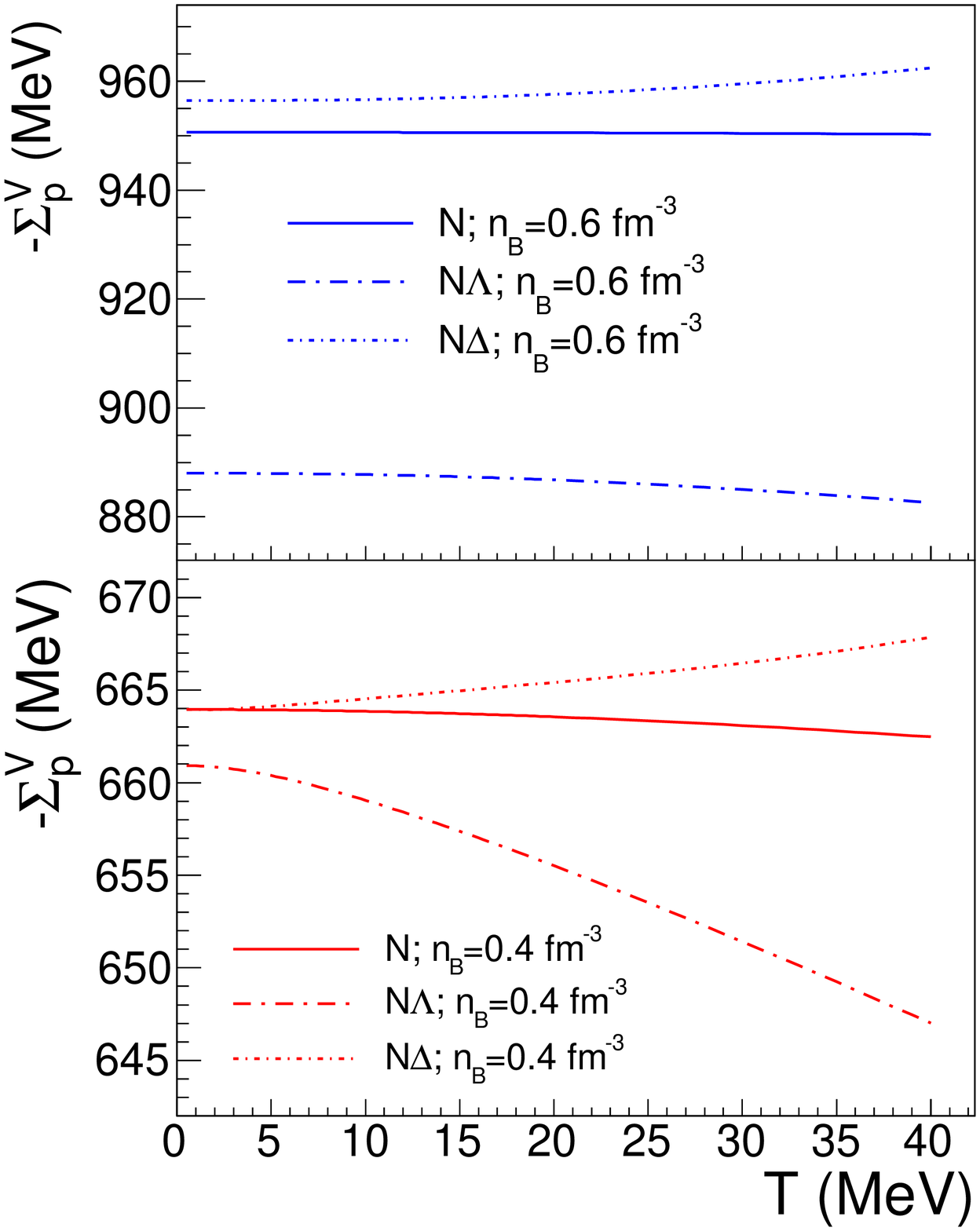}
\caption{Vector self energies of neutrons and protons, eq. (\ref{eq:Sv}), as function of $T$
  in nuclear matter $\left(N \right)$ as well as
  $N \Lambda$ and $N \Delta$-matter with $Y_Q=0.3$
  at $n_B=0.4~{\rm fm}^{-3}$ and $0.6~{\rm fm}^{-3}$.
}
\label{fig:vecselfen} 
\end{figure}

The $T$-dependence of vector self-energies, eq. (\ref{eq:Sv}),
at constant density is represented in Fig. \ref{fig:vecselfen};
only the behavior of dominant species is illustrated.
One notes that $-\Sigma^V_{n,p}$ strictly follow the $T$-evolution of dominant
mesonic field $\omega$, see Fig. \ref{fig:fields}.
The strongest $T$-dependence is obtained for $N \Lambda$ matter at the lowest
considered density, where chemical composition changes steeply.

\begin{figure}
\includegraphics[width=0.49\columnwidth]{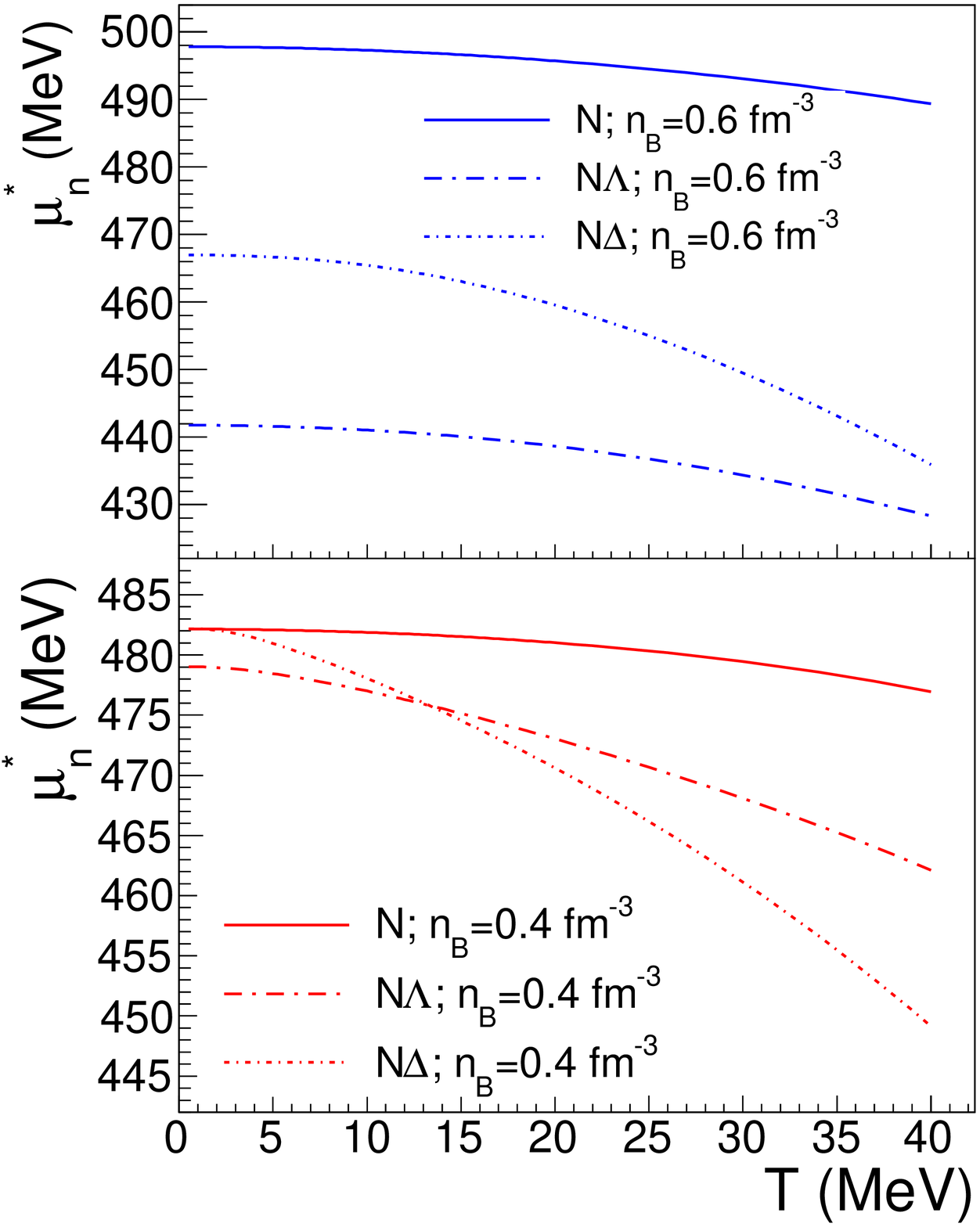}
\includegraphics[width=0.49\columnwidth]{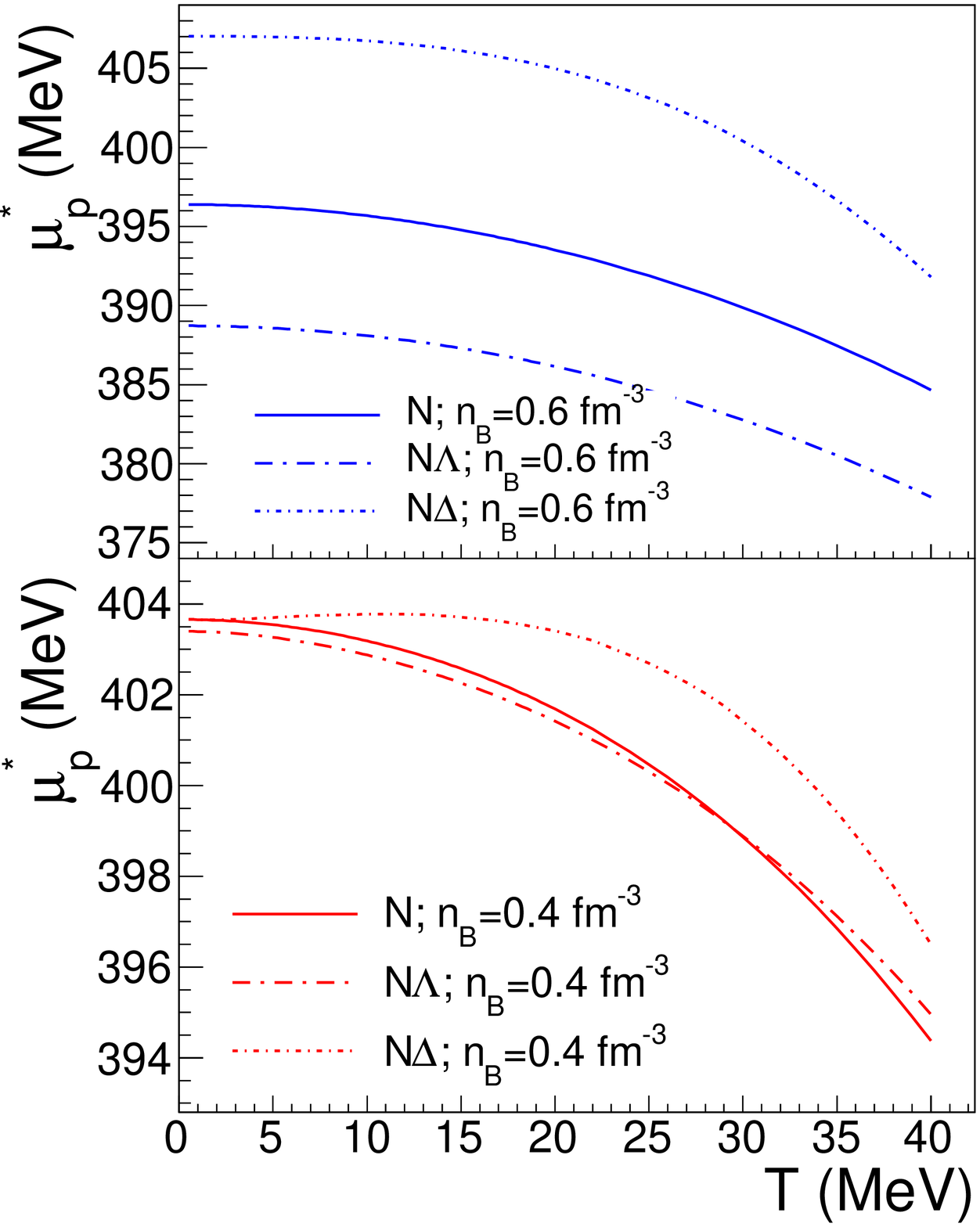}
\caption{The same as in Fig. \ref{fig:mus} for
  neutron and proton effective chemical potentials, eq.(\ref{eq:mueff}).
}
\label{fig:mueff} 
\end{figure}

The $T$-evolution of neutron and proton chemical potentials and effective chemical potentials
is illustrated in Figs. \ref{fig:mus} and, respectively, \ref{fig:mueff}.
With the exception of relatively abundant particles whose abundances augment with $T$,
as is the case of protons in $N\Delta$-matter at $n_B=0.4~{\rm fm}^{-3}$,
(effective) chemical potentials diminish with temperature.
For neutrons, which represent the dominant species, a strong correlation exists between
abundances and effective chemical potential.
As in all cases considered here matter is neutron rich, $\mu_n>\mu_p$.

\begin{figure}
\includegraphics[width=0.49\columnwidth]{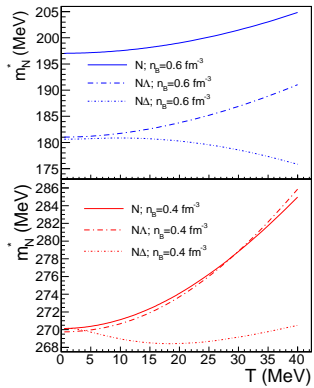}
\caption{$T$-dependence of nucleon Dirac effective mass,
  eq.(\ref{eq:meffD}), for the
  cases considered in Fig. \ref{fig:Xi}.
}
\label{fig:Diracmass} 
\end{figure}

Eq. (\ref{eq:meffD}) states that the $T$- and $n_B$-dependence of the
Dirac effective mass of any baryon is governed by the corresponding
dependencies of $\bar \sigma$.
$m^*_{n/p}(T,n_B)$ are illustrated in Fig. \ref{fig:Diracmass}.
Fig. \ref{fig:fields} shows that
$\Delta$-admixed matter singles out by $\bar \sigma$ increasing both with temperature
and density.
As such, contrary to what happens in nucleonic matter
and $\Lambda$-admixed nuclear matter, in $\Delta$-admixed nuclear matter
the Dirac effective mass of nucleons decreases with density and,
under specific thermodynamic conditions, also with temperature.
The validity limit of the model is reached when nucleon effective Dirac masses
vanish. For a parameter study of the maximum baryonic densities reachable
in $N\Delta$-matter, see Sec. \ref{sec:ND}.

\begin{figure*}
  \begin{center}
\includegraphics[width=0.5\columnwidth]{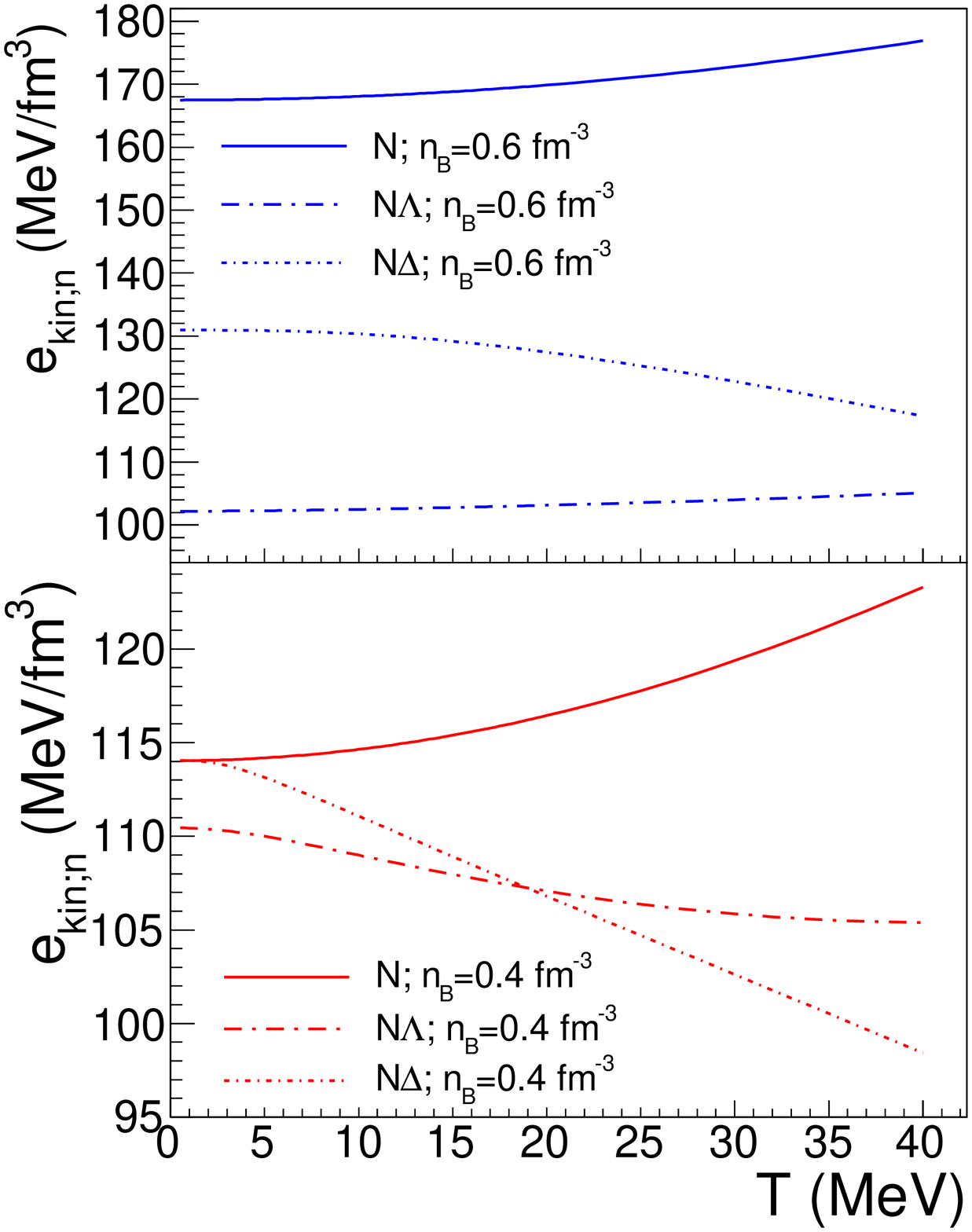}
\includegraphics[width=0.5\columnwidth]{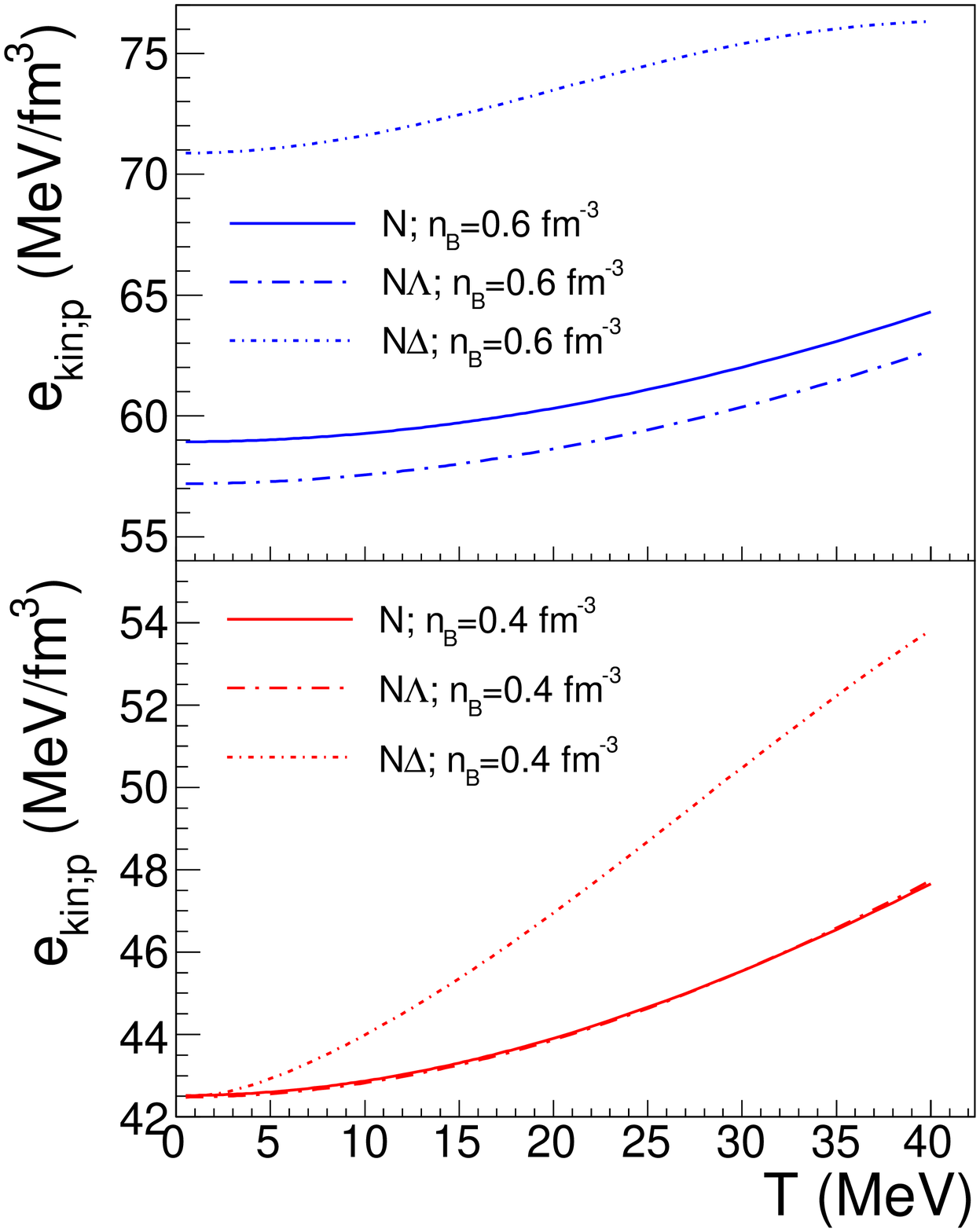}
\includegraphics[width=0.5\columnwidth]{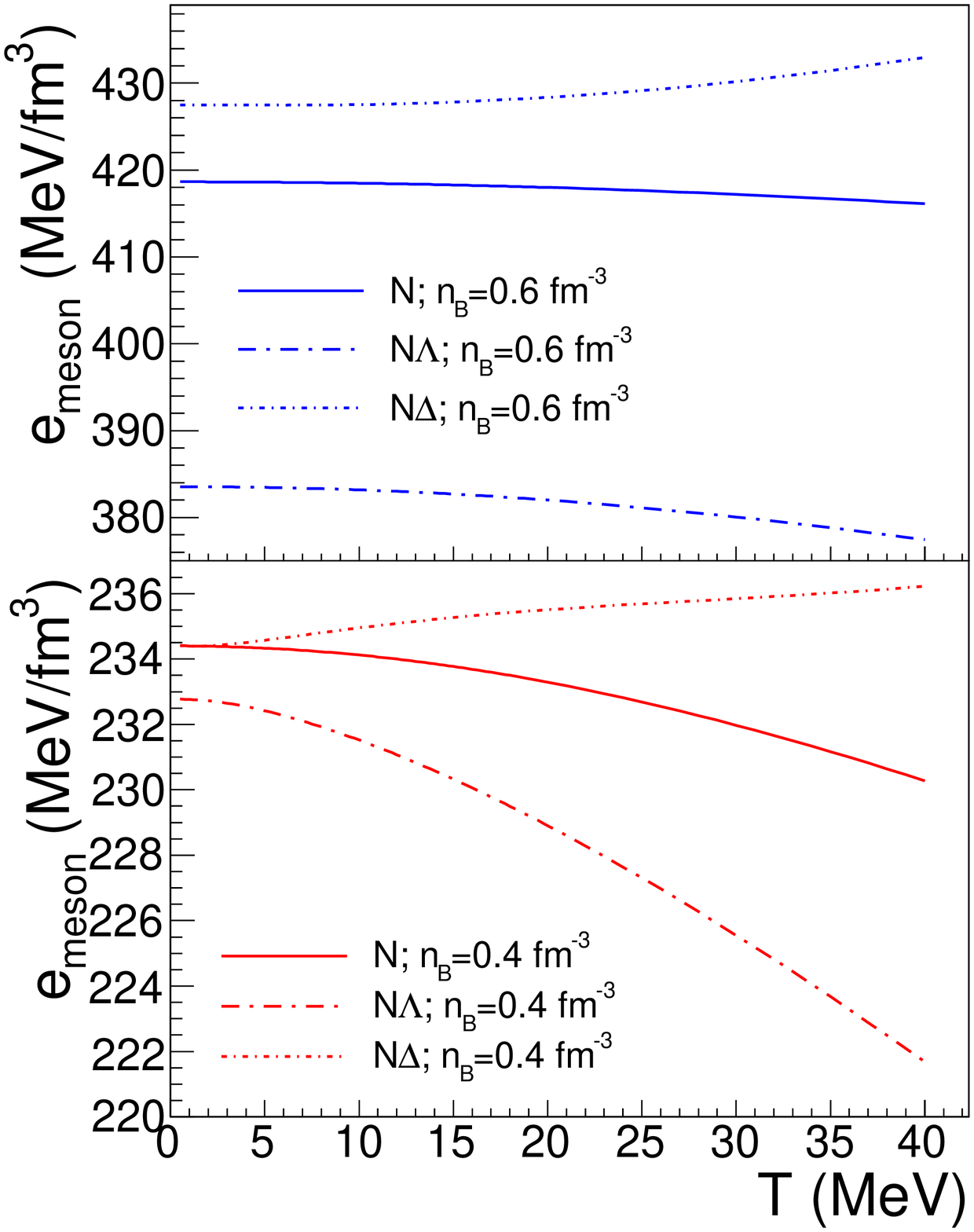}
\end{center}
  \caption{$T$-dependence of different terms in eq. (\ref{eq:ebar}).
    Results corresponding to $N$-, $N\Lambda$- and $N \Delta$-matter
    with $Y_Q=0.3$ at $n_B=0.4~{\rm fm}^{-3}$ and $0.6~{\rm fm}^{-3}$.    
}
\label{fig:suppl_en} 
\end{figure*}

\begin{figure*}
\includegraphics[width=0.5\columnwidth]{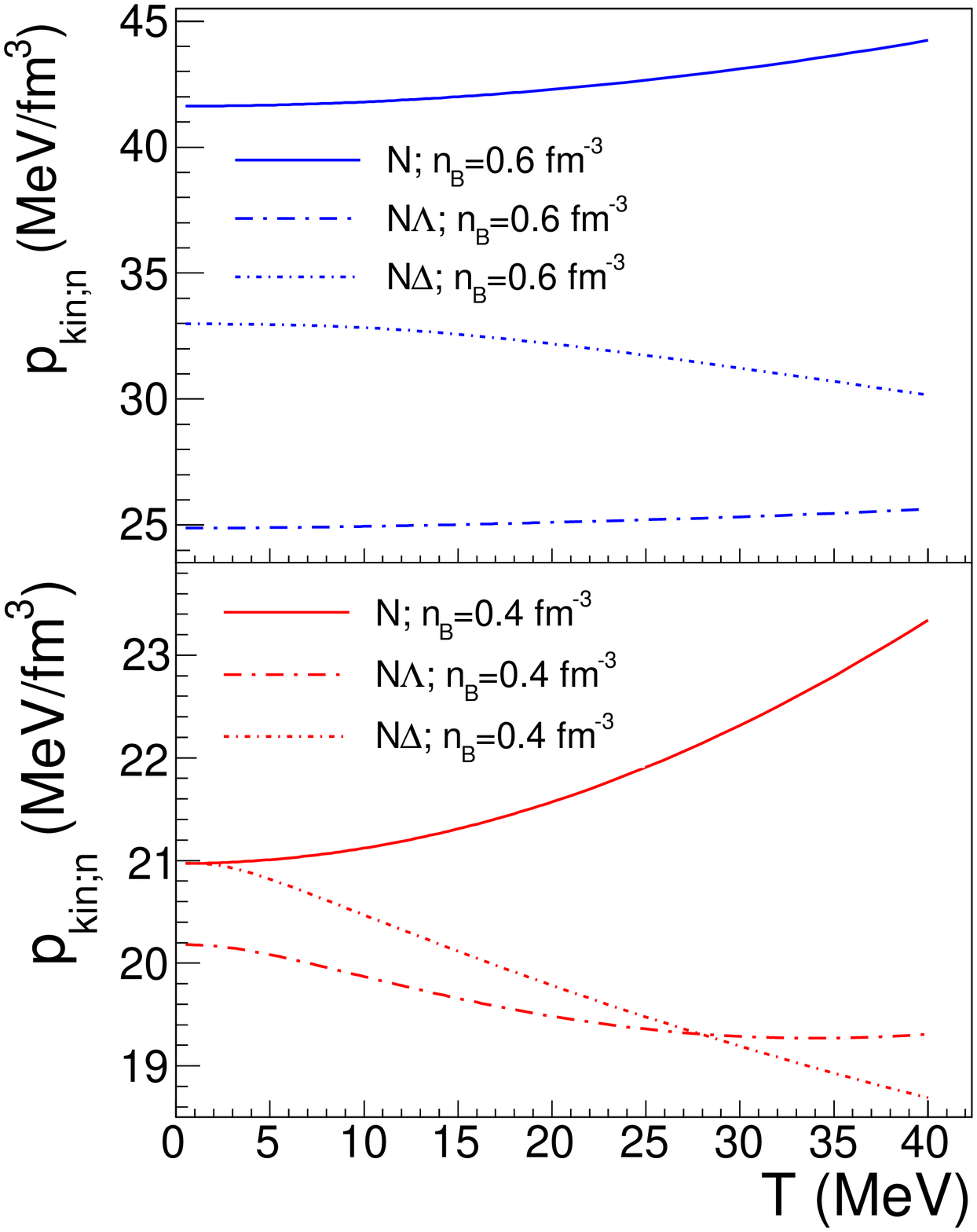}
\includegraphics[width=0.5\columnwidth]{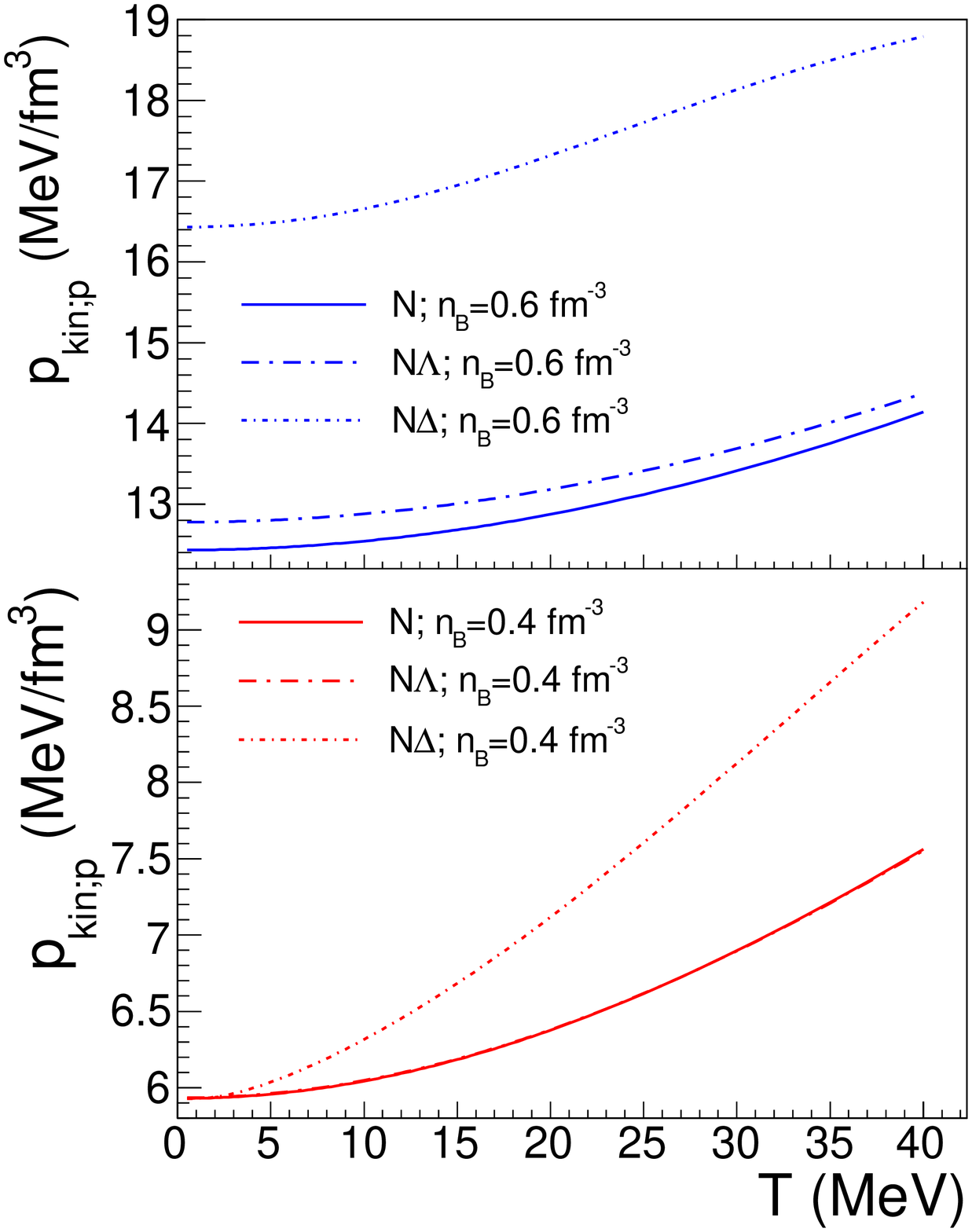}
\includegraphics[width=0.5\columnwidth]{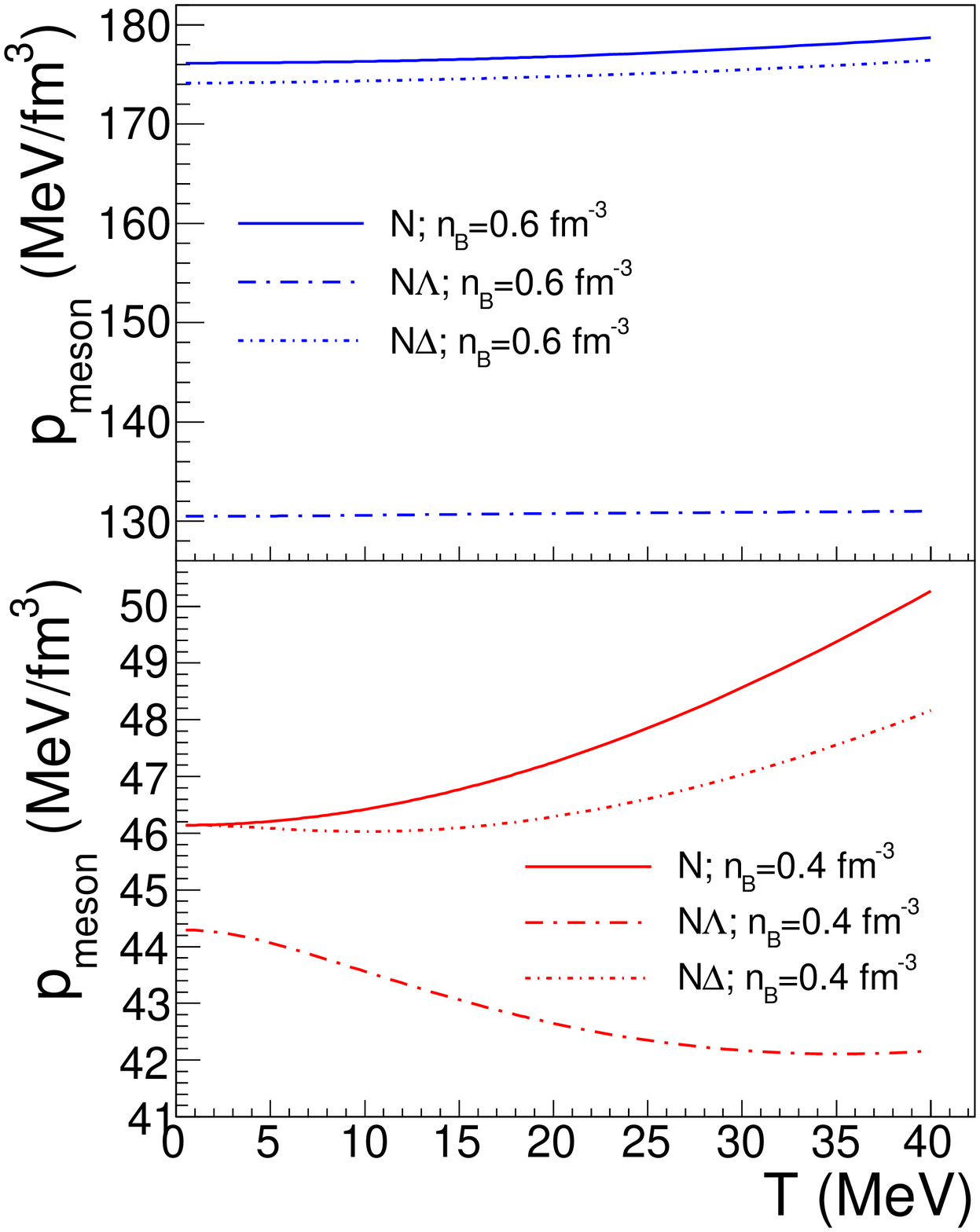}
\includegraphics[width=0.5\columnwidth]{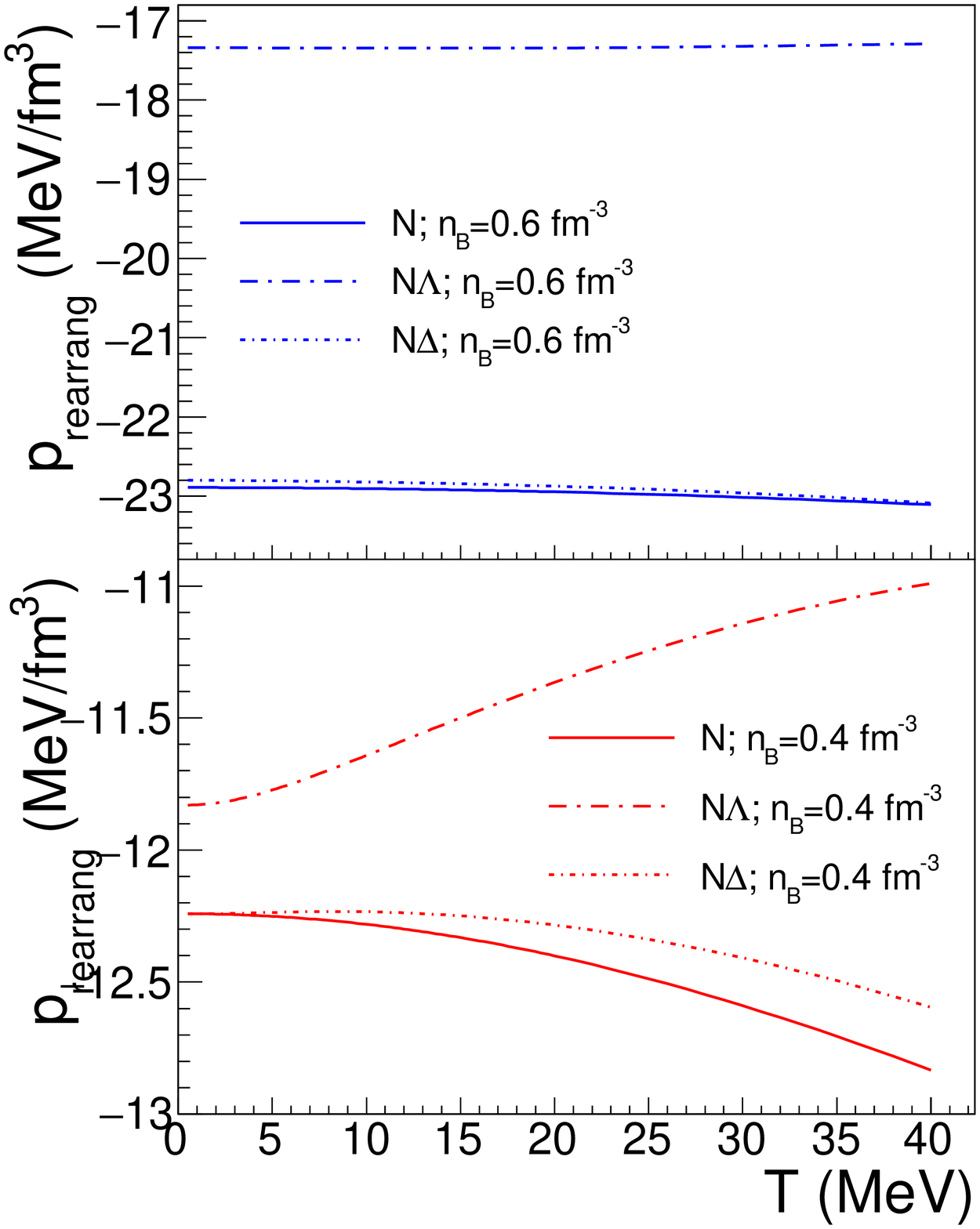}
\caption{The same as in Fig. \ref{fig:suppl_en} for different terms in
  eq. (\ref{eq:pbar}).
}
\label{fig:suppl_p} 
\end{figure*}

Contributions of different terms in eqs. (\ref{eq:ebar}) and (\ref{eq:pbar})
are illustrated in Figs. \ref{fig:suppl_en} and, respectively, \ref{fig:suppl_p}
as function of $T$ for different values of baryonic density and matter compositions.

In some cases, neutron and proton kinetic energy densities and pressures increase with $T$.
These situations correspond to contributions of high momentum states in eqs. (\ref{eq:ekin})
and (\ref{eq:pkin}) that
underscore the $T$-reduction of effective chemical potentials.
For $N\Lambda$-matter at $n_B=0.4~{\rm fm}^{-3}$ and $N\Delta$-matter the kinetic contribution
of neutrons manifests an opposite behavior. These situations correspond to the cases 
where $\mu^*_{n}(T)$ shows a steep decrease, see Fig. \ref{fig:mueff}.
Quite remarkably, with the exception of protons in $N\Lambda$ matter at $n_B=0.6~{\rm fm}^{-3}$,
the qualitative patterns of $e_{kin;n/p}(T)$ and $p_{kin;n/p}(T)$ as well
as the relative arrangement of curves corresponding to different mixtures and/or
baryonic densities is the same.

The $T$-dependence of $p_{meson}$ shows a complex behavior: at high densities where mesonic fields
show small variation with $T$, a tiny increase with $T$ is obtained;
for $N$- and $N\Delta$-matter with $n_B=0.4~{\rm fm}^{-3}$, $p_{meson}$ increases with $T$; in the
first (second) case this is mainly due to the decrease (increase) with $T$
of $\bar \sigma$ ($\bar \omega$), see Fig. \ref{fig:fields};
for $N\Lambda$ matter with $n_B=0.4~{\rm fm}^{-3}$, $p_{meson}$ has a non-monotonic behavior due
to the interplay between $\bar \sigma$ and $\bar \omega$, both decreasing with $T$.

Similarly to $p_{meson}$ at $n_B=0.6~{\rm fm}^{-3}$ and for the same reasons also
$e_{meson}$ at $n_B=0.6~{\rm fm}^{-3}$ shows a weak sensitivity to $T$. In what regards the behavior at $n_B=0.4~{\rm fm}^{-3}$,
one notes that $e_{meson}(T)$ increases (decreases) in $N\Delta$ ($N$ and $N\Lambda$) matter.
Fig. \ref{fig:fields} shows that this is the direct consequence of $T$-dependence of dominant
$\bar \sigma$ and $\bar \omega$ fields.

For fixed densities, the $T$-dependence of the rearrangement pressure depends on the
$T$-evolution of average mesonic fields.
Little (strong) $T$-dependence of these fields at high (low) densities,
see Fig. \ref{fig:fields}, explain the behavior shown in the right panel.

\begin{figure}
\includegraphics[width=0.49\columnwidth]{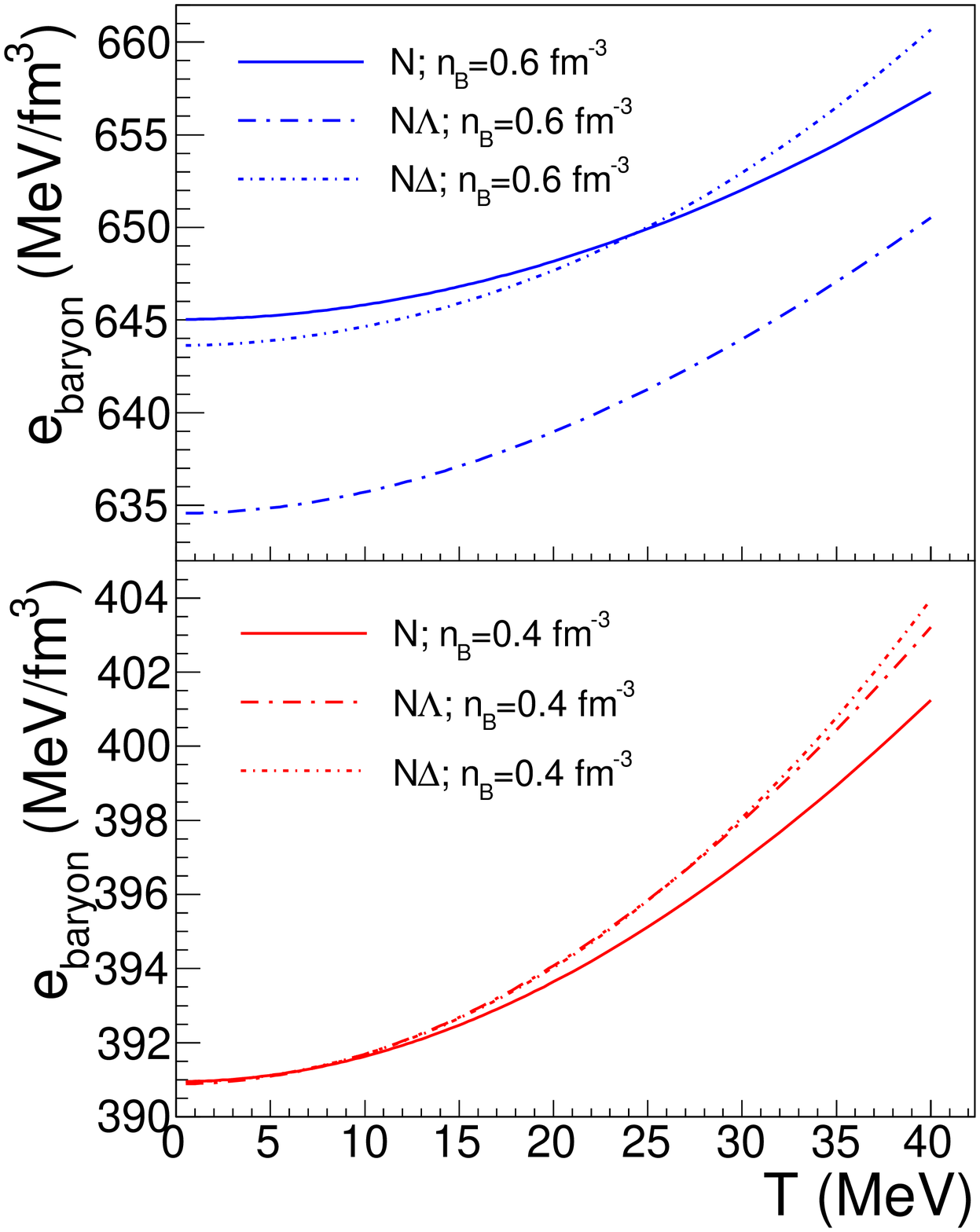}
\includegraphics[width=0.49\columnwidth]{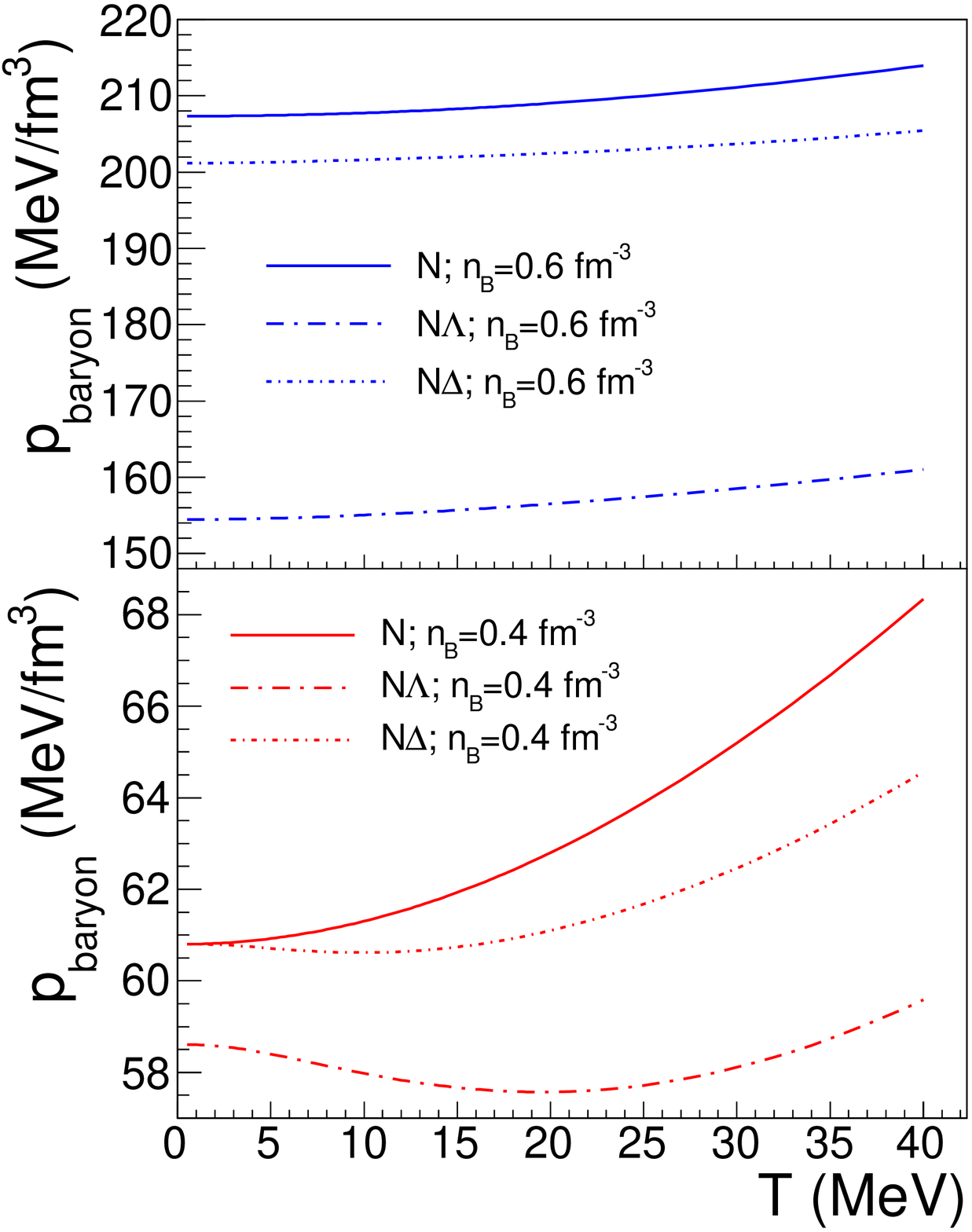}
\caption{$T$-dependence of energy density and pressure.
  Results corresponding to $N$-, $N\Lambda$- and $\Delta N$-matter
  with $Y_Q=0.3$
  at $n_B=0.4~{\rm fm}^{-3}$ and $0.6~{\rm fm}^{-3}$.
}
\label{fig:enp} 
\end{figure}

For the thermodynamic conditions under consideration here meson terms overshoot all other terms in $e_{baryon}$ and $p_{baryon}$.
Out of the different terms in eq. (\ref{eq:pbar}) 
the highest $T$-dependence is shown by $p_{meson}$.
This means that, to a large extend, $p_{baryon}(T)$ will reflect the
$T$-dependence of $p_{meson}$.
The confirmation is provided by the right panel in Fig. \ref{fig:enp}.
For $n_B=0.4~{\rm fm}^{-3}$ $p_{baryon}^{(N\Lambda)}(T)$ has a $U$-shape,
while in all other cases $p_{baryon}$ increases with $T$.
The left panels in Fig. \ref{fig:enp} show that, in spite of the
complex $T$-dependence of various terms entering eq. (\ref{eq:ebar}),
$e_{baryon}$ increases with $T$.

\section{Thermodynamic stability of $\Delta$-admixed nuclear matter: the case of $T=0$}
\label{sec:ND}

Within QCD the Delta-resonance is explained as a spin-isospin excitation
of the nucleon into a $\left(J,I \right)=\left(3/2, 3/2\right)$ configuration.
Its strong coupling with the continuum of pion-nucleon $p$-wave scattering states
explains its rapid decay in vacuum, within a half-life of about
$t_{1/2}\approx 10^{-23}~{\rm s}$. Due to the Pauli blocking, the $\Delta$s
are nevertheless considered to be stabilized when embedded in a medium.

The fact that, contrary to other baryons, $\Delta$-resonances are excited in
various types of reactions ~\cite{Lenske_PPNP_2018}, can in principle
be exploited to constrain the interaction potential with nucleons.
The data are nevertheless scarce such that the consequences of the onset
of $\Delta$s in neutron stars are typically evaluated by allowing
to the coupling constants of $\Delta$ to various mesonic fields to span wide ranges
\cite{Drago_PRC_2014,Kolomeitsev_NPA_2017}. Herein the same modus operandi will be adopted.

Refs. \cite{Cai_PRC_2015,Zhu_PRC_2016} show that, if no other exotic degree of freedom is considered,
nucleation of $\Delta$s reduces the maximum mass of NS by up to $\approx 0.5M_{\odot}$
and radii of intermediate mass NS by up to $\approx 2~{\rm km}$.
Both effects are due to the pressure decrease that accompanies the onset of any extra
particle degree of freedom. The reduction of maximum mass reflects the reduction of pressure at
baryonic densities equal to the values of the maximum mass configurations.
The reduction of radii of intermediate mass NS reflects the evolution of pressure at the power $1/4$ at
lower densities, typically $n_{sat} \lesssim n_B \lesssim 2n_{sat}$ \cite{Lattimer_PhysRep_2007}.
If also hyperons are accounted for the only notable effect is the reduction of NS radii \cite{Drago_PRC_2014,Kolomeitsev_NPA_2017,Li_PLB_2018,Ribes_2019,Li_ApJL_2019,Raduta_MNRAS_2020}.
The explanation resides in that, due to their attractive potential, $\Delta$s start nucleating at
lower densities than the hyperons.
The magnitude of these modifications nevertheless depends on the underlying
nucleon effective interaction, strength of $\Delta N$ potential and $\Delta$s effective mass.

Most recently Ref. \cite{Raduta_PLB_2021} has highlighted a series of other modifications.
The first consists in the fact that, if the $\Delta N$ interaction potential is attractive enough,
$\Delta$-admixed nuclear matter becomes thermodynamically instable. These instabilities persist
at finite temperatures and out of the $\beta$-equilibrium condition, which might in principle impact
astrophysical scenarios like CCSN, collapse into a BH and BNS mergers \cite{Peres_PRD_2013,Aloy_MNRAS_2019}.
The second finding regards the shift of the nucleonic dUrca threshold to low densities, with
obvious impact on the thermal evolution of isolated and accreting NS.
This modification arises because $\Delta^-$, which is the first $\Delta$ to be populated
at $\beta$-equilibrium, partially supresses the neutralizing electron gas which, in turn, regulates
the relative abundances of neutrons and protons.
Though also these modifications depend on the strength of the $\Delta$ effective
interaction they manifest over domains of the parameter space compatible with present 
astrophysical constraints.

In this section we shall investigate the following aspects:
i) the dependence of the $\Delta$-potential in nuclear matter and
ii) the maximum value of baryonic particle number density on $\Delta$ couplings to mesonic fields,
iii) thermodynamic instabilities of cold $N \Delta$ and $N \Delta e$ matter.
For ii) and iii) the limiting values $Y_Q=0.01$ and 0.5 will be considered.
As in Ref. \cite{Raduta_PLB_2021} hyperons and muons are disregarded for simplicity.
As \cite{Drago_PRC_2014,Kolomeitsev_NPA_2017,Li_PLB_2018,Ribes_2019,Li_ApJL_2019}
we shall allow the strength functions to cover wide domains of values,
$0.90 \leq x_{\sigma \Delta} \leq 1.50$,
$-0.20 \leq x_{\sigma \Delta}-x_{\omega \Delta} \leq 0.20$.
For brevity $x_{\rho \Delta}=1$.
For nucleons the DD2 \cite{DD2} effective interaction is employed; note that
this interaction is different than the one used in \cite{Raduta_PLB_2021}.

\begin{figure}
  \begin{center}
    \includegraphics[width=0.79\columnwidth]{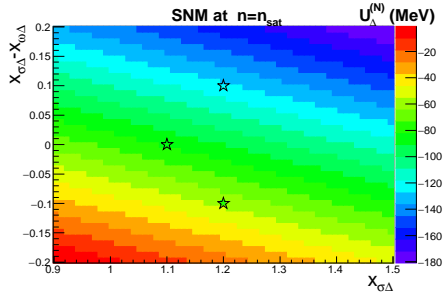}
  \end{center}
  \caption{Single-particle potential of $\Delta$ isobars 
    in saturated symmetric nuclear matter, $U_{\Delta}^{(N)}(n_{sat})$, 
    for $x_{\rho\Delta}=1$.
    The symbols correspond to the following sets of $(x_{\sigma\Delta},x_{\omega\Delta})$:
    (1.1,1.1), (1.2, 1.1), (1.2, 1.3) for which $U_{\Delta}^{(N)}=-83~{\rm MeV}$, -124 MeV and -57 MeV.
  }
\label{fig:UDN} 
\end{figure}

\begin{figure}
  \begin{center}
    \includegraphics[width=0.79\columnwidth]{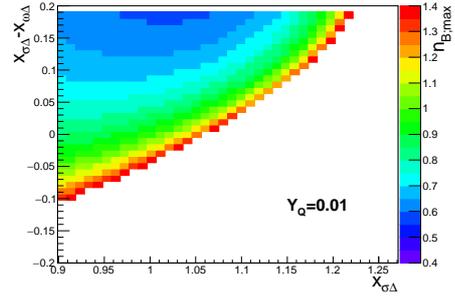}
    \includegraphics[width=0.79\columnwidth]{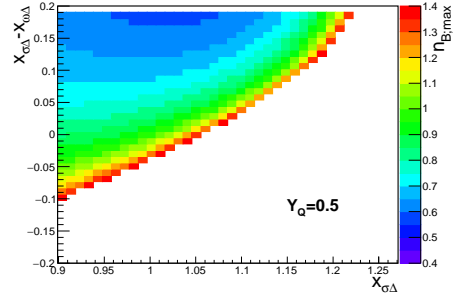}
  \end{center}
  \caption{Maximum reachable value of $n_B$ (in ${\rm fm}^{-3}$)
    in cold $\Delta$-admixed
    nuclear matter with $Y_Q=0.01$ (top) and $Y_Q=0.5$ (bottom),
    determined by vanishing nucleon Dirac effective mass.
  }
  \label{fig:nbmax} 
\end{figure}

\begin{figure}
  \begin{center}
    \includegraphics[width=0.79\columnwidth]{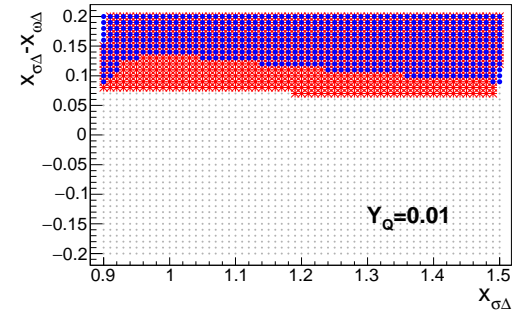}
    \includegraphics[width=0.79\columnwidth]{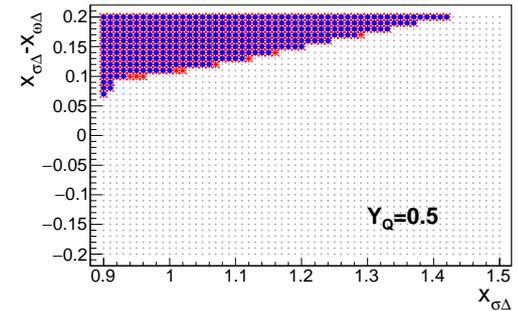}
  \end{center}
  \caption{Parameter sets which allow for $\Delta$-driven spinodal instabilities
    in cold $\left( N, \Delta \right)$ (red)
    and $\left( N, \Delta, e \right)$ (blue)
    matter with $Y_Q=0.01$ (top) and $Y_Q=0.5$ (bottom).
  }
  \label{fig:instab} 
\end{figure}

\begin{figure}
  \begin{center}  
    \includegraphics[width=0.49\columnwidth]{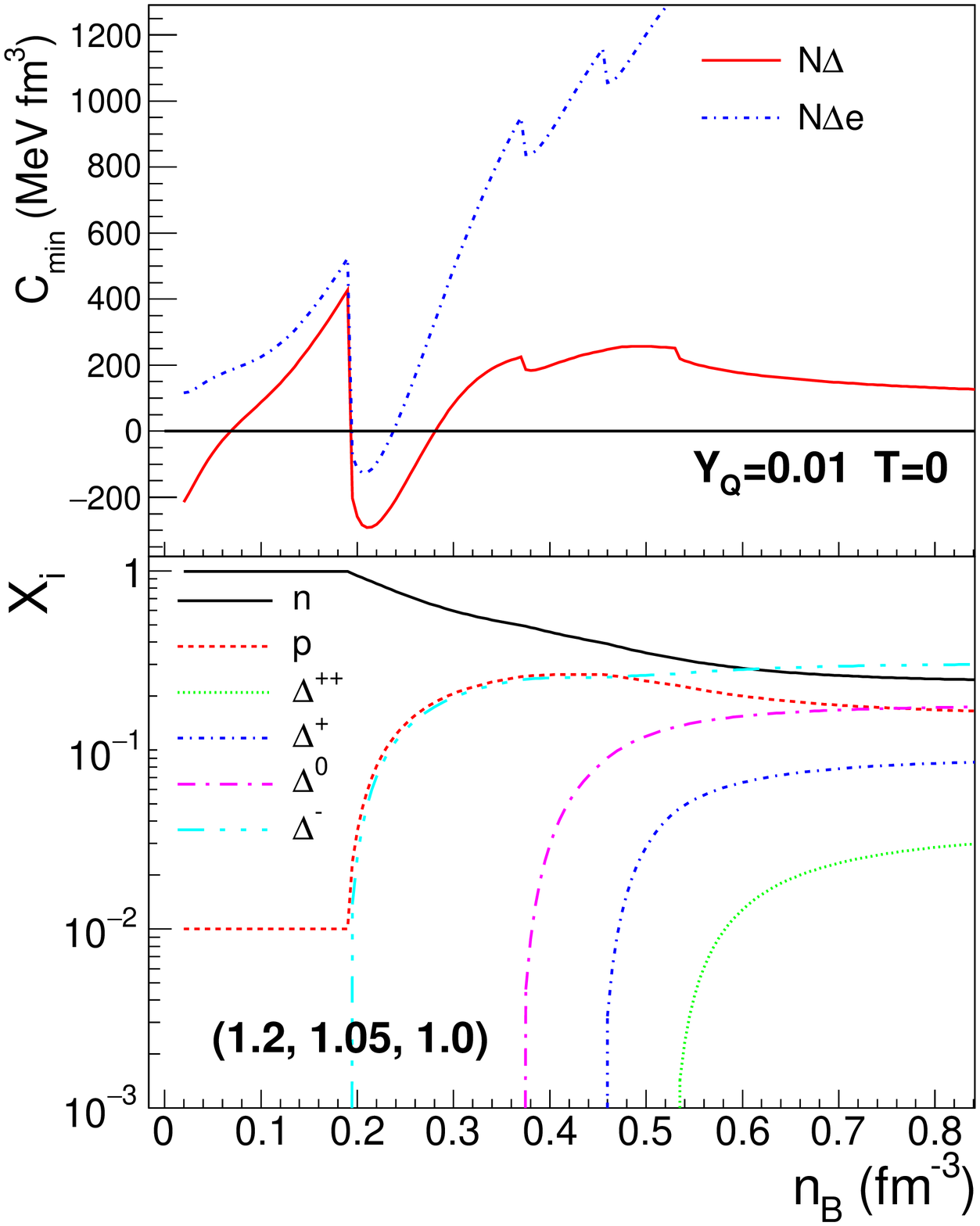}
    \includegraphics[width=0.49\columnwidth]{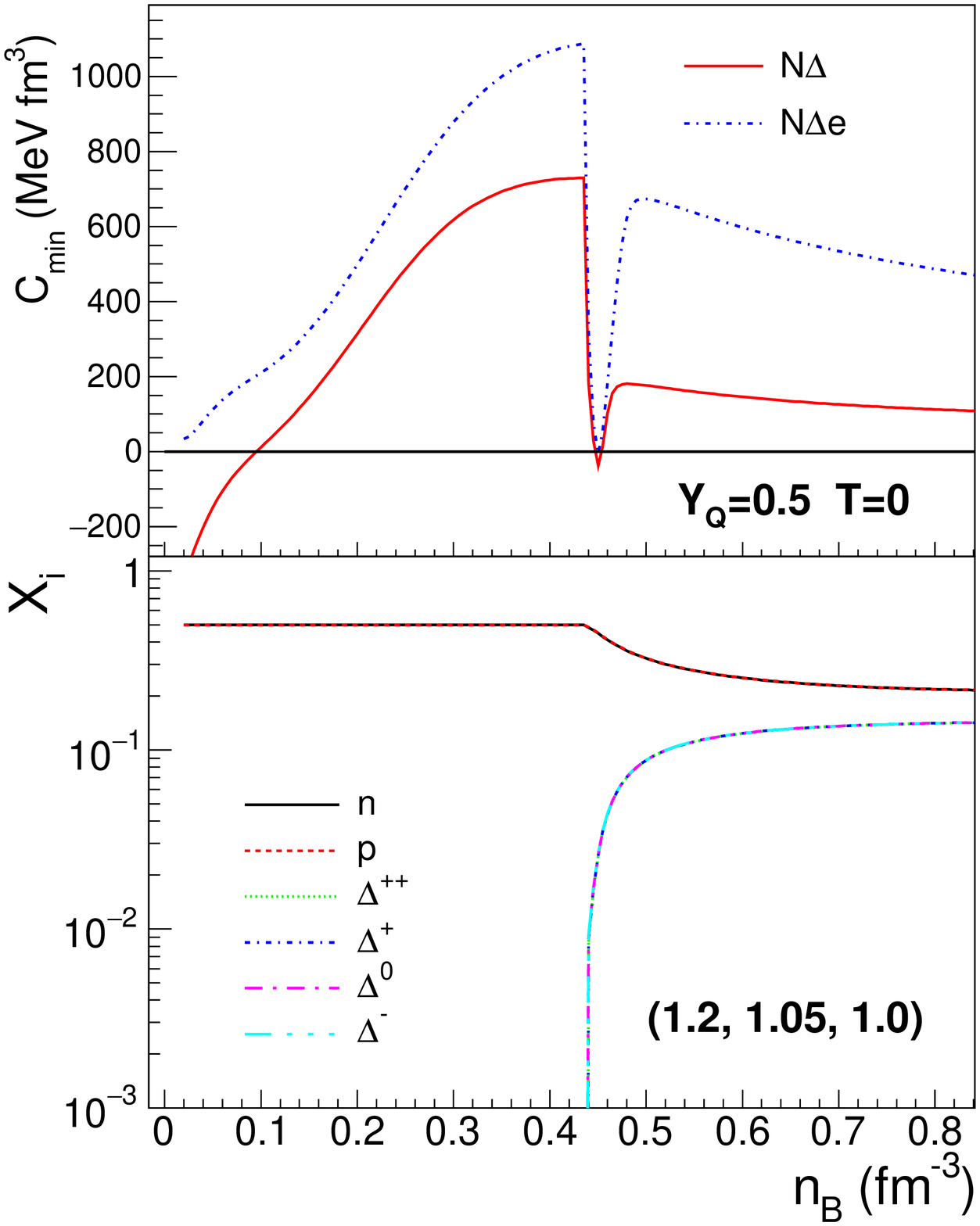}
  \end{center}
  \caption{
    Top: Minimum value of the curvature matrix as function of
    $n_B$ in cold $\Delta$-admixed nuclear matter;
    results corresponding to pure baryonic matter (baryonic matter plus
    electrons) are illustrated with red (blue) lines.
    Bottom: Relative particle abundances as function of
    $n_B$.
    Left (right) panels correspond to $Y_Q=0.01$ ($Y_Q=0.5$).
    The values of coupling constants are mentioned on the figure.
  }
  \label{fig:Cmin} 
\end{figure}

\subsection{$U_{\Delta}^{(N)}$ potentials}
\label{ssec:UDN}

The well depth of the $\Delta$-potential in nuclear matter writes~\footnote{Eq. (\ref{eq:UDN}) corrects eq. (1)
  in \cite{Raduta_PLB_2021}, where the rearrangement term has been omitted.
  The calculations in \cite{Raduta_PLB_2021} have been nevertheless performed with the correct expression.}
\begin{equation}
  U_{\Delta}^{(N)}=-g_{\sigma \Delta} \bar \sigma+g_{\omega \Delta}
  \bar \omega + g_{\rho \Delta} t_{3} \bar \rho+\Sigma_R,
  \label{eq:UDN}
  \end{equation}
where $t_3$ is the third component of the isospin, with the convention $t_{3\Delta^{++}}=3/2$.
[Herein the baryonic particle number density dependence of
  the potential, coupling constants, rearrangement term and mesonic fields has been omitted.]
Eq. (\ref{eq:UDN}) shows that large values of $g_{\sigma \Delta}$
and low values of $g_{\omega \Delta}$ entail more attractive potentials.
It also shows that in neutron rich matter, where $\bar \rho<0$,
the potential felt by positively charged isobars
is larger in absolute values than the one felt by negative isobars.
In symmetric nuclear matter, where $\bar \rho=0$, all isobars experience
the same potential.

Experimental data from pion-nucleus scattering and pion photo-production, electron scattering
on nuclei and electromagnetic excitation of the $\Delta$-baryons,
compiled by \cite{Drago_PRC_2014,Kolomeitsev_NPA_2017}, have been translated into a $\Delta$-potential
at rest in nuclear matter slightly more attractive than the one of nucleons,
$-30~{\rm MeV}+U_{N}^{(N)} \lesssim U_{\Delta}^{(N)} \lesssim U_{N}^{(N)}$.
Recent analysis of inclusive quasi-elastic electron scattering data on nuclear targets
\cite{Bodek_EPJC_2020} led to $U_{\Delta}^{(N)} \approx 1.5 U_{N}^{(N)}$, which corresponds to the
maximum attraction boundary cited above.
Comparison between experimental pion production data in energetic heavy ion collisions and
results of a quantum molecular transport model have been exploited in order to extract, in addition
to the isoscalar $\Delta$ potential, also the isovector component \cite{Cozma_EPJA_2021}.
The results reveal correlations of these two components with the value of Landau effective mass of
$\Delta$ and slope of the symmetry energy.

Fig. \ref{fig:UDN} illustrates the range of values spanned by $U_{\Delta}^{(N)}$
in saturated symmetric nuclear matter.
It comes out that $U_{\Delta}^{(N)}(n_{\rm sat})$ is degenerate with respect to
these two coupling constants and for all considered combinations
$\left( x_{\sigma \Delta}, x_{\omega \Delta} \right)$ the potential is attractive.
The values taken by $U_{\Delta}^{(N)}(n_{\rm sat})$ when the coupling constants take the values
employed by the three new general purpose EoS models proposed in this paper, see Sec. \ref{sec:Models},
are signaled by symbols.
For $(1.1,1.1,1.0)$ $U_{\Delta}^{(N)}$ falls in the domain
$-30~{\rm MeV}+U_{N}^{(N)} \lesssim U_{\Delta}^{(N)} \lesssim U_{N}^{(N)}$ \cite{Drago_PRC_2014,Kolomeitsev_NPA_2017},
while more (less) attraction is obtained for $(1.2,1.1,1.0)$ [$(1.2,1.3,1.0)$].

\subsection{Maximum baryonic particle number density}

Refs. \cite{Spinella-PhD,Lavagno_2019} showed that, for wide domains of the parameter set,
the maximum mass of $\Delta$-admixed NS is much lower than the one corresponding to nucleonic NS built upon
the same effective nucleonic interaction.
\cite{Raduta_PLB_2021} showed that this does not stem from the softening of the $P(e)$ EoS upon nucleation
of $\Delta^-$ but from the inability to reach baryon densities high enough to sustain massive NS,
due vanishing nucleon Dirac effective masses. For the $\Delta$-induced modifications of $m^*_N$,
see Sec. \ref{sec:HeavyBaryons}. EoS models falling in this situation will obviously not meet the lower
bound on NS maximum mass \cite{Antoniadis2013} and should be ruled out. 

Maximum values of $n_B$ allowed by different combinations of
$\left( x_{\sigma \Delta},x_{\omega \Delta}\right)$ but smaller than (the arbitrarily chosen value)
$1.4~{\rm fm}^{-3}$ are depicted in Fig. \ref{fig:nbmax}.
The panels corresponding to $Y_Q=0.5$ and 0.01 are very similar,
as expected due to the weak sensitivity of the $\bar \sigma$ to the isospin asymmetry.
The color levels show that high enough values of $n_{B;max}$ may be reached for any value of
$x_{\sigma \Delta}$ provided that $x_{\omega \Delta}$ is tuned accordingly.
Some of the thus selected sets of parameters will though not agree with constraints on $U_{\Delta}^{(N)}$,
see Sec. \ref{ssec:UDN}.

\subsection{Thermodynamic (in)stabilities}

Sufficiently attractive interaction potentials are known to lead to thermodynamic instabilities.
The best documented example in nuclear physics corresponds to sub-saturated nuclear matter where
a liquid-gas like phase transition is predicted by a large variety of models, 
including phenomenological non-relativistic \cite{Ducoin_NPA_2006,Ducoin_NPA_2007,Rios_NPA_2010} and 
relativistic \cite{Ducoin_PRC_2008,Typel_EPJA_2014} mean field models 
and microscopic approaches \cite{Rios_PRC_2008,Carbone_PRC_2018}.
This phase transition is nevertheless not expected to play a role in NS or core-collapse supernovae
as it is suppressed by the neutralizing electron gas.
According to \cite{Gulminelli_PRC_2012,Gulminelli_PRC_2013,Oertel_JPG_2015,Torres_PRC_2017}
strangeness-driven instabilities may manifest in supra-saturated matter and, contrary to the
case discussed above, are not significantly altered by electrons such that they might be explored along
the $\beta$-equilibrium path as well as out of $\beta$-equilibrium in the neutrino transparent
matter.
In lack of sufficient knowledge of effective baryon interactions in strangeness $S=1$ and 2 channels this
possibility nevertheless remains speculative.
Ref. \cite{Lavagno_2019,Raduta_PLB_2021} proved that CDFT models allow
for $\Delta$-driven instabilities in NS matter. These instabilities might occur for strength
parameters in accord with available data and extend over a narrow density domain at
densities slightly exceeding $n_{sat}$. 

Spinodal instabilities manifest as convexity anomalies of the thermodynamic potential
and are mathematically signaled by negative values of the curvature matrix
$C_{i,j}=\partial \mu_i/\partial n_j$, where $\mu$ and $n$ stand for the chemical potential and
number density of the two conserved charges $i,j=B,Q$.

We extend here the investigation performed in \cite{Raduta_PLB_2021} to matter with
fixed charged fraction. Properties of purely baryonic matter and mixture of baryons and
electrons with $Y_Q=Y_e$ are confronted in Fig. \ref{fig:instab}.
The cases $Y_Q=0.01$ and 0.5 are considered.
$\Delta$-driven instabilities exist for both systems; the parameter sets which allow
for them are more numerous for $Y_Q=0.01$ than for $Y_Q=0.5$.
This first result is counter-intuitive as the domain of thermodynamic instabilities in
dilute nuclear matter shrinks as $Y_Q$ diminishes.
It can nevertheless be understood considering that production of $\Delta$s, in particular
$\Delta^-$, is favored by small $Y_Q$-values.
Indeed, Fig. \ref{fig:Cmin} shows that the $\Delta$ onset occurs at $n_B \approx 0.2~{\rm fm}^{-3}$
($n_B \approx 0.42~{\rm fm}^{-3}$) for $Y_Q=0.01$ ($Y_Q=0.5$).
Moreover at large densities, $\Delta$s are relatively more important in matter with low $Y_Q$-values.
Fig. \ref{fig:instab} also shows that the neutralizing electron gas has almost no consequence
at $Y_Q=0.5$ while at $Y_Q=0.01$ it suppresses the instabilities for
$0.05 \lesssim x_{\sigma\Delta}-x_{\omega\Delta} \lesssim 0.1$.
Having in mind the behavior of dilute nuclear matter also this result is surprising.
It can nevertheless be explained considering that the surface of the thermodynamic potential
is altered more by the early and strong production of $\Delta$s in matter with low $Y_Q$, see bottom
panels of Fig. \ref{fig:Cmin}. 
Finally the density domains affected by instabilities are wider in matter with low $Y_Q$.

We note that, when the contribution of the electron gas is accounted for, none of the models of
$Y\Delta$-admixed matter introduced in this paper manifests instabilities.
At variance with this pure baryonic matter is stable [unstable] for $\left(1.1, 1.1, 1.0 \right)$
and $\left(1.2, 1.1, 1.0 \right)$
[$\left( 1.2, 1.3, 1.0 \right)$]. 

\section{Finite temperature behavior}
\label{sec:FiniteT}

In this section we shall analyze thermal properties of a bunch of selected
models accounting for various particle d.o.f. The behavior of baryonic matter
and/or stellar matter, which means with lepton and photon gases contributions
included, will be considered. Properties of pure baryonic matter are inferred
from the corresponding tables on \textsc{CompOSE} database or, when these are
not available, by subtracting from the values corresponding to stellar matter
the contribution of lepton and photon gases.

For each quantity we shall start by commenting the behavior of purely nucleonic matter
seen in Paper I.

\subsection{Energy density and pressure}
\label{ssec:ethpth}

Thermal effects on thermodynamic state variables like energy density and pressure
are conveniently highlighted by considering the difference between
the finite temperature and the zero temperature quantity
\begin{equation}
  X_{\mathit{th}}=X(n_B, Y_e, T)-X(n_B, Y_e, 0)~.
  \label{eq:Xth}
\end{equation}

\begin{figure*}
  \begin{center}
    \includegraphics[width=0.99\textwidth]{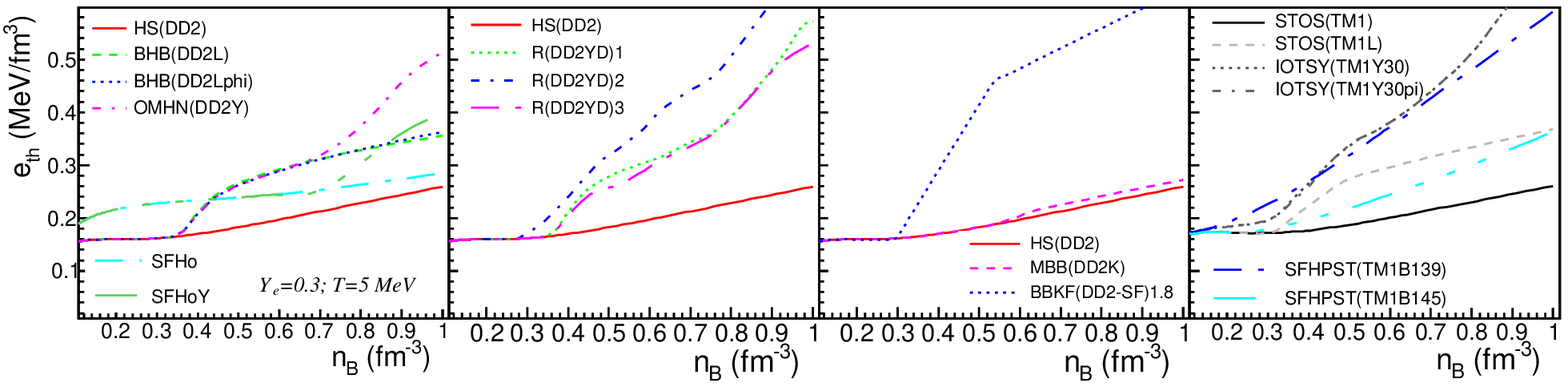}
    \includegraphics[width=0.99\textwidth]{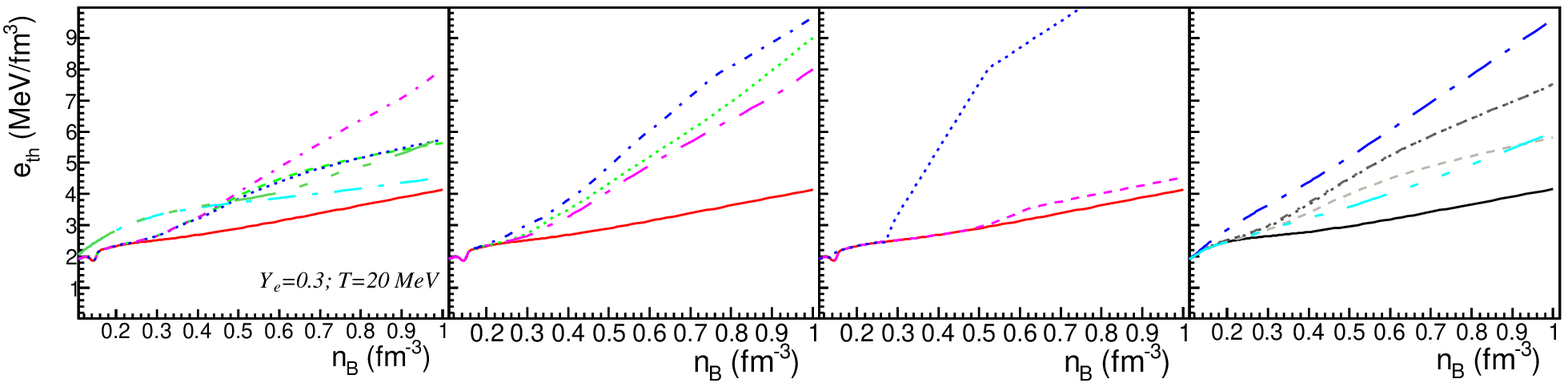}
    \includegraphics[width=0.99\textwidth]{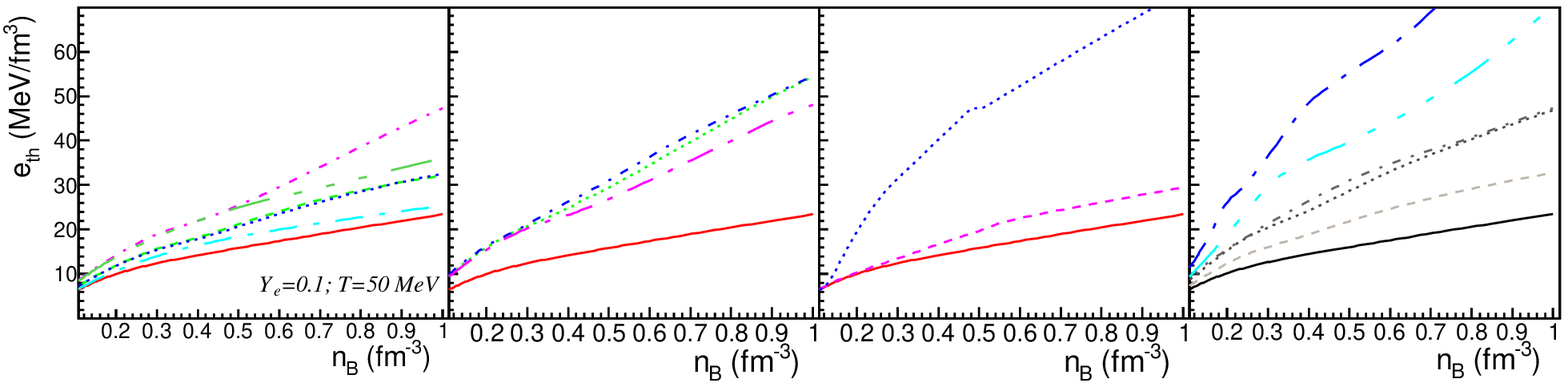}
   \end{center}
  \caption{Baryon (quark) contribution to the thermal energy density $e_{\mathit{th}}$,
  eq.~(\ref{eq:Xth}), as function of baryon number density for
  ($Y_e=0.3$, $T$=5 MeV) (top), ($Y_e=0.3$, $T$=20 MeV) (middle)
  and ($Y_e=0.1$, $T$=50 MeV) (bottom).
  The results are depicted for various EoS models.
  The indices 1, 2 and 3 of R(DD2YDelta) models refer to the following
  sets of couplings of $\Delta$ to $\sigma$, $\omega$ and $\rho$ mesonic fields:
  (1.1, 1.1, 1.0), (1.2, 1.1, 1.0), (1.2, 1.3, 1.0).
  }
  \label{fig:eth}
\end{figure*}

\begin{figure*}
  \begin{center}
    \includegraphics[width=0.99\textwidth]{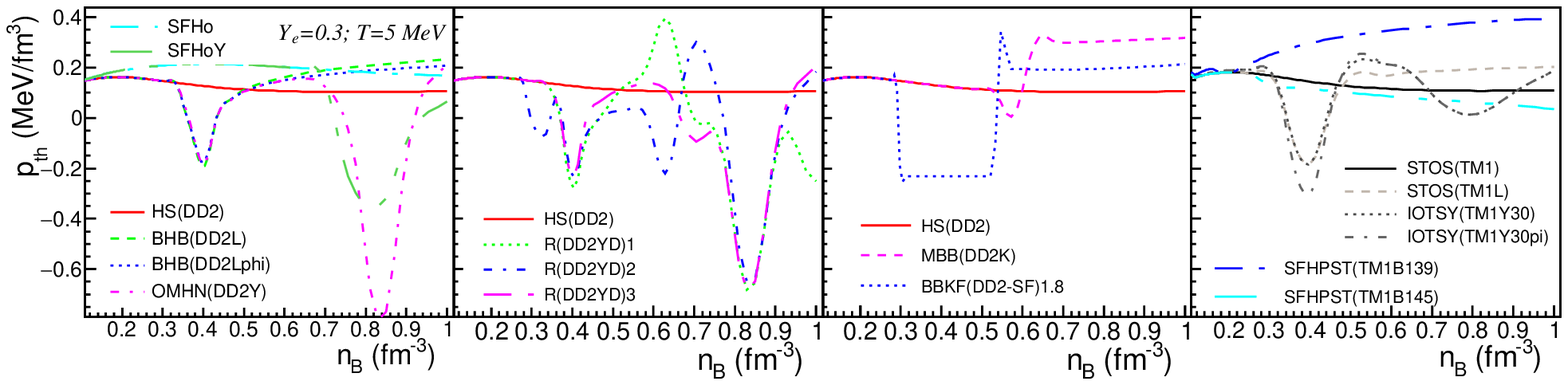}
    \includegraphics[width=0.99\textwidth]{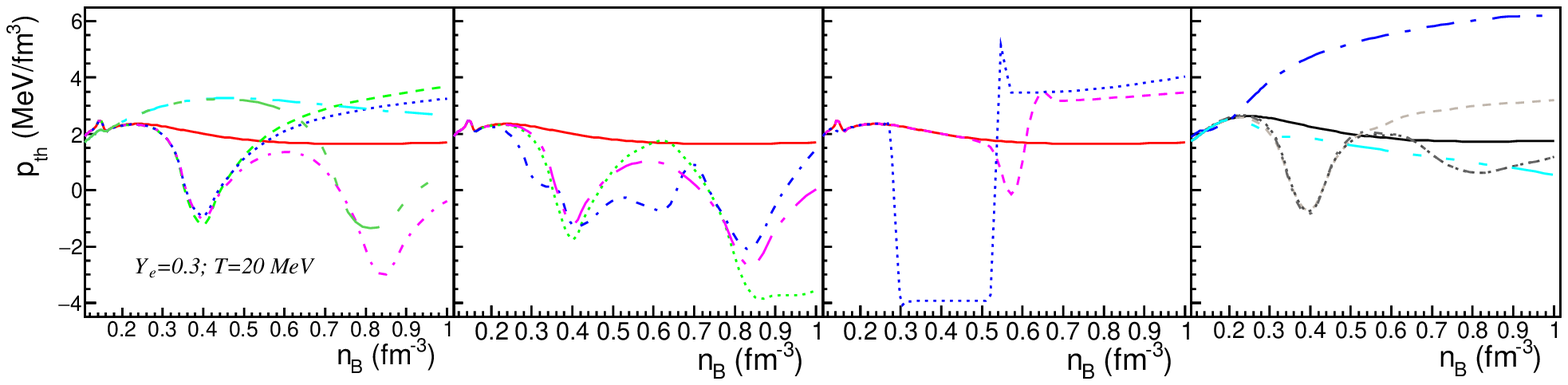}
    \includegraphics[width=0.99\textwidth]{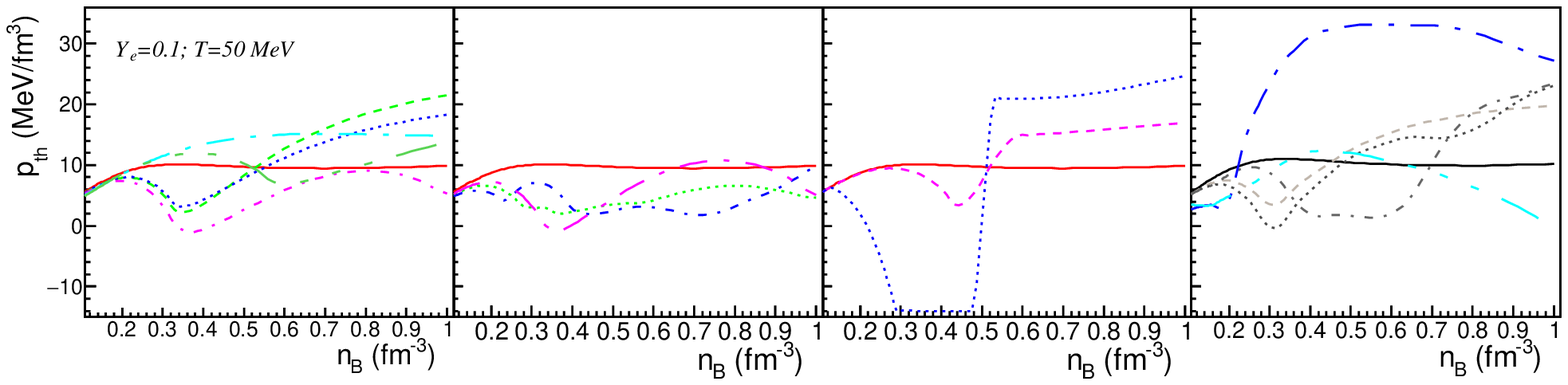}
  \end{center}
  \caption{Same as fig. \ref{fig:eth} for the baryon (quark) contribution to the thermal pressure.
  }
  \label{fig:pth}
\end{figure*}

In Paper I we have shown that in the limit of vanishing density the baryonic
thermal energy density $e_{th}$ and pressure $p_{th}$
approach zero and have no EoS-dependence, which means that whenever the interactions are
weak the system recovers the ideal gas behavior; beyond this limit a strong
EoS-dependence exists; at $n_B\gtrsim 1~{\rm fm}^{-3}$ the dispersion between non-relativistic
(relativistic) models is large (small); the relative significance of thermal effects
is more important at low densities than at high densities; pressure is more sensitive
to $T$ than energy; non-relativistic and relativistic models manifest different patters
of the $n_B$-dependence of thermal energy density and pressure.  The latter feature
is attributable to the different expressions of single particle energies in the two classes
of models together with the different density dependencies of the effective masses they depend on.
In agreement with previous findings in \cite{Constantinou_PRC_2014,Constantinou_PRC_2015}
Paper I showed that in non-relativistic models $e_{th}$ depends on the Landau effective mass
while $p_{th}$ depends on both Landau effective mass and its density dependence.
In relativistic models the correlation between $e_{th}$ and $p_{th}$ and Dirac effective mass appears
in that models with similar (different) density dependencies of this quantity produce similar (different)
thermal effects. This feature is visible also in Figs. \ref{fig:eth} and \ref{fig:pth}, where
thermal effects in HS(DD2) are similar to those in STOS(TM1) and different from those
in SFHo.

Figs. \ref{fig:eth} and \ref{fig:pth} further analyze the role of exotic particle d.o.f.,
hyperons, $\Delta$-resonances, $\pi$, $K^-$ and quarks, on baryonic thermal energy density
$e_{th}(n_B)$ and pressure $p_{th}(n_B)$ at various thermodynamic conditions.
Predictions of models allowing for exotica are confronted with those of purely
nucleonic counterparts.

Fig. \ref{fig:eth} shows that, similarly to what happens for purely nucleonic matter,
also in matter which accounts for extra particle d.o.f. $e_{th}$ augments with $n_B$ and $T$.
Nucleation of exotic particles makes $e_{th}$ increase with respect
to its value in nucleonic matter and the larger the number of particle d.o.f.
the more significant this increase is.
Indeed, for whatever $T$ and $n_B$,
$e_{th}$ is larger for R(DD2YDelta) than for OMHN(DD2Y), which in turn is larger than for BHB(DD2L),
which in turn is larger than for HS(DD2);
predictions of $e_{th}$ by MBB(DD2K) and HS(DD2);
IOTSY(TM1Y30pi), IOTSY(TM1Y30), STOS(TM1L) and STOS(TM1);
SFHPST(TM1b139) and STOS(TM1) obviously comply to the same rule.
The steep (smooth) increase of exotic particle abundances at low (high) temperature results
in an abrupt (smooth) variation of $e_{th}(n_B)$.
Confrontation of predictions of MBB(DD2K)~\cite{Malik_EPJA_2021} with those of
HS(DD2)~\cite{Hempel_NPA_2010} shows that significant effects related to $K^-$
are obtained only at temperatures of the order of several tens MeV.
Confrontation of predictions of IOTSY(TM1Y30pi) and IOTSY(TM1Y30)~\cite{Ishizuka_JPG_2008} shows that even at
high temperatures the pions play little role.
Within BBKF(DD2-SF)1.8~\cite{Bastian_PRD_2021} $e_{th}$ in quark matter is very high
and, in the mixed phase, increases linearly with $n_B$;
the fact that the temperature affects more the quark matter than its
hadron counterpart is due to the low mass of quarks.
In BBKF(DD2-SF) models~\cite{Bastian_PRD_2021} pure hadron and quark phases in coexistence have equal values of $T$,
$P$, $\mu_B$ and $Y_Q$ (instead of $\mu_Q$).
The latter feature explains that along trajectories of constant-$Y_Q$ quantities calculated in the coexistence
phase as linear combinations of values at the borders, as it is the case of $e_{th}$, manifest linear behaviors.
In what regards the width and localization of the phase coexistence domain, no significant $T$-dependence is observed.
Relying on a Glendenning construction \cite{Glendenning_PRD_1992} SFHPST models~\cite{Sagert_JPG_2010}
  manifest other features than those previously seen for BBKF(DD2-SF) models~\cite{Bastian_PRD_2021}.
  In the mixed phase $e_{th}$ is not a linear function of $n_B$;
  straightforward identification of density domains populated by different phases is not possible;
  the curves corresponding to SFHPST models resemble qualitatively the curves of models with heavy baryons,
which feature no phase transition.
Results of BBKF(DD2-SF)1.8 and SFHPST are similar only in the very high values of $e_{th}$ obtained at
$T=50~{\rm MeV}$. We also note that the model with small value of the bag constant, which allows for an
early transition to quark matter, manifests also larger thermal contributions.

We have seen in Fig. \ref{fig:enp} that under specific thermodynamic conditions 
$p_{baryon}^{(N\Lambda)}$ and $p_{baryon}^{(N\Delta)}$ decrease with $T$, which translates
into negative values for the baryonic component of $p_{th}$.
Figs. \ref{fig:pth} confirms that nucleation of exotic species results in a strong
reduction of the baryonic component of $p_{th}$ and that under specific conditions this
quantity becomes negative.
Competition with increasing values of baryonic chemical potential, which favor higher
$p_{th}$-values, explains the strongly fluctuating behavior of this quantity
over the considered density domain for all EoS models which account for heavy baryons and $K^-$.
The quark phase in BBKF(DD2-SF)1.8~\cite{Bastian_PRD_2021} is characterized also by large values of
$p_{th}$, in addition
to the large values of $e_{th}$ discussed earlier and, with the exception of a narrow domain neighboring
the hadron-quark coexistence, $p_{th}$ increases with $n_B$.
Equality of temperature, pressure and $Y_Q$ between pure hadron and quark phases, assumed by the (approximate) Gibbs
construction, explains that in the coexistence domain $p_{th;baryon}(n_B)$ curves at constant-$T$ feature a plateau.
The narrow peak at the border between phase coexistence and quark phase is likely a numerical artifact.
No plateau is present in the mixed phase domain of SFHPST-models. This is due to the fact that
mechanical equilibrium is imposed not among pure phases at the boundaries of the phase coexistence
region but between different phases in the coexistence domain.
The evolution with $n_B$ is smooth but not necessarily monotonic.
As for $e_{th}$, models with small bag constants lead to large thermal effects on $p_{th}$.
Overall $p_{th}$ manifests strong dependence on EoS, temperature and baryonic density.
The only situation in which a particle is seen to play a little role corresponds
to pions at temperatures of the order of a few tens MeV or lower.

\subsection{Thermal index}
\label{ssec:Gammath}

The strong EoS- and density dependence of $e_{th}$ and $p_{th}$ seen in Paper I
spurred us to further investigate the reliability of the so-called $\Gamma_{th}$-law,
customarily employed in numerical simulations \cite{Hotokezaka_PRD_2013,Bauswein_PRD_2010,Endrizzi_PRD_2018,Camelio2019,Weih_PRL_2020}
in order to supplement cold EoS with finite temperature contributions.
The behavior of the $\Gamma$-factor defined as
\begin{equation}
  \Gamma_{th}=1+\frac{p_{th}}{e_{th}}~,
  \label{eq:Gammath}
\end{equation}
was analyzed separately for baryonic and stellar matter.
In each case values of realistic EoS models have been compared with limiting
values of idealized systems as well as with the domain of values, $1.5 \leq \Gamma_{th} \leq 2$,
employed in simulations.
The conclusions of Paper I dealing with nucleonic models can be summarized as follows:
i) at high temperatures and low densities, where the ideal gas behavior is recovered,
$\Gamma_{th;baryon} \to 5/3$, where $5/3$ is the classical limit for a dilute gas,
ii) relativistic models provide for high $n_B$-values, $\Gamma_{th;baryon} \approx 4/3$, where
$4/3$ corresponds to ultra-relativistic gases,
iii) at high densities predictions of relativistic and non-relativistic models differ
qualitatively and quantitatively,
iv) over wide but EoS-dependent domains of $(n_B,T)$, $\Gamma_{th;tot}>2$ or
$\Gamma_{th;tot}<1.5$.


\begin{figure*}
  \begin{center}
    \includegraphics[width=0.99\textwidth]{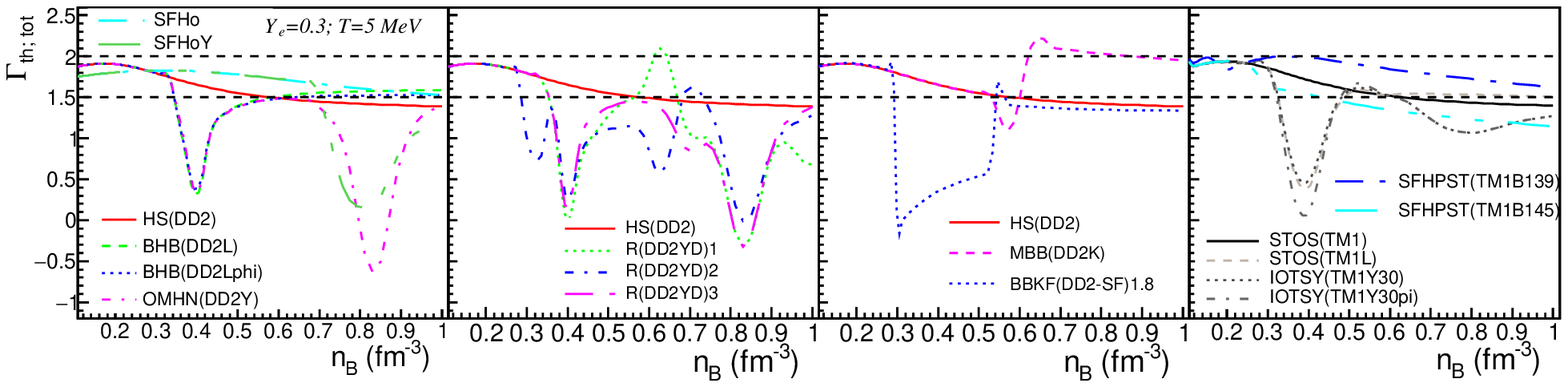}
    \includegraphics[width=0.99\textwidth]{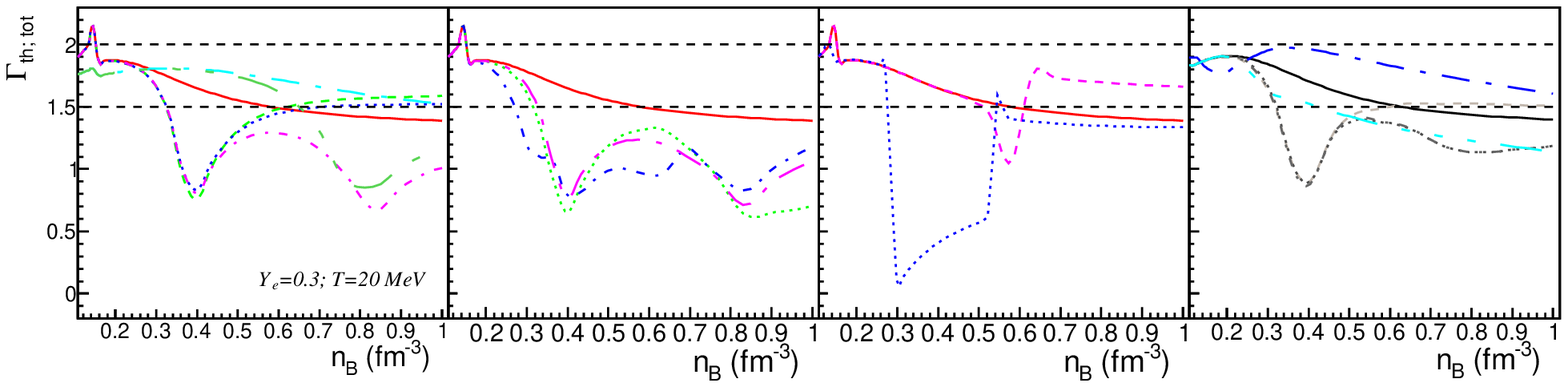}
    \includegraphics[width=0.99\textwidth]{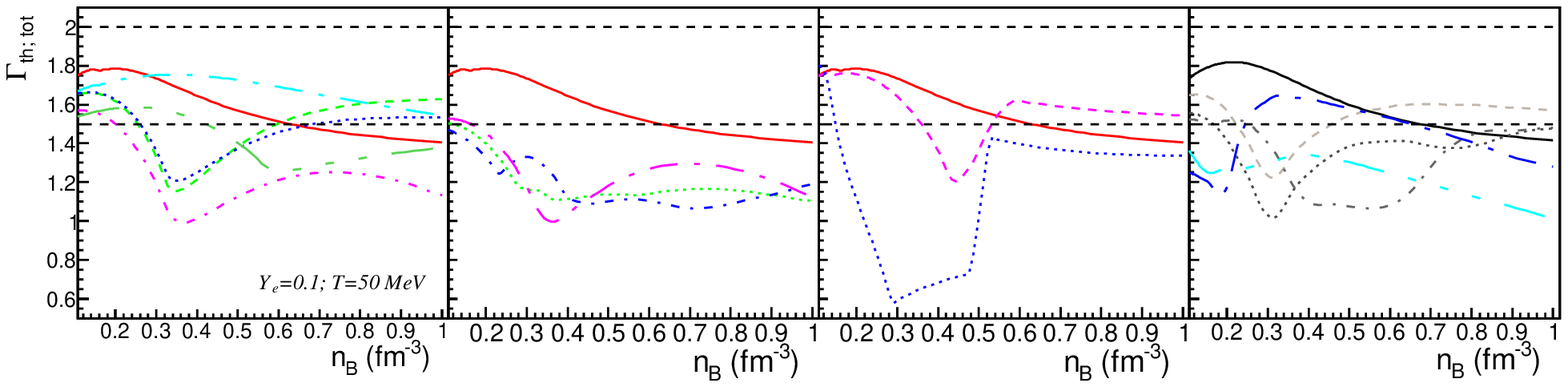}
  \end{center}
  \caption{$\Gamma_{\mathit{th}}$, eq.(\ref{eq:Gammath}), as function of baryon number density at
    ($T$=5 MeV, $Y_e=0.3$), ($T$=20 MeV, $Y_e=0.3$), ($T$=50 MeV, $Y_e=0.1$).
    Results corresponding to stellar matter (contributions of lepton and photon gases are
    included).
    Dashed horizontal lines mark the values $\Gamma_{\mathit{th}}=1.5$ and 2.
  }
  \label{fig:Gammathtot}
\end{figure*}

The behavior of the $\Gamma_{th}$ factor for stellar matter is investigated
in Fig. \ref{fig:Gammathtot}. Predictions corresponding to 
different EoS models and various thermodynamic conditions are considered.
As in Paper I, we obtain a strong EoS-, $T$- and $n_B$-dependence
of $\Gamma_{th;tot}$. The minima in $p_{th}(n_B)$, see Fig. \ref{fig:pth}, lead to one or several
minima in $\Gamma_{th;tot}(n_B)$; the negative values of $p_{th}$ lead, for the lowest considered
temperature, to negative values of $\Gamma_{th;tot}$.
The hadron-quark coexistence region of BBKF(DD2-SF)1.8 singles out by $\Gamma_{th}$-values
significantly smaller than those obtained in pure hadron and quark phases.
The SFHPST-models show rather smooth but non-monotonic $\Gamma_{th;tot}(n_B)$ curves.

The upper value $\Gamma_{th}=2$ used in simulations is exceeded only over narrow density domains
and only by some of the models allowing for $\Delta$, $K^-$ and quarks;
at variance with this all models predict $\Gamma_{th}<1.5$ over broad ranges of densities, 
where 1.5 is the lower limit assumed in simulations.
The departure from $1.5 \leq \Gamma_{th}$ is more significant for exotic matter
than it was for nucleonic matter.

\subsection{Chemical potentials}
\label{ssec:mus}

\begin{figure*}
  \begin{center}
    \includegraphics[width=0.99\textwidth]{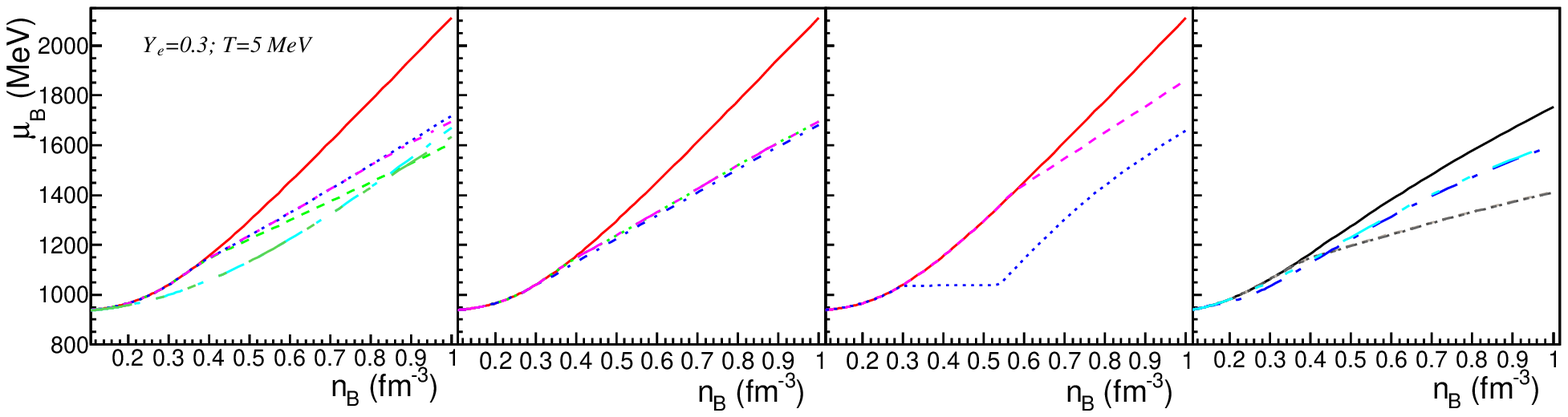}
    \includegraphics[width=0.99\textwidth]{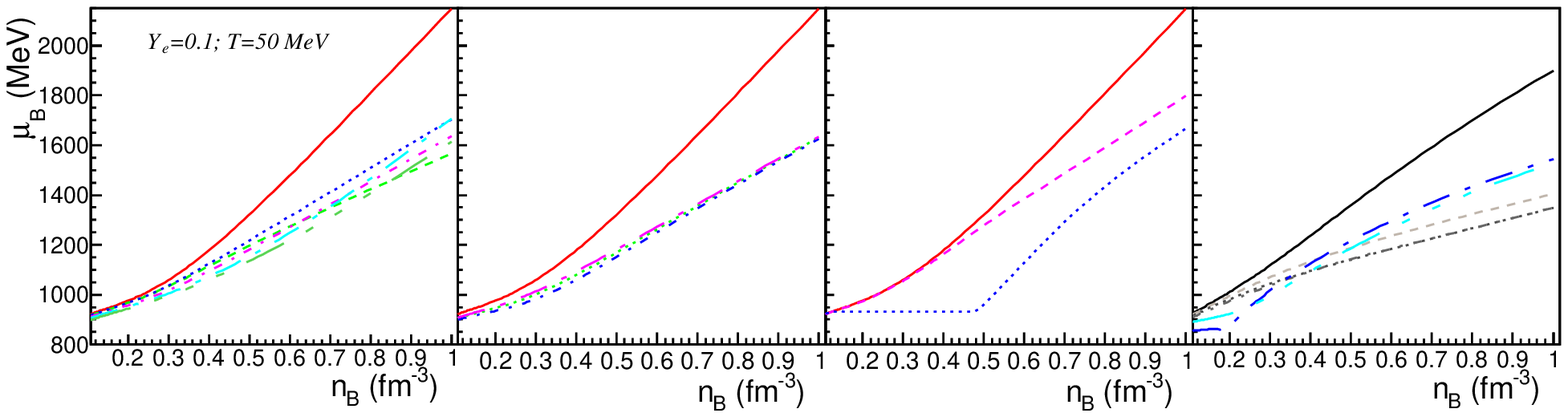}
  \end{center}
  \caption{Baryon chemical potential as function of baryon number density at
    ($T$=5 MeV, $Y_e=0.3$) and ($T$=50 MeV, $Y_e=0.1$). The key legend is as in Fig. \ref{fig:eth}. 
  }
  \label{fig:muB}
\end{figure*}

\begin{figure*}
  \begin{center}
    \includegraphics[width=0.99\textwidth]{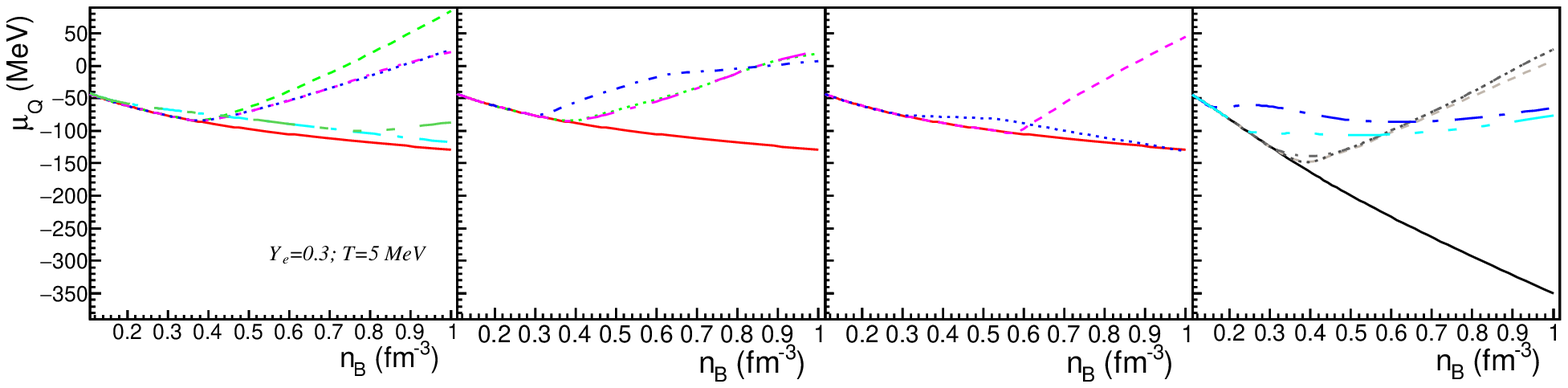}
    \includegraphics[width=0.99\textwidth]{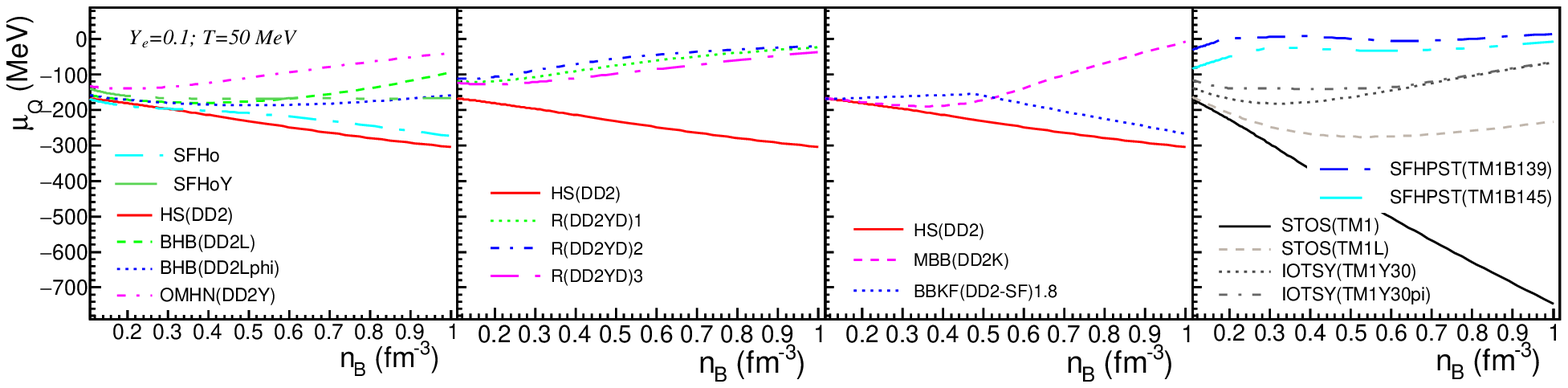}
  \end{center}
  \caption{The same as in Fig. \ref{fig:muB} for charge chemical potential.
  }
  \label{fig:muQ}
\end{figure*}

The impact of composition changes on baryon and charge chemical potentials is
addressed in Figs. \ref{fig:muB} and \ref{fig:muQ}.
Fig. \ref{fig:muB} confirms what we have seen in Fig. \ref{fig:mus} (left panel)
for the simpler cases of $N\Lambda$ and $N\Delta$ matter. Namely that exotic
particles are created at the cost of neutrons, which make $\mu_B=\mu_n$ decrease.
At low temperatures, predictions of OMHN(DD2Y) are identical to those of BHB(DD2Lphi) and
differ from those of BHB(DD2L); this suggests that the repulsion mediated by the exchange of $\phi$ mesons
dominates over modifications induced by $\Sigma$s and $\Xi$s.
Predictions of other hyperonic models based on the same nucleonic EoS
model are very similar (slightly different) at low (high) $T$.
At both low and high temperatures the three models corresponding to $NY\Delta$-matter provide
almost the same $\mu_B(n_B)$ in spite of notable differences in chemical composition.
The equality of baryon chemical potentials and $Y_Q$-values in the hadron and quark phases imposed by the
approximate Gibbs construction in BBKF(D2-SF)1.8 explains the plateau in the
curves belonging to this model;
by increasing $T$ this plateau gets wider and shifts to lower densities;
the information thus extracted from $\mu_B(n_B)$ at constant $T$ corrects
the one previously extracted from the less thermodynamically relevant $e_{th}$ and $p_{th}$
as function of $n_B$ at constant $T$, see Figs. \ref{fig:eth} and \ref{fig:pth}.
No particular structure suggestive for phase coexistence manifests  
 in the curves of SFHPST-models.

The partial replacement of neutrons by exotic particles also makes $|\mu_Q|$ decrease, see
Fig. \ref{fig:muQ}.
The most dramatic modifications are obtained for models with large-$L$ values.
As with $\mu_B(n_B)$, at $T=5~{\rm MeV}$ the predictions of OMHN(DD2Y) coincide with those of BHB(DD2Lphi) and
differ from those of BHB(DD2L).
At $T=50~{\rm MeV}$ and $Y_Q=0.1$
the largest departure from the predictions of nucleonic models is obtained for the
following models: 
OMHN(DD2Y), the three R(DD2YD)-models; IOTSY(TM1Y), IOTSY(TM1Ypi); MBB(DD2K);
the two SFHPST-models.
For the first listed models the explanation relies in the presence of
negatively charged exotic species.
For SFHPST(TM1) the explanation relies in the negatively charged strange quark.
The fact that in BBKF(D2-SF)1.8 the phase construction is performed at constant
$Y_Q$ explains why $\mu_Q$ is not constant in the hadron-quark coexistence region.
The similar charge fractions of the coexisting phases~\cite{Bastian_PRD_2021} nevertheless makes that
the slope of $\mu_Q(n_B)$ is small. 
As for $\mu_B(n_B)$,  $\mu_Q(n_B)$ of SFHPS-models feature no plateau.

We also note that in warm $Y$-, $Y\Delta$- and $K^-$-admixed matter
with densities exceeding a certain value $\mu_Q>0$, which means that $n_n<n_p$. 
Over two density domains also SFHPST(TM1B139) predicts that for $T=50~{\rm MeV}$ and $Y_Q=0.1$
$\mu_Q \gtrsim 0$; whenever this is the case the number of up quarks exceeds the one of down quarks.

Before leaving this section let us note that, for nucleonic models, EoS stiffness and density
dependence of the symmetry energy impact the high density values of baryon and charge chemical potentials.
Indeed, the highest (smallest) value of $\mu_B$ corresponds to DD2 (TM1), which provides the stiffest
(softest) model.
Similarly, for isospin asymmetric nuclear matter DD2 (TM1), with a low (high) $L$-value, provides small
(large) values of $|\mu_Q|$.

\subsection{Entropy, specific heats, adiabatic index, speed of sound}
\label{ssec:entropy}

\begin{figure*}
  \begin{center}
    \includegraphics[width=0.99\textwidth]{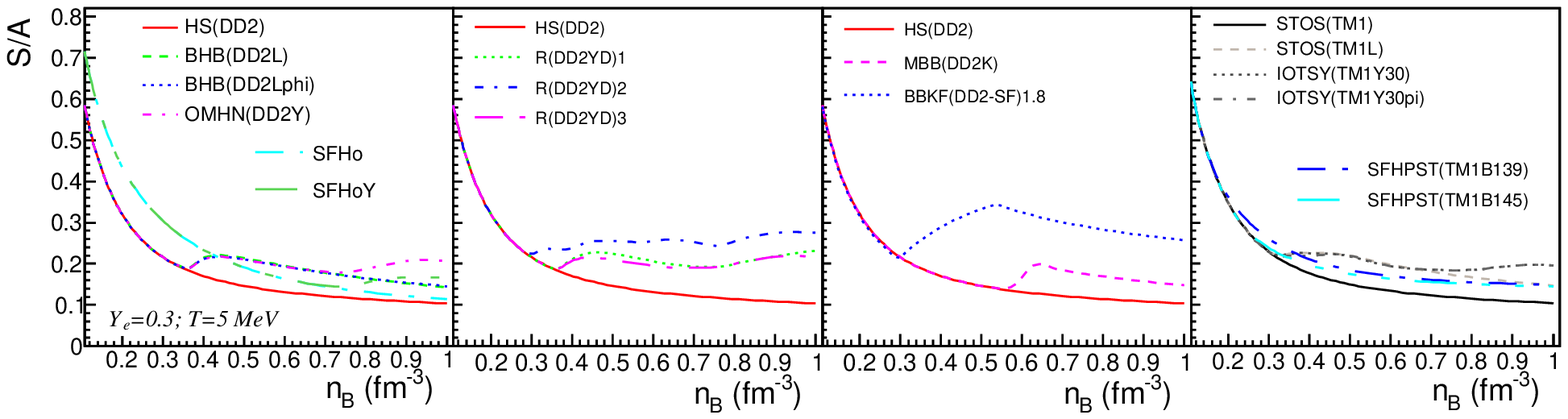}
    \includegraphics[width=0.99\textwidth]{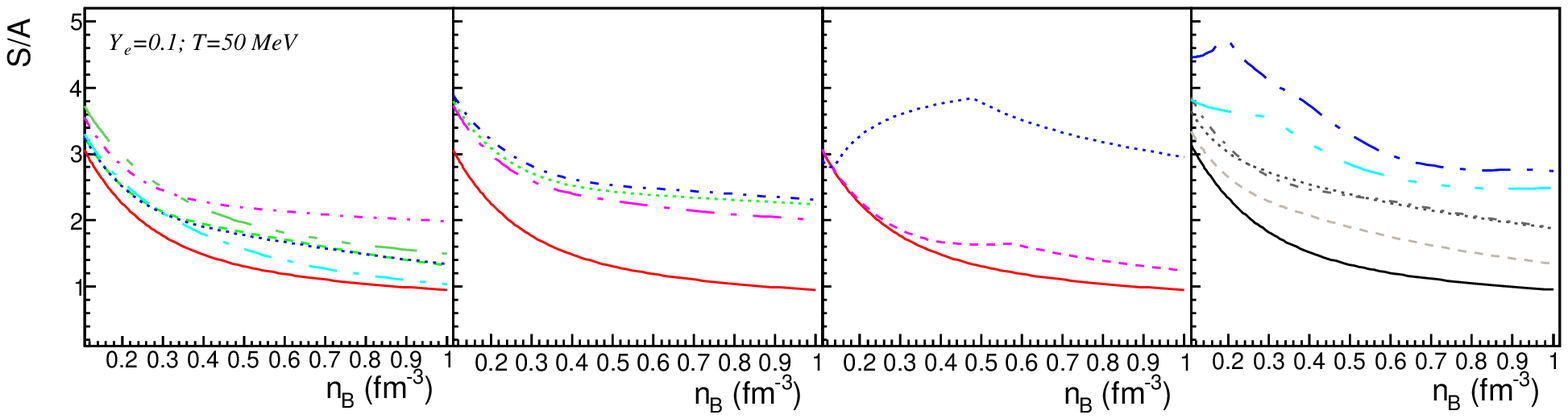}
  \end{center}
  \caption{Baryon (quark) contribution to the entropy per baryon as function
  of baryon number density for ($T$=5 MeV, $Y_e=0.3$) and
  ($T$=50 MeV, $Y_e=0.1$). 
  }
  \label{fig:SperA}
\end{figure*}

Further insight into the impact of exotic particle d.o.f. on
thermodynamic quantities is offered in Fig. \ref{fig:SperA}.
It shows the baryonic entropy per baryon for ($T$=5 MeV, $Y_e=0.3$) and
($T$=50 MeV, $Y_e=0.1$). As already seen and commented at length in Paper I,
at fixed $T$ in models with nucleonic d.o.f. $S/A$ decreases with $n_B$.
The same holds true for models with heavy baryons and/or $K^-$, $\pi$ at high temperatures.
At fixed $T$ and $n_B$, $S/A$ increases with the number of particle species.
The steep increase in the abundance of newly populated species,
as it happens with heavy baryons and $K^-$ at low $T$, results in bumps of $S/A(n_B)$.
As it was the case with $e_{th}(n_B)$ pions do not alter $S/A(n_B)$ when added to hyperons
within IOTSY with $U_{\Sigma}^{(N)}=30~{\rm MeV}$.
The non-monotonic $S/A(n_B)$ in BBKF(DD2-SF)1.8 results from the first order transition between
the hadron and quark phases, the latter being characterized by much larger values of $S/A$. 
Curves belonging to SFHPST-models are qualitatively similar to those produced by models
with heavy baryons. 
The dispersion between the different curves increases
with $T$ as so does the dispersion between chemical compositions.
At high-$T$ models accounting for quarks deviate the most from the predictions of the
nucleonic counterparts.

\begin{figure*}
  \begin{center}
    \includegraphics[width=0.99\textwidth]{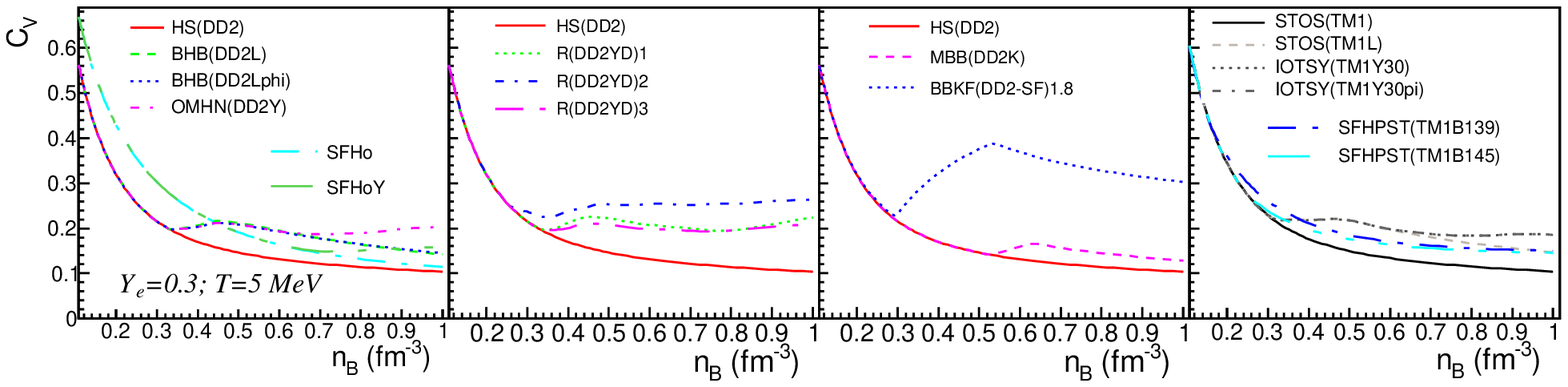}
    \includegraphics[width=0.99\textwidth]{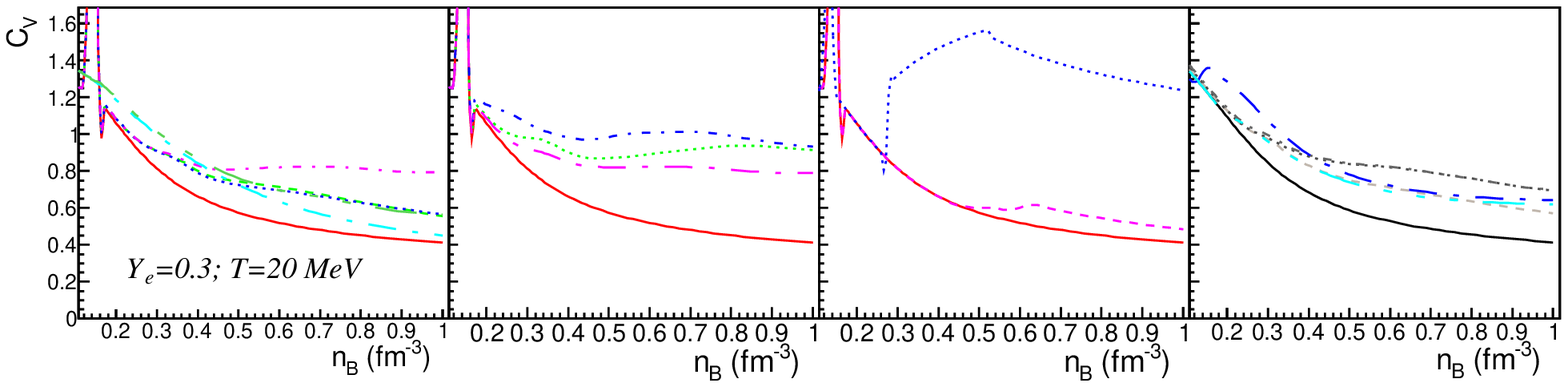}
    \includegraphics[width=0.99\textwidth]{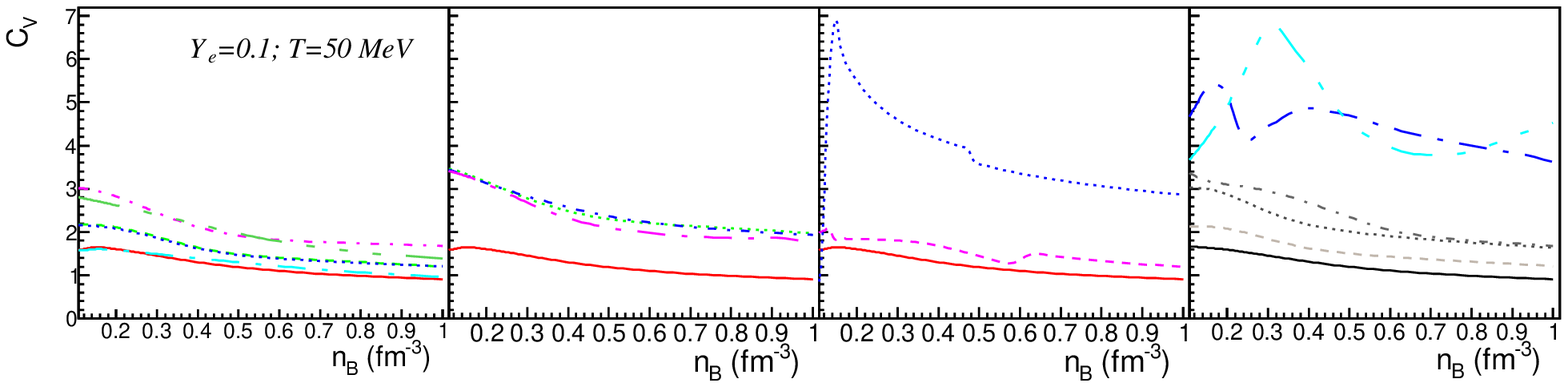}
  \end{center}
  \caption{Specific heat at constant volume, $C_V$, as function of baryon
    number density for baryon (quark) matter for
    ($T$=5 MeV, $Y_e=0.3$), ($T$=20 MeV, $Y_e=0.3$) and ($T$=50 MeV, $Y_e=0.1$).
  }
  \label{fig:Cv}
\end{figure*}

The specific heat at constant volume is defined as
\begin{equation}
  C_V=T \frac{\partial \left( S/A \right)}{\partial T}|_{V,\{N_i\}}~.
  \label{eq:cv}
\end{equation}

Fig. \ref{fig:Cv} illustrates the evolution of $C_V$
as function of baryon number density for different thermodynamic conditions.
As easy to anticipate based of the definition, eq. (\ref{eq:cv}), $C_V(n_B)$
replicates the behavior of $S/A(n_B)$.
Under specific thermodynamic conditions models which employ for the sub-saturation density
domain the extended NSE calculations in \cite{Hempel_NPA_2010}
show discontinuities over $n_{sat}/2 \lesssim n_B \lesssim n_{sat}$.
These are numerical artifacts of the way in which the transition from clusterized
to homogeneous matter was dealt with and have no physical ground.

\begin{figure*}
  \begin{center}
    \includegraphics[width=0.99\textwidth]{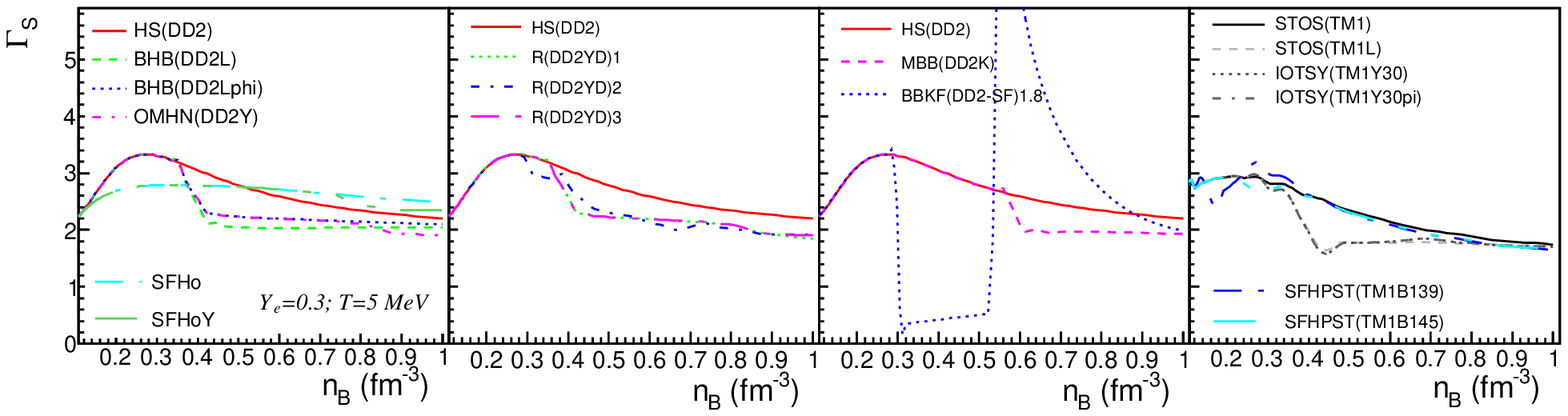}
    \includegraphics[width=0.99\textwidth]{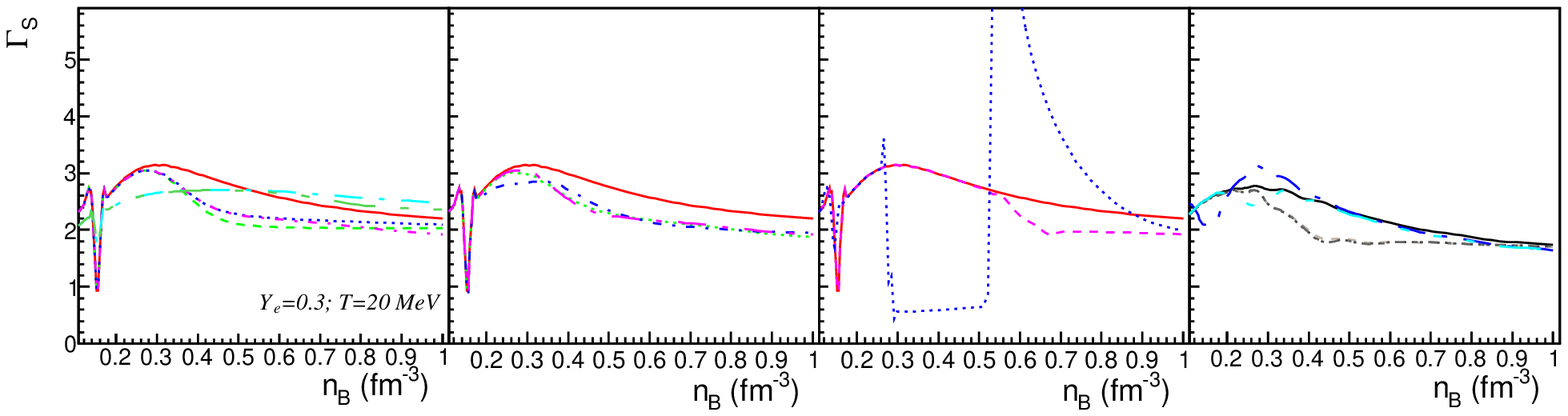}
    \includegraphics[width=0.99\textwidth]{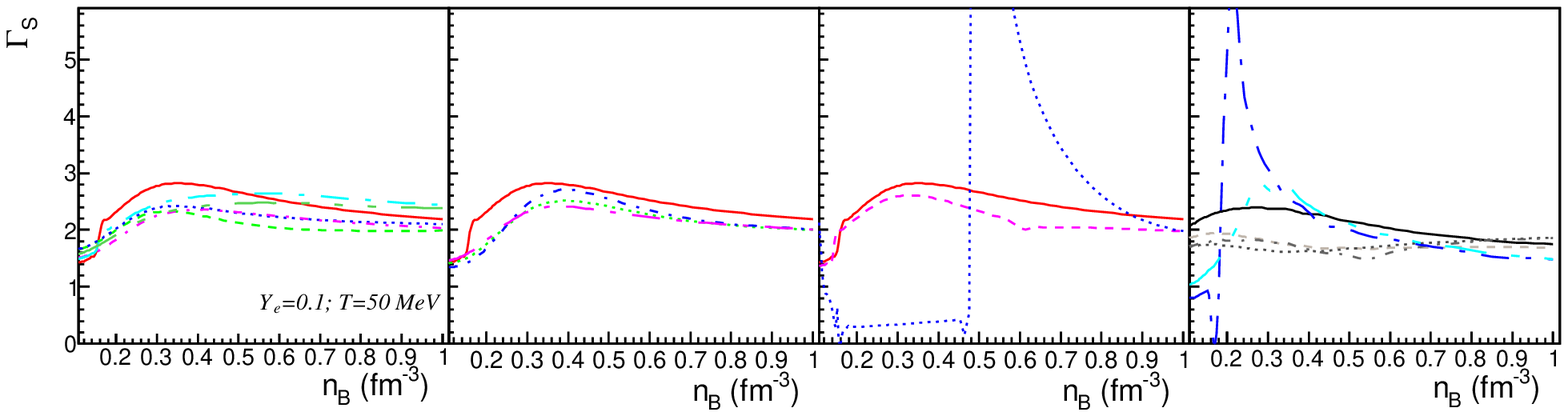}
  \end{center}
  \caption{$\Gamma_S$, eq.~(\ref{eq:GS}), as function of baryon number density
  for ($T$=5 MeV, $Y_e=0.3$),
  ($T$=20 MeV, $Y_e=0.3$) and ($T$=50 MeV, $Y_e=0.1$).
  Results corresponding to stellar matter, as predicted by various EoS models.
  }
  \label{fig:GammaS}
\end{figure*}

The behavior of the adiabatic index, defined as
\begin{equation}
  \Gamma_{S}=\frac{\partial \ln P}{\partial \ln n_B}|_{S}=
  \frac{C_P}{C_V} \frac{n_B}{P} \frac{\partial P}{\partial n_B}|_T~,
   \label{eq:GS}
\end{equation}
is illustrated in Fig.~\ref{fig:GammaS} for stellar matter,
\textit{i.e.} with the contributions of leptons and photons included.
The predictions by various EoS models at different thermodynamic conditions are
considered. $\Gamma_S$ gives an indication about the stiffness of the
EoS in all processes occuring at constant entropy.

With the exception of the transition density from clusterized to homogeneous matter
$\Gamma_S(n_B)$ of nucleonic EoS models is a smooth function.
As discussed in Paper I, nucleonic models manifest a strong dependence of $\Gamma_S$
on EoS-model and $n_B$.
Fig.~\ref{fig:GammaS} shows that at low (high) $T$, where abundances of heavy baryons and
$K^-$ increase steeply (slowly) with baryonic density,
EoS models which account for these particles provide $\Gamma_S(n_B)$ with sudden (smooth)
changes of slope.
The EoS softening produced by hyperons, $\Delta$s and $K^-$
leads to $\Gamma_S^{exotic}<\Gamma_S^{nucleonic}$.
Predictions of IOTSY(TM1Y30) and IOTSY(TM1Y30pi) are identical or, for the highest $T$, very similar
to those of STOS(TM1L).
Also similar are the predictions of the three models that account for $Y\Delta$
introduced in this work.
This suggests that modifications in heat capacities, see Fig. \ref{fig:Cv},
are canceled out by modifications in $\left(\partial \ln P/\partial\ln n_B\right) |_T$.
EoS-models with transitions to quark matter show different features depending on how
the transition was dealt with.
In the hadron-quark coexistence BBKF-models present a plateau. It stems from the mechanical equilibrium
between phases in the Gibbs construction.
The high peak at densities slightly exceeding the high density border of phase coexistence
is due to the stiff behavior of $P(e)|_T$.
At temperatures of the order of a few tens MeV or lower, $\Gamma_S(n_B)$ predicted
by SFHPST-models resembles the one of models with heavy baryons and $K^-$.
At temperatures of the order of several tens MeV, $\Gamma_S(n_B)$ of SFHPST(TM1B139)
has a high peak at $n_B \approx 1-2 n_{sat}$.
Quite remarkably, at the highest densities the scattering between different models
is limited.

\begin{figure*}
  \begin{center}
    \includegraphics[width=0.99\textwidth]{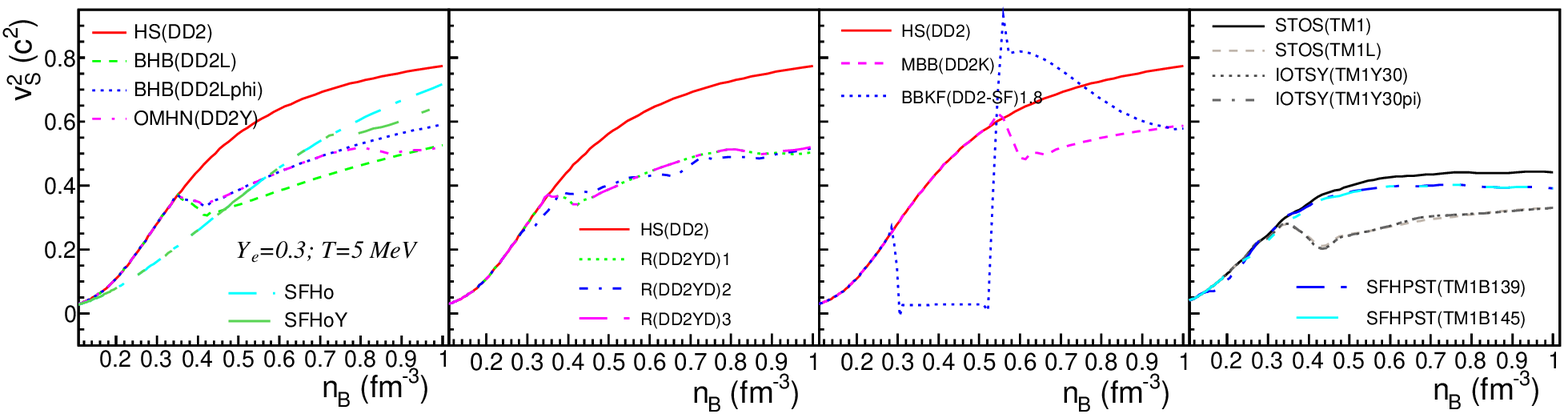}
    \includegraphics[width=0.99\textwidth]{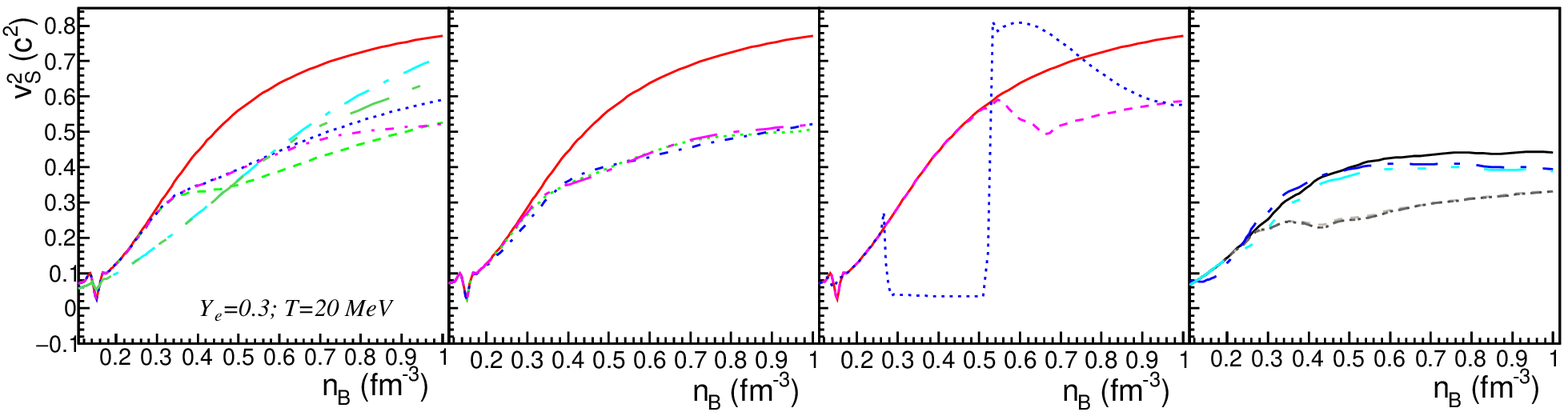}
    \includegraphics[width=0.99\textwidth]{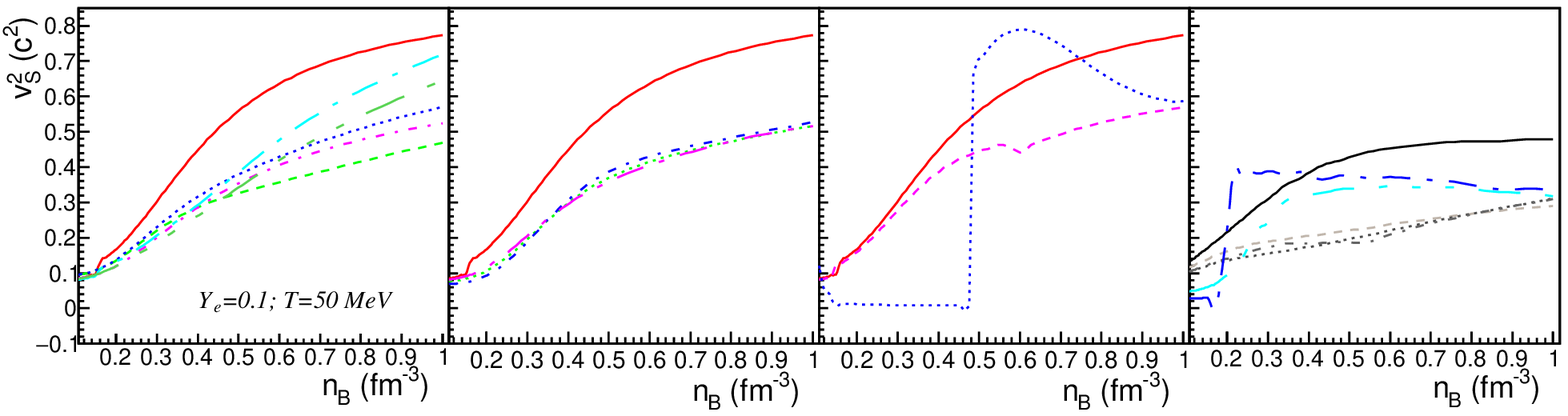}
  \end{center}
  \caption{Speed of sound squared, eq. (\ref{eq:cs2}), as function of
  baryon number density in stellar matter for various EoS models.
  The considered thermodynamic conditions are:
  ($T$=5 MeV, $Y_e=0.3$), ($T$=20 MeV, $Y_e=0.3$) and 
  ($T$=50 MeV, $Y_e=0.1$). 
  }
  \label{fig:cs2}
\end{figure*}

A key quantity in dynamical numerical simulations is the speed of sound $v_S$.
In units of $c$, the speed of light, it is given by
\begin{equation}
  v_S^2=\frac{dP}{de}|_{S,A,Y_e}=\Gamma_S \frac{P}{e+P}~.
  \label{eq:cs2}
\end{equation}

Fig. \ref{fig:cs2} illustrates its behavior as function of baryon
number density within the different EoS models. Different
thermodynamic conditions, indicated in the figures, are considered.
The three nucleonic models show a strong increase of $v_S$ with $n_B$ and limited
sensitivities to the other thermodynamic variables, $T$ and $Y_Q$.
The behavior of $v_S(n_B)$ in DD2 and TM1 differs qualitatively from the one in SFHo:
while in DD2 and TM1 the increase gets attenuated at $n \approx 4 n_{sat}$,
no such effect is seen for SFHo.  
This situation is attributable to the mixed coupling terms in the latter model.
At high densities DD2 provides for the speed of sound a value almost twice larger than
the one provided by the softer TM1 model.
The CDFT formulation prevents violation of causality.
In all circumstances models allowing for hyperons, $\Delta$s, $\pi$ and $K^-$ provide
smaller $v_S(n_B)$-values than their nucleonic counterparts.
As it was the case with previously considered quantities, sudden modifications in
abundances of exotic species are translated into sudden modifications of the speed of sound.
The $T$-dependence of $v_S$ remains small as in the considered cases
hyperons, $\Delta$s, $\pi$ and $K^-$ are sub-dominant.
As for $\Gamma_S$ to which $v_S$ is related, chemical equilibrium among hadron and quark
phases in BBKF-models leads to vanishing values of the speed of sound.
Moreover, in the pure quark phase, BBKF predicts that $v_S$ decreases with $n_B$.
The only other case in Fig. \ref{fig:cs2} where such a behavior is obtained corresponds to SFHPST(TM1B139) and
SFHPST(TM1B145) at the highest considered temperature.
Detailed investigation in \cite{Aloy_MNRAS_2019} probes that the same holds true if, instead
of constant-$T$ curves, one considers constant $S/A$-curves.
We nevertheless stress that also the nucleonic EoS SRO(SkAPR) \cite{SRO_PRC_2017} provides for the
  squared speed of sound decreasing evolution with $n_B$ \cite{Raduta_EPJA_2021};
  this means that one should not a priori consider such a behavior as a signature of quark matter.
Predictions of IOTSY(TM1Y30) and IOTSY(TM1Y30pi) are identical or very close to those of STOS(TM1L).
A very small dispersion is obtained also among the predictions of the three models
with hyperons and $\Delta$s.

\section{Conclusions}
\label{sec:Concl}

Within CDFT we have constructed three general purpose EoS models
with hyperons and $\Delta$-resonances ready to use in astrophysical
simulations and made them available on the \textsc{CompOSE}
repository (\url{https://compose.obspm.fr/}); 
to our knowledge these are the first publicly available 3D EoS
databases which account for the $\Delta$-quadruplet.
The domains of temperature, $0.1~{\rm MeV} \leq T \leq 100 ~{\rm MeV}$,
baryonic particle number density,
$10^{-12} ~{\rm fm}^{-3} \leq n_B \leq 1.1 ~{\rm fm}^{-3}$
and charge fraction $0.01 \leq Y_Q \leq 0.6$ for which
data are provided make them suitable for studies of CCSN and BNS.
Also available are databases corresponding to purely baryonic matter, which
may be used for heavy ion collision studies.
Our models have been built such as to comply with experimental data from
nuclear and hyper-nuclear physics; ab initio calculations of neutron matter;
astrophysical observations of cold compact stars and - at the same time -
partially account for uncertainties related to the population
of $\Delta$s.

As previously discussed in \cite{Raduta_PLB_2021}, for certain combinations
of coupling constants of the $\Delta$ to mesonic fields, the nucleation
of $\Delta$s may result in thermodynamic instabilities or limit the validity
domain of the model by preventing the baryonic number density to exceed
values of the order of $2-3 n_{sat}$.
The latter issue arises because the Dirac effective mass of nucleons vanishes.
The present work complements the study performed in \cite{Raduta_PLB_2021},
and which focuses on cold $\beta$-equilibrated matter and persistence of $\Delta$-driven
instabilities at finite-$T$,
by considering cold $\Delta$-admixed nuclear matter with extreme values of charge
fraction $Y_Q=0.01$ and 0.5. Present results show that the parameter space which allows 
for instabilities gets wider as $Y_Q$ diminishes and that, for matter with $Y_Q=0.5$,
purely baryonic and stellar matter exhibit instabilities for the same values of
$(x_{\sigma \Delta}, x_{\omega \Delta})$. Both findings are in contrast with the phenomenology
of dilute nuclear matter and understandable considering the $\Delta$'s abundances.

We have then reviewed the general purpose EoS models with exotic d.o.f. presently available
on \textsc{CompOSE}. The information provided for each model includes
the nucleonic effective interaction together with the values of the nuclear matter parameters; 
the exotic d.o.f.; for CDFT models, information on various mesonic fields and values of the
hyperon and $\Delta$ well depth potentials in symmetric saturated matter on which some
meson coupling constants have been tuned; properties of cold $\beta$-equilibrated NS
and compliance with constraints from compact star observations.

The role of hyperons, $\Delta$s, pions, kaons and quarks as well as the consequences
of the way in which the hadron-quark phase transition was dealt with have been investigated
by analyzing the thermal properties for a number of models; wide ranges of
baryonic number density $n_{sat} \lesssim n_B \leq 1~{\rm fm}^{-3}$,
temperature $5~{\rm MeV} \leq T \leq 50~{\rm MeV}$ have been considered for
charge franctions $Y_Q=0.1$ and 0.3.
In agreement with previous studies we show that extra d.o.f. modify the values of
baryon and charge chemical potentials and the modification depends upon the employed
nucleonic effective interaction;
for fixed values of $T$, $Y_Q$ and $n_B$ the entropy per baryon increases with the number of
species; the behavior of $C_V(n_B)$ replicates the one of $S/A(n_B)$;
EoS softening upon nucleation of exotica results in decreased values of the speed of sound;
quark matter is characterized by $v_S^2$ decreasing with $n_B$.
Other results concern the thermal energy density and pressure and the thermal index.
Specifically we show that exotic d.o.f. result in thermal energy densities enhanced with respect
to what is obtained in purely nucleonic matter;
a strongly oscillatory behavior of the thermal pressure and thermal index and,
under specific conditions, negative values of $p_{th}$ and $\Gamma_{th;tot}$.
The latter feature suggests that the use of the $\Gamma$-law for supplementing
cold EoS with thermal components is even less advisable for exotic matter than for nucleonic matter.
Last but not least we have shown that the way in which the transition from hadronic to quark
matter is dealt with impacts on a series of quantities like $p_{th}$, $\Gamma_{th}$, $\mu_B$, $\Gamma_S$,
$v_S^2$. BBKF-models show - in the phase coexistence domain -
plateaus for all the above listed quantities while no such behavior is obtained in the case of
SFHPST-models.

{\bf Data Availability Statement} his manuscript has no associated data
or the data will not be deposited.

[Authors’ comment: This manuscript
has no associated data as that all data are already available on the
\textsc{CompOSE} site {\url{https://compose.obspm.fr/}}].

\begin{acknowledgements}
  The author gratefully acknowledges discussions with M. Oertel,
  C. Providencia and A. Sedrakian.
  Support from a grant of the Ministry of Research,
  Innovation and Digitization, CNCS/CCCDI – UEFISCDI, Project
  No. PN-III-P4-ID-PCE-2020-0293 is also acknowledged.
  This work has been partially funded by the European COST Action CA16214 PHAROS
  ''The multi-messenger physics and astrophysics of neutron stars''. 
\end{acknowledgements}

\begin{appendix}{Appendix A}

  See Table \ref{tab:grids}.
  
\begin{table}
  \caption{Domains of temperature, baryonic number density and
    charge fraction covered
    by the six R(DD2YDelta) EoS tables proposed in this work
    and the corresponding numbers of mash points.}
  \label{tab:grids}
 \begin{tabular}{llll}
   \hline\noalign{\smallskip}
   & $T$ & $n_B$ & $Y_Q$ \\
   & [MeV] & [${\rm fm}^{-3}$] & \\
   \noalign{\smallskip}\hline\noalign{\smallskip}
   Number of points & 76 & 302 & 60 \\
   Minimum & 0.1 & $10^{-12}$ & 0.01 \\
   Maximum & 100 & 1.1 & 0.60 \\
   \noalign{\smallskip}\hline\noalign{\smallskip}
\end{tabular}
\end{table}

\end{appendix}  

\bibliographystyle{spphys}       
\bibliography{ComposeII_v2.bib}   

\end{document}